\documentclass[12pt]{article}
\usepackage{graphicx} 

\usepackage[margin=2cm]{geometry}

\usepackage{tikz}

\usepackage{amsmath}
\usepackage{amsfonts}

\usepackage[parfill]{parskip}

\usepackage{algorithm}
\usepackage{algpseudocode}

\usepackage{subcaption}

\usepackage{hyperref}

\usepackage{bm}

\usepackage{natbib}
\usepackage[font=bf]{caption}

\algnewcommand\algorithmicswitch{\textbf{switch}}
\algnewcommand\algorithmiccase{\textbf{case}}

\algdef{SE}[SWITCH]{Switch}{EndSwitch}[1]{\algorithmicswitch\ #1\ \algorithmicdo}{\algorithmicend\ \algorithmicswitch}%
\algdef{SE}[CASE]{Case}{EndCase}[1]{\algorithmiccase\ #1}{\algorithmicend\ \algorithmiccase}%
\algtext*{EndSwitch}%
\algtext*{EndCase}%

\newtheorem{theorem}{Theorem}
\newtheorem{proposition}[theorem]{Proposition}%

\title{Non-centering for discrete-valued state transition models: an application to ESBL-producing \textit{E. coli} transmission in Malawi}
\author{James Neill,\textsuperscript{$1,\ast$} Rebecca Lester,\textsuperscript{$2,\ast$} Winnie Bakali,\textsuperscript{$3$} Gareth Roberts,\textsuperscript{$4$} \\ Nicholas Feasey,\textsuperscript{$5$} Lloyd A. C. Chapman\textsuperscript{$1,\dagger$} and Chris Jewell\textsuperscript{$1,\dagger$}}
\date{\footnotesize \textsuperscript{1}School of Mathematical Sciences, Lancaster University, Lancaster, LA1 4YF, UK, \textsuperscript{2}Division of Infection and Immunity, University College London, London, WC1E 6BT, UK, \textsuperscript{3}Malawi-Liverpool Wellcome Research Programme, Kamuzu University of Health Sciences, Blantyre, Malawi, \textsuperscript{4}Department of Statistics, University of Warwick, Coventry, CV4 7AL, UK, and \textsuperscript{5}School of Medicine, University of St Andrews, North Haugh, St Andrews, KY16 9TF, UK \\ ~ \\ Corresponding author: \href{email:j.neill@lancaster.ac.uk}{j.neill@lancaster.ac.uk} \\ $\ast$ Joint first authors. $\dagger$ Joint last authors.}

\begin{document}

\maketitle

\begin{abstract}
Infectious disease transmission is often modelled by discrete-valued stochastic state-transition processes. Due to a lack of complete data, Bayesian inference for these models often relies on data-augmentation techniques. These techniques are often inefficient or time consuming to implement. We introduce a novel data-augmentation Markov chain Monte Carlo method for discrete-time individual-based epidemic models, which we call the Rippler algorithm. This method uses the transmission model in the proposal step of the Metropolis-Hastings algorithm, rather than in the accept-reject step. We test the Rippler algorithm on simulated data and apply it to data on extended-spectrum beta-lactamase (ESBL)-producing \textit{E. coli} collected in Blantyre, Malawi. We compare the Rippler algorithm to two other commonly used Bayesian inference methods for partially observed epidemic data, and find that it has a good balance between mixing speed and computational complexity.
\end{abstract}


\section{Introduction} \label{sec_intro}


Stochastic state-transition models of infectious disease transmission are a major pillar of evidence-based disease policy information. In an individual-based epidemic model, the population of interest is partitioned into a discrete set of states that individuals can transition between (where the rate of transition between states for one individual can depend on the current state of other individuals). When fitted to data using statistically principled methods, these models can be used to deduce relevant drivers of infection that allow control interventions to be applied appropriately and efficiently.

Epidemic models fit into the class of hidden Markov models, where, conditional on a set of initial conditions $\bm{x}_0$, parameters $\bm{\theta}$, and the state-transition process $G$, $\bm{X} \sim G(\bm{x}_0, \bm{\theta})$ is latent, but subject to a noisy observation process $H$ such that $\bm{Y} \sim H(\bm{X}, \bm{\phi})$, where $\bm{\phi}$ are a set of parameters such as the sensitivity and specificity of a diagnostic test.  For inference purposes, given that $\bm{X}$ is latent, we require to marginalise the likelihood $L(\bm{Y} | \bm{\theta}, \bm{\phi})$. Denoting the distribution functions for $G$ and $H$ as $g(\bm{X}|\bm{\theta})$ and $h(\bm{Y}|\bm{X},\bm{\phi})$ respectively, we have
\begin{equation}\label{eq:marginal-likelihood}
L(\bm{Y} | \bm{\theta}, \bm{\phi}) \propto \int_{\bm{X}} h(\bm{Y} | \bm{X}, \bm{\phi})g(\bm{X} | \bm{\theta}) ~\mathrm{d}\bm{X}.
\end{equation}
In general, the integral in Equation \ref{eq:marginal-likelihood} is intractable, since the latent variables $\bm{X}$ (individuals' unobserved disease status) occupies a high-dimensional and highly-correlated state space.  This makes the use of frequentist methods difficult, such that Bayesian methods have become the toolset of choice for estimating the model parameters and latent variables jointly.

The most commonly used data-augmentation Markov chain Monte Carlo (MCMC) method is reversible-jump MCMC \citep{o1999bayesian, gibson1998estimating, chapman2020inferring, jewell2009bayesian, bridgen2024bayesian}. This method has low computational complexity, but explores the latent variable space slowly, particularly as the number of individuals in the model increases. Reversible-jump MCMC also becomes difficult to implement for more complicated transmission models (such as models where there are multiple paths individuals can take through the compartments \citep{chapman2020inferring}, or models where individuals can be in the same compartment for multiple disjoint periods of time \citep{spencer2015super}). Alternatively, sequential Monte Carlo techniques such as particle MCMC \citep{endo2019introduction} and SMC\textsuperscript{2} \citep{rimella2023inference} can be used for data-augmentation. These methods are useful for small populations but often inefficient and slow for large populations.

Other techniques can produce samples directly from the posterior distribution. The forward filtering backwards sampling (FFBS) algorithm \citep{carter1994gibbs, chib1996calculating} samples from the full conditional distribution of the model, but is extremely computationally costly (due to the large latent variable space for individual-based models). The individual FFBS algorithm \citep{touloupou2020scalable} samples from the full conditional distribution of a single individual's latent variables, conditional on the latent variables of all other individuals. This method explores the space efficiently and at reduced computational cost compared with the standard FFBS algorithm, but its computational cost still scales poorly with the number of compartments of the model.

Another option is approximate Bayesian computation \citep{minter2019approximate, luciani2009epidemiological}. This method only requires being able to simulate from the transmission model to function, which is useful when the model likelihood is unknown or intractable. However, it only provides samples from an approximation to the true posterior distribution (with the accuracy of the approximation relying on the user's choice of summary statistics).

There is therefore scope to introduce an exact inference method that balances the lower computational cost of reversible-jump MCMC with the more efficient mixing of individual FFBS, and that is more straightforward to implement than reversible-jump MCMC for more complex transmission models.

In this paper, we aim to fit extended-spectrum beta-lactamase (ESBL)-producing \textit{E. coli} transmission data to an individual-based, discrete-time model of colonisation. To fit our model we introduce a novel data-augmentation MCMC algorithm, taking a non-centred approach \citep{jewell2009bayesian}. While there exists non-centred methodology \citep{neal2005case} and model-based proposals \citep{pooley2015using, wang2025bayesian} for continuous-time models, there has been less work in discrete time. Our algorithm proposes a new latent variable state by making a small change in an individual's colonisation state(s), and then simulating forwards using the transmission model. This small change ripples through the rest of the latent variable space, so we call this algorithm the Rippler method. The Rippler algorithm is described in full for our model in Section \ref{section_rippler}, but can be applied to any discrete-time individual-based epidemic model.


ESBL-producing \textit{E. coli} are antimicrobially resistant (AMR) pathogens. AMR pathogens are those for which antimicrobial drugs are significantly less effective than expected. The spread of AMR pathogens is a major and increasing public health concern worldwide. \cite{murray2022global} estimate that more than 1.2 million people died from AMR pathogens in 2019, and the \cite{world2015global} states that antimicrobial resistance ``threatens the very core of modern medicine and the sustainability of an effective, global public health response to the enduring threat from infectious diseases''. 

Our work aims to describe the spread of AMR bacteria within a population using an individual-based stochastic epidemic model. We are motivated by AntiDOTE, a previously collected longitudinal study of third generation cephalosporin-resistant \textit{Escherichia coli} (3GC-R \textit{E. coli}). This is an ESBL-producing Enterobacteria \citep{lester2021third} which has been identified by the World Health Organisation as a major threat to pubic health \citep{essack2017antimicrobial}.  AntiDOTE aimed to identify risk factors for colonisation of individuals by 3GC-R \textit{E. coli}, and thereby suggest potential interventions to reduce the spread of such bacteria in the population.

AntiDOTE collected longitudinal samples from randomly-sampled households in urban Blantyre, Malawi, between April 2018 and June 2019.  The core dataset contains the ESBL-colonisation status of 478 individuals as measured by microbiological testing of stool samples taken at baseline, one-month, three-months, and six-months.  Individuals were grouped into 110 households, ranging in size from a single individual to a maximum of 11 individuals (Figure S1).  The original study design randomised household locations spatially over the entirety of urban Blantyre resulting in a sparsely sampled region.  In this respect, spatial correlation was not of primary concern.  Moreover, the logistics of sample-collection meant that household testing schedules were inevitably staggered in time (Figure S2).  Individual-level covariates were collected as part of the study, including age, sex, and HIV status.

These data were collected in accordance with the AntiDOTE research ethics and governance approval from the Malawi College of Medicine Research Ethics Committee (protocol number P.10/17/2299), and by Liverpool School of Tropical Medicine Research Ethics Committee (protocol number 17-063).  Ethics approval for this secondary data analysis was granted by Lancaster University Faculty of Science and Technology Research Ethics Committee (review reference FST-2023-4087).

\section{Model} \label{section_model}


In this section we construct a model to represent the acquisition and loss of 3GC-R \textit{E. coli} from an individual's stool sample.  Importantly, since these bacteria are part of an individual's microbiota, we refer to an individual carrying 3GC-R \textit{E. coli} as being ``colonised'', and non-carriers as ``uncolonised''.  Below, we first define a dynamical model of colonisation and de-colonisation, before defining an observation model that accounts for the data in the AntiDOTE study.

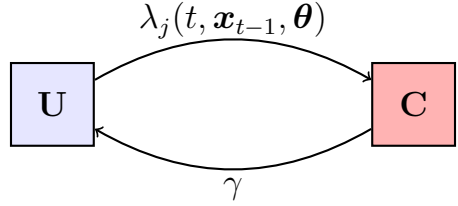
\begin{figure}[h]
    \centering
    \begin{center}
        \begin{tikzpicture}[thick, scale=1.6, every node/.style={scale=1.1}]
            \node[thick, draw, rectangle, minimum size=1cm, fill=blue!10] (U) at (0,0) {\textbf{U}};
            \node[thick, draw, rectangle, minimum size=1cm, fill=red!30] (C) at (3,0) {\textbf{C}};
            \draw[->] (U) edge [bend left] (C);
            \draw[->] (C) edge [bend left] (U);
            \node (lambda) at (1.5,0.7) {$\lambda_j(t, \bm{x}_{t-1},\bm{\theta})$};
            \node (gamma) at (1.5,-0.7) {$\gamma$};
        \end{tikzpicture}
    \end{center}
    \caption{A graphical representation of the UC model for individual $j$ at time $t$, who is either colonised (in state C) or uncolonised (in state U).}
    \label{UC_model}
\end{figure}

At any given point in time, we assume that individuals belong to exactly one of two mutually-exclusive categories: colonised (C), or uncolonised (U). We refer to this model as the UC model. The rate at which individuals move from state U to state C (colonisation) changes over time, and is different for each individual (i.e. an individual-based model). For individual $j$ at time $t$, the colonisation pressure is $\lambda_j(t, \bm{x}_{t-1},\bm{\theta})$, where $\bm{\theta}$ are the model parameters and $\bm{x}_{t-1}$ is the vector of colonisation states at time $t-1$. The rate at which individuals move from C to U (de-colonisation) is a constant $\gamma$, which we treat as known. A diagram of the model is shown in Figure \ref{UC_model}. This model is used to represent the true state of the population when performing inference, and is used to simulate the data for testing the Rippler algorithm in Section \ref{subsection_simulation}.

Let $x_{t,j}$ be the colonisation status of individual $j$ in time-step $t$: $x_{t,j}=1$ if the individual is colonised, and $x_{t,j}=0$ if the individual is uncolonised. We assume a fixed population size of $N$ individuals and a duration of $T$ time-steps.

Each individual is assumed to be initially colonised with some probability $p_0$, so for every $j \in \{1,\dots,N\}$ we sample \begin{equation}
    x_{0,j} \sim \text{Bernoulli}(p_0). \label{transmission_model_0}
\end{equation} 
Since we model in discrete time, for every time-step $t \in \{1,\dots,T\}$ of length $\Delta$ and individual $j \in \{1,\dots,N\}$ we sample \begin{equation}
    x_{t,j} | \bm{\theta},\bm{x}_{t-1} \sim \begin{cases} \text{Bernoulli}(1-e^{-\lambda_j(t, \bm{x}_{t-1},\bm{\theta})\Delta }) & \text{if } x_{t-1,j}=0, \\ \text{Bernoulli}(e^{-\gamma\Delta}) & \text{if } x_{t-1,j}=1, \end{cases} \label{transmission_model_full}
\end{equation} 
recursively.  We collect these results in the $(T+1) \times N$ matrix ${\bm{X} = (x_{t,j} : {t \in \{0,\dots,T\}}, {j \in \{1,\dots,N\}})}$.


Given the transmission outcome $\bm{X}$, we now introduce an observation outcome $\bm{Y}$ according to a hidden Markov model setup. The observation process identifies time-points at which individuals are sampled, and accounts for the sensitivity and specificity of the 3GC-R \textit{E. coli} test. We express $\bm{Y}$ as a $(T+1) \times N$ matrix. Each element $y_{t,j}$ is a single test result for individual $j$ at time $t$: $y_{t,j}=1$ if they test positive for 3GC-R \textit{E. coli}, and $y_{t,j}=0$ if they test negative for 3GC-R \textit{E. coli}. This model is used to represent how the test data is determined (given the true state of the population) when performing inference, and is used to simulate the data for testing the Rippler algorithm in Section \ref{subsection_simulation}.

Let $s_e, s_p \in [0,1]$ be the test sensitivity and specificity respectively. We assume that $s_e$ and $s_p$ are known values for identifiability. This means that a given test result $y_{t,j}$ will not always accurately represent the true colonisation status of individual $j$ at time $t$ (we may have $y_{t,j} \neq x_{t,j}$). For each $t \in \{1,\dots,T\}$ and $j \in \{1,\dots,N\}$, where individual $j$ is tested for 3GC-R \textit{E. coli}, we simulate the result $y_{t,j}$ by sampling \begin{equation}
    y_{t,j} | x_{t,j} \sim \begin{cases} \text{Bernoulli}(s_e) & \text{if } x_{t,j}=1, \\ \text{Bernoulli}(1-s_p) & \text{if } x_{t,j}=0. \end{cases} \label{obersation_model}
\end{equation} We let $y_{t,j}$ be undefined if individual $j$ is not tested at time $t$. 

\section{Inference: The Rippler Algorithm} \label{section_rippler}

\subsection{Overview}

In this section we develop a novel non-centered MCMC algorithm to estimate the parameters $\bm{\theta}$, given longitudinal diagnostic test information $\bm{Y}$. In this hidden Markov model, the colonisation status of every individual in every week, $\bm{X}$, is unobserved, and the task is therefore to marginalise over it in order to obtain an unbiased estimate of $\bm{\theta}$.  As previously mentioned, Bayesian inference is the approach of choice, treating $\bm{X}$ as part of the parameter space and aiming to estimate the joint distribution $\pi(\bm{\theta}, \bm{X} | \bm{Y})$.  However, this is complicated by the inherent dependence between elements of $\bm{X}$, and between $\bm{X}$ and $\bm{\theta}$.

We approach this problem by proposing an orthogonalising ``non-centered'' re-parameterisation of $\bm{X}$ inspired by the so-called Sellke construction, resulting in a parameter space that is \emph{a priori} independent both with respect to the epidemic process and the model parameters \citep{andersson2012stochastic}.  We then employ this re-parameterisation in a Metropolis-within-Gibbs algorithm that alternately draws from the parameters $\bm{\theta}$ and non-centered latent variables $\bm{U}$, transforming $\bm{U} \leftrightarrow \bm{X}$ on-demand when required.

\subsection{Non-centering the epidemic state space}

When simulating our transmission model, we repeatedly sample from Bernoulli distributions (Equations \ref{transmission_model_0} and \ref{transmission_model_full}) with some parameter $p \in [0,1]$. The full model simulation is shown in Algorithm \ref{alg_sim} (and is used to initialise the colonisation matrix $\bm{X}$ in the inference algorithm). To non-centre $\bm{X}$ with respect to $\bm{\theta}$, we observe that $x \sim \text{Bernoulli}(p)$ is equivalent to $x=\mathbf{I}(u<p)$, where $u \sim \text{Unif}(0,1)$. In principle, this means we could sample the matrix \[\bm{U}\overset{\text{iid}}{\sim} \text{Unif}(0,1),\] \textit{a priori} and simulate $\bm{X}$ as a deterministic function of $\bm{U}$ and $\bm{\theta}$. This function is described in Algorithm \ref{alg_deterministic}, which is a modified version of Algorithm \ref{alg_sim} that replaces the sampling of random numbers with $\bm{U}$ as an input. Thus \textit{a priori} $\bm{U}$ is independent of $\bm{\theta}$ and elements of $\bm{U}$ are independent of each other.

In principle, we could now invoke a Metropolis-within-Gibbs algorithm that alternately updates $\bm{\theta}$ and $\bm{U}$. However, in practise, updating $\bm{\theta}$ in the non-centred parameterisation is infeasible. We find that proposing a new value of $\bm{\theta}$ while keeping $\bm{U}$ constant results in a large change to the transmission outcome $\bm{X}$. This large change moves the transmission outcome away from the observed data $\bm{Y}$ to the extent that the proposed $\bm{\theta}$ is almost always rejected in the Metropolis-Hastings accept-reject step. 

Thus the parameters $\bm{\theta}$ must be updated in the centred parameterisation (keeping $\bm{X}$ constant). Consequently, when proposing an update for the epidemic state space we must move between the parameterisations in the following three steps: moving to the non-centred parameterisation ($\bm{X} \rightarrow \bm{U}$), proposing new non-centred latent variables ($\bm{U} \rightarrow \bm{U}^\ast$), and moving back to the centred parameterisation ($\bm{U}^\ast \rightarrow \bm{X}^\ast$). After these three steps we perform a Metropolis-Hastings accept-reject step to determine whether $\bm{X}^\ast$ is accepted.

\begin{algorithm}[t]
\caption{Model Simulation}
\label{alg_sim}
\begin{algorithmic}[1]
\Require Parameters $\bm{\theta}$, de-colonisation rate $\gamma$, initial probability of colonisation $p_0$.
\For{$j \in \{1,\dots,N\}$}
    \State Sample $u \sim \text{Unif}(0,1)$.
    \State If $u<p_0$, let $x_{0,j}=1$. \Comment{Determining if individual $j$ is initially colonised.}
    \State Otherwise let $x_{0,j}=0$.
\EndFor
\State Calculate $p_{\text{CU}} = 1-e^{-\gamma\Delta}$.
\For{$(t,j) \in \{1,\dots,T\} \times \{1,\dots,N\}$}
    \State Sample $u \sim \text{Unif}(0,1)$.
    \Switch{$x_{t-1,j}$}
        \Case{$0$}
            \State Calculate $p_{\text{UC}} = 1-e^{-\lambda_j(t, \bm{x}_{t-1},\bm{\theta})\Delta}$.
            \State If $u<p_{\text{UC}}$, let $x_{t,j}=1$. \Comment{Determining if a colonisation event occurs.}
            \State Otherwise let $x_{t,j}=0$.
        \EndCase
        \Case{$1$}
            \State If $u<p_{\text{CU}}$, let $x_{t,j}=0$. \Comment{Determining if a de-colonisation event occurs.}
            \State Otherwise let $x_{t,j}=1$.
        \EndCase
    \EndSwitch
\EndFor
\end{algorithmic}
\end{algorithm}

\begin{algorithm}[t]
\caption{$f(\bm{U},\bm{\theta}) = \bm{X}$}
\label{alg_deterministic}
\begin{algorithmic}[1]
\Require Random numbers $\bm{U}$, parameters $\bm{\theta}$, de-colonisation rate $\gamma$, initial probability of colonisation $p_0$.
\For{$j \in \{1,\dots,N\}$}
    \State If $u_{0,j}<p_0$, let $x_{0,j}=1$. \Comment{Checking if individual $j$ was initially colonised.}
    \State Otherwise let $x_{0,j}=0$.
\EndFor
\State Calculate $p_{\text{CU}} = 1-e^{-\gamma\Delta}$.
\For{$(t,j) \in \{1,\dots,T\} \times \{1,\dots,N\}$}
    \Switch{$x_{t-1,j}$}
        \Case{$0$}
            \State Calculate $p_{\text{UC}} = 1-e^{-\lambda_j(t, \bm{x}_{t-1},\bm{\theta})\Delta}$.
            \State If $u_{t,j}<p_{\text{UC}}$, let $x_{t,j}=1$. \Comment{Checking if a colonisation event occurred.}
            \State Otherwise let $x_{t,j}=0$.
        \EndCase
        \Case{$1$}
            \State If $u_{t,j}<p_{\text{CU}}$, let $x_{t,j}=0$. \Comment{Checking if a de-colonisation event occurred.}
            \State Otherwise let $x_{t,j}=1$.
        \EndCase
    \EndSwitch
\EndFor
\end{algorithmic}
\end{algorithm}

\begin{algorithm}[h!]
\caption{Proposal Bounds for $\bm{U}|\bm{\theta},\bm{X}$}
\label{alg_a_b}
\begin{algorithmic}[1]
\Require Transmission outcome $\bm{X}$, parameters $\bm{\theta}$, de-colonisation rate $\gamma$, initial probability of colonisation $p_0$.
\For{$j \in \{1,\dots,N\}$}
    \Switch{$x_{0,j}$}
        \Case{$1$}
            \State Let $a_{0,j}=0$ and $b_{0,j}=p_0$. \Comment{Proposal bounds if individual $j$ was initially colonised.}
        \EndCase
        \Case{$0$}
            \State Let $a_{0,j}=p_0$ and $b_{0,j}=1$. 
        \EndCase
    \EndSwitch
\EndFor
\State Calculate $p_{\text{CU}} = 1-e^{-\gamma\Delta}$.
\For{$(t,j) \in \{1,\dots,T\} \times \{1,\dots,N\}$}
    \Switch{$(x_{t-1,j}, x_{t,j})$}
        \State Calculate $p_{\text{UC}} = 1-e^{-\lambda_j(t, \bm{x}_{t-1},\bm{\theta})\Delta}$.
        \Case{$(0, 1)$}
            \State Let $a_{t,j}=0$ and $b_{t,j}=p_{\text{UC}}$. \Comment{Proposal bounds if a colonisation event occurred.}
        \EndCase
        \Case{$(0, 0)$}
            \State Let $a_{t,j}=p_{\text{UC}}$ and $b_{t,j}=1$.
        \EndCase
        \Case{$(1, 0)$}
            \State Let $a_{t,j}=0$ and $b_{t,j}=p_{\text{CU}}$. \Comment{Proposal bounds if a de-colonisation event occurred.}
        \EndCase
        \Case{$(1, 1)$}
            \State Let $a_{t,j}=p_{\text{CU}}$ and $b_{t,j}=1$.
        \EndCase
    \EndSwitch
\EndFor
\end{algorithmic}
\end{algorithm}

\subsection{Latent variables update}

 We now give details on the steps taken to propose new latent variables $\bm{X}^\ast$ from the current latent variables $\bm{X}$:

\textbf{Step 1:} $\bm{X} \rightarrow \bm{U}$. We first propose a matrix of random numbers $\bm{U}$: each entry of $\bm{U}$ is sampled uniformly from the widest possible range of values that is consistent with $\bm{X}$ (given the current value of $\bm{\theta}$). For example, if $x_{t-1,j}=1$ and $x_{t,j}=0$ (a de-colonisation event), then we sample $u_{t,j} \sim \text{Unif}(0,p_{\text{CU}})$. These ranges $(a_{t,j},b_{t,j})$ are calculated using Algorithm \ref{alg_a_b}. This deterministic algorithm is an inverse version of the model simulation (Algorithm \ref{alg_deterministic}), calculating the probability bounds from the transmission events (rather than the transmission events from the probabilities). 

\textbf{Step 2:} $\bm{U} \rightarrow \bm{U}^\ast$. The next move is to propose a new value of $\bm{U}$ by altering one entry in the matrix. We randomly choose $t \in \{0,\dots,T\}$ and $j \in \{1,\dots,N\}$, and propose a new value $u^{\ast}_{t,j}$. All other entries in $\bm{U}^{\ast}$ are equal to their corresponding entries in $\bm{U}$. A simple approach to proposing $u^{\ast}_{t,j}$ is to choose $t$ and $j$ uniformly, and then to sample from $\text{Unif}(0,1)$. However, this method of proposing a new $\bm{U}$ can often produce `null moves': changes to $\bm{U}$ that do not change $\bm{X}$ (so $\bm{X}^{\ast}=\bm{X}$). These moves will always be accepted, but without actually changing the current underlying epidemic. Instead, we sample a new value of an entry of $\bm{U}$ uniformly from the largest possible range that will result in a change to $\bm{X}$ -- this is the range $(a_{t,j},b_{t,j})^\mathsf{C}$, where we define ${(a_{t,j},b_{t,j})^\mathsf{C} = (0,1)~ \backslash ~(a_{t,j},b_{t,j})}$. We also choose $t \in \{0,\dots,T\}$ and $j \in \{1,\dots,N\}$ non-uniformly, selecting each pair $(t,j)$ with probability proportional to $1+a_{t,j}-b_{t,j}$ (the size of the range that $u_{t,j}^{\ast}$ can be sampled from).

\textbf{Step 3:} $\bm{U}^\ast \rightarrow \bm{X}^\ast$. We then propose $\bm{X}^{\ast}$ from $\bm{U}^{\ast}$ using the deterministic function given in Algorithm \ref{alg_deterministic}. Using $\bm{U}^{\ast}$ as the random numbers for calculating the new latent variable matrix $\bm{X}^{\ast}$ via Algorithm \ref{alg_deterministic} can produce a slightly different result than if we had used $\bm{U}$, potentially affecting multiple individuals in one move. We call this method the \textit{Rippler algorithm}, as the change `ripples' forward in time through the epidemic.

The new latent variables $\bm{X}^{\ast}$ are accepted with probability \[\alpha(\bm{X},\bm{X}^{\ast}) = \min\left\{1,\frac{\pi(\bm{X}^{\ast}|\bm{\bm{\theta},\bm{Y}})}{\pi(\bm{X}|\bm{\bm{\theta},\bm{Y}})}\frac{q(\bm{X}|\bm{\theta},\bm{X}^{\ast})}{q(\bm{X}^{\ast}|\bm{\theta},\bm{X})}\right\}.\] 

\begin{proposition}
    The latent variable acceptance rate simplifies to \[\alpha(\bm{X},\bm{X}^{\ast}) = \min\left\{1,\frac{\pi(\bm{Y}|\bm{X}^{\ast},\bm{\theta})}{\pi(\bm{Y}|\bm{X},\bm{\theta})} \frac{q(\bm{U}|\bm{\theta},\bm{U}^\ast)}{q(\bm{U}^\ast|\bm{\theta},\bm{U})}\right\},\] where $q(\bm{U}^\ast|\bm{\theta},\bm{U})$ is the proposal distribution of the new random number matrix $\bm{U}^\ast$ (given the parameters $\bm{\theta}$ and current random number matrix $\bm{U}$).
\end{proposition}

\textit{Proof} Let the distribution of each of the three steps of the latent variable proposal be $q(\bm{U}|\bm{\theta},\bm{X})$, $q(\bm{U}^{\ast}|\bm{\theta},\bm{U})$, and $q(\bm{X}^{\ast}|\bm{\theta},\bm{U}^{\ast})$ respectively. Since we know $\pi(\bm{X}|\bm{\theta},\bm{Y}) \propto \pi(\bm{\theta},\bm{X}|\bm{Y})$, we use Bayes' theorem to split the acceptance rate into its component terms: \[\alpha(\bm{X},\bm{X}^{\ast}) = \min\left\{1,\frac{\pi(\bm{Y}|\bm{X}^{\ast},\bm{\theta})\pi(\bm{X}^{\ast}|\bm{\theta})}{\pi(\bm{Y}|\bm{X},\bm{\theta})\pi(\bm{X}|\bm{\theta})}\frac{q(\bm{U}^{\ast}|\bm{\theta},\bm{X}^{\ast})q(\bm{U}|\bm{\theta},\bm{U}^{\ast})q(\bm{X}|\bm{\theta},\bm{U})}{q(\bm{U}|\bm{\theta},\bm{X})q(\bm{U}^{\ast}|\bm{\theta},\bm{U})q(\bm{X}^{\ast}|\bm{\theta},\bm{U}^{\ast})}\right\}.\] 
    
We will now show that $\pi(\bm{X}|\bm{\theta})q(\bm{U}|\bm{\theta},\bm{X}) = 1$ and $\pi(\bm{X}^{\ast}|\bm{\theta})q(\bm{U}^{\ast}|\bm{\theta},\bm{X}^{\ast}) = 1$. During the first part of the proposal step, we sample each entry of $\bm{U}$ uniformly from the widest possible range of values that will produce $\bm{X}$ (given the current value of $\bm{\theta}$), which are calculated using Algorithm \ref{alg_a_b}. Let ${UC(\bm{X}) = \{ (t,j) : x_{t-1,j}=0, x_{t,j}=1 \}}$ be the set of colonisation events. For each colonisation event, the random number needed must be between ${a_{t,j}=0}$ and ${b_{t,j}=1-e^{-\lambda_j(t, \bm{x}_{t-1},\bm{\theta})\Delta}}$, and so we propose $u_{t,j}$ from ${\text{Unif}(0,1-e^{-\lambda_j(t, \bm{x}_{t-1},\bm{\theta})\Delta})}$. This means the probability density of this proposal is \[\frac{1}{1-e^{-\lambda_j(t, \bm{x}_{t-1},\bm{\theta})\Delta}}.\]

We can similarly define ${UU(\bm{X}) = \{ (t,j) : x_{t-1,j}=0, x_{t,j}=0 \}},$ where we must propose from $\text{Unif}(1-e^{-\lambda_j(t, \bm{x}_{t-1},\bm{\theta})\Delta},1)$, ${CU(\bm{X}) = \{ (t,j) : x_{t-1,j}=1, x_{t,j}=0 \}},$ where we must propose from $\text{Unif}(0,1-e^{-\gamma\Delta})$, and $CC(\bm{X}) = \{ (t,j) : x_{t-1,j}=1, x_{t,j}=1 \},$ where we must propose from ${\text{Unif}(1-e^{-\gamma\Delta},1)}$. For each individual's initial state, we must propose from ${\text{Unif}(0,p_0)}$ if they are initially colonised and from ${\text{Unif}(p_0,1)}$ if they are initially uncolonised. Let ${N_0(\bm{X})=\sum_{j=1}^Nx_{0,j}}$. This means the full proposal density is \begin{align} \begin{split}
    q(\bm{U}|\bm{\theta},\bm{X}) &= \prod_{(t,j) \in UC(\bm{X})} \frac{1}{1-e^{-\lambda_j(t, \bm{x}_{t-1},\bm{\theta})\Delta}} \prod_{(t,j) \in UU(\bm{X})} \frac{1}{e^{-\lambda_j(t, \bm{x}_{t-1},\bm{\theta})\Delta}}
    \\ &~ \hspace{0.5cm} \frac{1}{(1-e^{-\gamma\Delta})^{|CU(\bm{X})|}} \frac{1}{e^{-\gamma\Delta|CC(\bm{X})|}} \frac{1}{p_0^{N_0(\bm{X})}}\frac{1}{(1-p_0)^{N-N_0(\bm{X})}}. \label{U|theta,X}
\end{split} \end{align}

From our transmission model (Equations \ref{transmission_model_0} and \ref{transmission_model_full}) we have \begin{align} \begin{split}
    \pi(\bm{X}|\bm{\theta}) &= \prod_{(t,j) \in UC(\bm{X})} (1-e^{-\lambda_j(t, \bm{x}_{t-1},\bm{\theta})\Delta}) \prod_{(t,j) \in UU(\bm{X})} e^{-\lambda_j(t, \bm{x}_{t-1},\bm{\theta})\Delta}
    \\ &~ \hspace{0.5cm} (1-e^{-\gamma\Delta})^{|CU(\bm{X})|}  e^{-\gamma\Delta|CC(\bm{X})|}p_0^{N_0(\bm{X})}(1-p_0)^{N-N_0(\bm{X})}. \label{X|theta}
\end{split} \end{align}

Multiplying Equation \ref{U|theta,X} and Equation \ref{X|theta} we see that $\pi(\bm{X}|\bm{\theta})q(\bm{U}|\bm{\theta},\bm{X}) = 1$. Therefore by symmetry we have $\pi(\bm{X}^{\ast}|\bm{\theta})q(\bm{U}^{\ast}|\bm{\theta},\bm{X}^{\ast}) = 1$ as required. 

We also consider the final part of the proposal step. During this step, we deterministically calculate $\bm{X}^{\ast}$ from $\bm{U}^{\ast}$ and $\bm{\theta}$ using Algorithm \ref{alg_deterministic}. This means $q(\bm{X}^{\ast}|\bm{\theta},\bm{U}^{\ast}) = 1$ when $\bm{X}^{\ast}=f(\bm{U}^{\ast},\bm{\theta})$ and $q(\bm{X}^{\ast}|\bm{\theta},\bm{U}^{\ast}) = 0$ otherwise. Similarly, for the reverse move, we have $q(\bm{X}|\bm{\theta},\bm{U}) = 1$ when $\bm{X}=f(\bm{U},\bm{\theta})$ and $q(\bm{X}|\bm{\theta},\bm{U}) = 0$ otherwise, and so the result follows. \hfill $\square$

To calculate the probability of accepting the proposed latent variables we must determine $\pi(\bm{Y}|\bm{X},\bm{\theta})$ and $q(\bm{U}^{\ast}|\bm{\theta},\bm{U})$. We can determine $\pi(\bm{Y}|\bm{X},\bm{\theta})$ from our observation model (Equation \ref{obersation_model}). Let ${TP(\bm{X},\bm{Y}) = |\{ (t,j) : x_{t,j} = 1, y_{t,j} = 1 \}|}$, the number of true positive test results. We similarly define the number of false negative, false positive and true negative test results: $FN(\bm{X},\bm{Y}) = |\{ (t,j) : x_{t,j} = 1, y_{t,j} = 0 \}|$, ${FP(\bm{X},\bm{Y}) = |\{ (t,j) : x_{t,j} = 0, y_{t,j} = 1 \}|}$, and $TN(\bm{X},\bm{Y}) = |\{ (t,j) : x_{t,j} = 0, y_{t,j} = 0 \}|$. Then we have \begin{equation*}
    \pi(\bm{Y}|\bm{X},\bm{\theta}) = s_e^{TP(\bm{X},\bm{Y})} (1-s_e)^{FN(\bm{X},\bm{Y})} s_p^{FP(\bm{X},\bm{Y})} (1-s_p)^{TN(\bm{X},\bm{Y})}. \label{Y|X,theta}
\end{equation*}

When proposing $\bm{U}^\ast$ from $\bm{U}$, we randomly choose $t' \in \{0,\dots,T\}$ and $j' \in \{1,\dots,N\}$, selecting each pair $(t',j')$ with probability proportional to the size of the range that $u_{t',j'}^{\ast}$ can be sampled from. We then propose $u_{t',j'}^{\ast}$ from ${\text{Unif}(a_{t',j'},b_{t',j'})^\mathsf{C}}$, where ${(a_{t',j'},b_{t',j'})^\mathsf{C} = (0,1)~ \backslash ~(a_{t',j'},b_{t',j'})}$. This is the widest range of values such that ${x^{\ast}_{t',j'} = 1 - x_{t',j'}}$.

The probability of choosing to change $u_{t',j'}^{\ast}$, rather than any other entry in the matrix, is \[\frac{1+a_{t',j'}-b_{t',j'}}{\sum_{t=0}^T\sum_{j=1}^N (1+a_{t,j}-b_{t,j})}.\] The probability density of the forward move is $1/(1+a_{t',j'}-b_{t',j'})$, and so \begin{equation*}
    q(\bm{U}^{\ast}|\bm{\theta},\bm{U}) = \frac{1}{\sum_{t=0}^T\sum_{j=1}^N (1+a_{t,j}-b_{t,j})}. \label{U^ast|theta,U}
\end{equation*}

\subsection{Parameters update}

For the parameter step we use an adaptive random walk Metropolis (RWM), jointly updating all parameters together. We sample $\bm{\theta}^{\ast} \sim \text{Normal}(\bm{\theta},\kappa^2\bm{\Sigma})$, where $\kappa$ is a tuning value and $\bm{\Sigma}$ is an approximation to the covariance matrix of the parameters. We accept $\bm{\theta}^{\ast}$ with probability 
\[\alpha(\bm{\theta},\bm{\theta}^{\ast}) = \min\left\{1,\frac{\pi(\bm{\theta}^{\ast}|\bm{X},\bm{Y})}{\pi(\bm{\theta}|\bm{X},\bm{Y})}\right\}.\] The value of $\kappa$ is adaptively tuned based on whether or not $\bm{\theta}^{\ast}$ is accepted; we use the block adaptive method (algorithm 6) from \cite{sherlock2010random}. The value of $\bm{\Sigma}$ is adaptively calculated using the accelerated shaping algorithm from \cite{spencer2021accelerating}. To avoid the chain getting stuck at singular values of $\bm{\Sigma}$ we propose $\bm{\theta}^{\ast} \sim \text{Normal}(\bm{\theta},\bm{\Sigma}^{(0)})$ with some small probability $\varepsilon$ (where $\bm{\Sigma}^{(0)}$ has been chosen such that the $\bm{\theta}^{\ast}$ sampled this way have a high probability of being accepted). Following \cite{sherlock2010random} we set $\varepsilon=0.05$ and the initial value of $\kappa$ to 1.19 (i.e., 2.38 divided by 2, which is the square root of the number of parameters).

We combine these steps for the full Rippler algorithm in Algorithm \ref{alg_inference_noncentred}. Note that we repeat the latent variable update step multiple times in each iteration, since one update will only affect a small proportion of the latent variables.

\begin{algorithm}[h!]
\caption{Rippler Inference}
\label{alg_inference_noncentred}
\begin{algorithmic}[1]
\Require Number of iterations $K$, number of iterations for latent updates $K'$, starting values $\bm{\theta}^{(0)},\bm{X}^{(0)}$, prior distribution hyperparameters $\bm{\mu}$, de-colonisation rate $\gamma$, initial probability of colonisation $p_0$, test data $\bm{Y}$.
\State Let $\bm{\theta}=\bm{\theta}^{(0)},\bm{X}=\bm{X}^{(0)}$.
\For{$k \in \{1,\dots,K\}$}
    \State Sample $\bm{\theta} \sim \pi(\bm{\theta}|\bm{X},\bm{Y})$ using a suitable MCMC update.
    \For{$k' \in \{1,\dots,K'\}$}
        \State Calculate proposal bounds $a_{t,j}$ and $b_{t,j}$ for each $(t,j) \in \{0,\dots,T\} \times \{1,\dots,N\}$ using Algorithm \ref{alg_a_b}.
        \State Propose $\bm{U}$: sample $u_{t,j} \sim \text{Unif}(a_{t,j},b_{t,j})$ for each pair $(t,j)$.
        \State Randomly choose $(t',j') \in \{0,\dots,T\} \times \{1,\dots,N\}$, choosing each pair $(t',j')$ with probability \[\frac{1+a_{t',j'}-b_{t',j'}}{\sum_{t=0}^T\sum_{j=1}^N (1+a_{t,j}-b_{t,j})}.\]
        \State Define $(a_{t',j'},b_{t',j'})^\mathsf{C} = (0,1)~ \backslash ~(a_{t',j'},b_{t',j'})$.
        \State Propose $\bm{U}^\ast$: sample $u^{\ast}_{t',j'} \sim \text{Unif}(a_{t',j'},b_{t',j'})^\mathsf{C}$, and let $u^{\ast}_{t,j}=u_{t,j}$ for all $(t,j) \neq (t',j')$.
        \State Propose $\bm{X}^{\ast} = f(\bm{U}^{\ast},\bm{\theta})$ using Algorithm \ref{alg_deterministic}.
        \State Calculate \[\alpha(\bm{X},\bm{X}^{\ast}) = \min\left\{1,\frac{\pi(\bm{Y}|\bm{X}^{\ast},\bm{\theta})}{\pi(\bm{Y}|\bm{X},\bm{\theta})} \frac{q(\bm{U}|\bm{\theta},\bm{U}^\ast)}{q(\bm{U}^\ast|\bm{\theta},\bm{U})}\right\}.\]
        \State Let $\bm{X} = \bm{X}^{\ast}$ with probability $\alpha(\bm{X},\bm{X}^{\ast})$. Otherwise keep $\bm{X}$ constant.
    \EndFor
\EndFor
\end{algorithmic}
\end{algorithm}

\section{Application} \label{section_application}

\subsection{3GC-R \textit{E. coli} model}

To demonstrate our methodology we return to the example dataset presented in Section \ref{sec_intro}. The two main routes of transmission of 3GC-R \textit{E. coli} are identified in \cite{lester2021third} as from person-to-person within the whole population, and from person-to-person within individuals' households. We now define the colonisation pressure $\lambda_j(t,\boldsymbol{x}_{t-1},\boldsymbol{\theta})$ as a weighted sum of these two routes of transmission, and introduce the model parameters $\boldsymbol{\theta}$ that characterise the transmission.

We choose the time-step size $\Delta = 1$ week. Let $\beta_G$ be the global transmission rate parameter (per week) and $\beta_H$ be the relative within-household transmission parameter. When considering global transmission, we assume that individuals mix homogeneously within the population, and that transmission is frequency-dependent (proportional to the proportion of colonised individuals in the population). Following \cite{lewis2022colonization}, we also add a seasonal modifier to the global colonisation pressure, since the transmissibility of 3GC-R \textit{E. coli} in Malawi is dependent on the time of year, with higher transmissibility in the wet season. We define the seasonal modifier at time $t$ as ${\varsigma_t = 1 - \cos\left(2\pi(t+t_\ast)/T^{\ast}\right)}$, where $t_\ast$ is chosen to match the peak of $\varsigma_t$ with the peak of the wet season. Each time-step is one week, so we let $T^{\ast}=52$. Since the model is in discrete time, this is a piecewise-constant function  (we assume the value at the beginning of the time-step, and do not change it until the beginning of the next time-step). This means the global colonisation pressure on individual $j$ at time $t$ is \[\lambda^G_j(t, \bm{x}_{t-1},\bm{\theta}) = \beta_G\varsigma_{t-1}\frac{\sum_{i=1}^Nx_{t-1,i}}{N}.\]

The household mixing matrix $\bm{H}$ is defined as an $N \times N$ matrix where $h_{i,j}=1$ if individuals $i$ and $j$ live in the same household, and $h_{i,j}=0$ otherwise. Each individual is assumed to live in exactly one house. We assume that within-household transmission is density-dependent (proportional to the number of individuals in the household) and that individuals mix homogeneously within their household, so the within-household colonisation pressure on individual $j$ at time $t$ is \[\lambda^H_j(t, \bm{x}_{t-1},\bm{\theta}) = \beta_G\beta_H\sum_{i=1}^{N}h_{i,j}x_{t-1,i}.\] 

In addition, an individual's covariates could also influence their susceptibility to colonisation, so we introduce an additional term into the colonisation pressure on individual $j$: $e^{\bm{\delta} \cdot \bm{c}_j},$ where $\bm{\delta}$ are the covariate effect parameters and $\bm{c}_j$ is a vector of the covariates of individual $j$. We consider age and sex covariates, with age measured in years and sex considered as a binary variable (0 for female, 1 for male). We let $\delta_A$ be the age covariate effect parameter (per year) and let $\delta_S$ be the sex covariate effect parameter -- so $\bm{\delta}=(\delta_A,\delta_S)$. The covariates are properties of individuals, so will affect both global and within-household transmission.

Therefore the overall colonisation pressure on individual $j$ at time $t$ is \[\lambda_j(t, \bm{x}_{t-1},\bm{\theta}) = e^{\bm{\delta} \cdot \bm{c}_j}\left(\lambda^G_j(t, \bm{x}_{t-1},\bm{\theta}) + \lambda^H_j(t, \bm{x}_{t-1},\bm{\theta})\right),\] where $\bm{\theta}=(\beta_G,\beta_H,\delta_A,\delta_S)$.

We give our parameters $\bm{\theta} = (\beta_G,\beta_H,\delta_A,\delta_S)$ independent prior distributions, with hyperparameters ${\bm{\mu}=(\mu_G,\mu_H,\mu_A,\mu_S)}$. The prior distributions for $\beta_G$ are $\beta_H$ are exponential distributions with rates $\mu_G$ and $\mu_H$ respectively. The prior distributions for $\delta_A$ are $\delta_S$ are chosen as Laplace distributions with rates $\mu_A$ and $\mu_S$ respectively, since we do not assume directionality to the covariate effects \textit{a priori}, and to guard against over-fitting (by shrinking estimates towards 0). 


\subsection{Simulation study} \label{subsection_simulation}


We now simulate a 3GC-R \textit{E. coli} epidemic using the transmission model, and generate test data using the observation model. Using these test results, we assess the performance of the Rippler algorithm by attempting to recover the parameters and latent states.

The epidemic is simulated with parameters $\beta_G=0.1$, $\beta_H=1.5$, $\delta_A=0$, $\delta_S=0$. The seasonality offset term, $t_\ast$, is set to 17 weeks. The de-colonisation rate, $\gamma$, is set to 0.5 per week, and the initial probability of colonisation is assumed to be 0.3. We use the same population and timescale as the AntiDOTE data (including the household structure and covariates), giving us $N=478$ and $T=62$. 


The covariate data for age and sex have been centred around their respective means. We also simulate test data at the same times and on the same individuals as in the AntiDOTE data, totalling 1659 tests. Using a sensitivity of 0.8 and specificity of 0.99 we simulate new test results, resulting in 371 true positives, 15 false positives, 1148 true negatives, and 39 false negatives.

We run the MCMC algorithm for $K=100{,}000$ iterations, repeating the latent variable step 400 times within each iteration (to ensure proper mixing in the latent variable space). The first $5{,}000$ iterations are discarded as burn-in. We let $\bm{\theta}^{(0)}=(0.5,0.5,0,0)$, and then generate $\bm{X}^{(0)}$ from our transmission model using parameters $\bm{\theta}^{(0)}$. The rate of the prior distribution for each parameter is set to $0.001$, i.e.~a very uninformative prior. The trace plots for the posterior samples of the parameters are shown in Figure S5. 

\begin{figure}[h]
     \centering
     \hfill
     \begin{subfigure}[h]{0.45\textwidth}
         \centering
         \includegraphics[width=\textwidth]{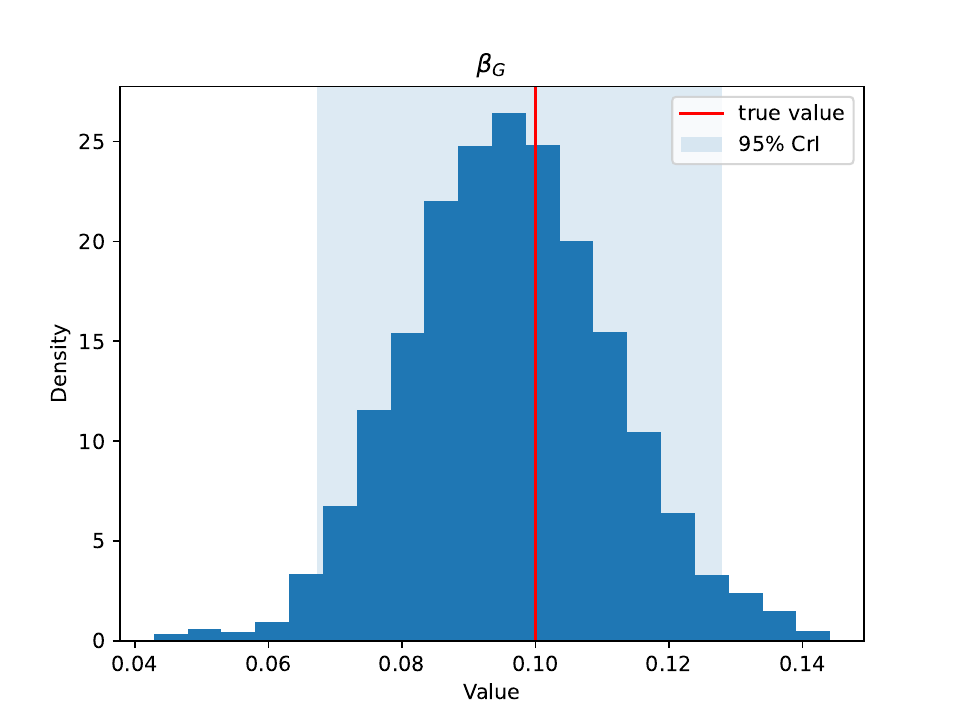}
     \end{subfigure}
     \hfill
     \begin{subfigure}[h]{0.45\textwidth}
         \centering
         \includegraphics[width=\textwidth]{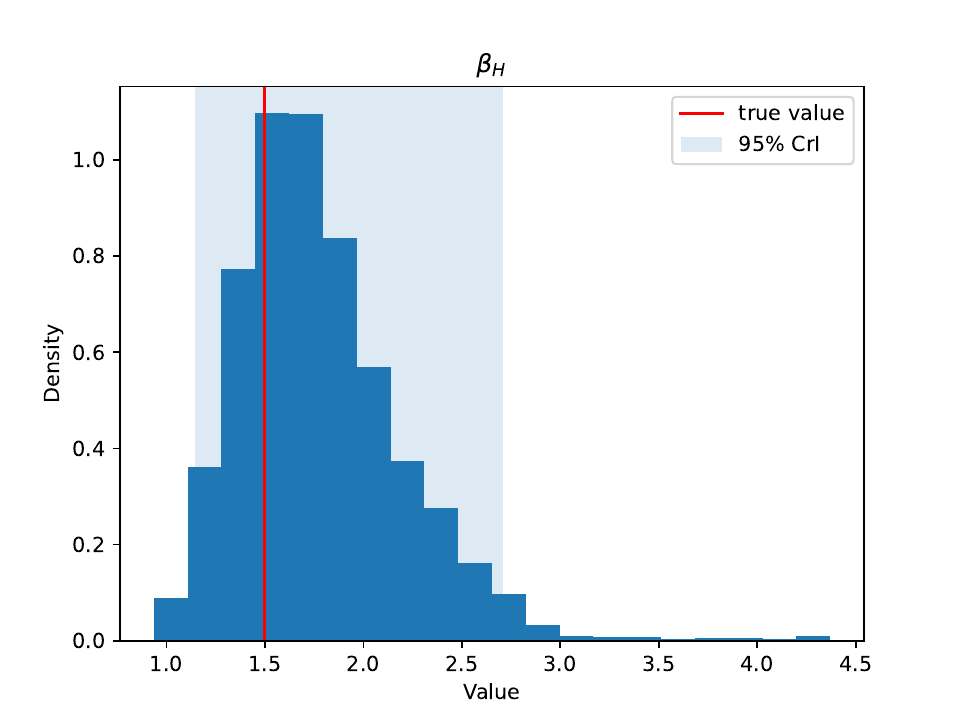}
     \end{subfigure}
     \hfill
     \\
     \hfill
     \begin{subfigure}[h]{0.45\textwidth}
         \centering
         \includegraphics[width=\textwidth]{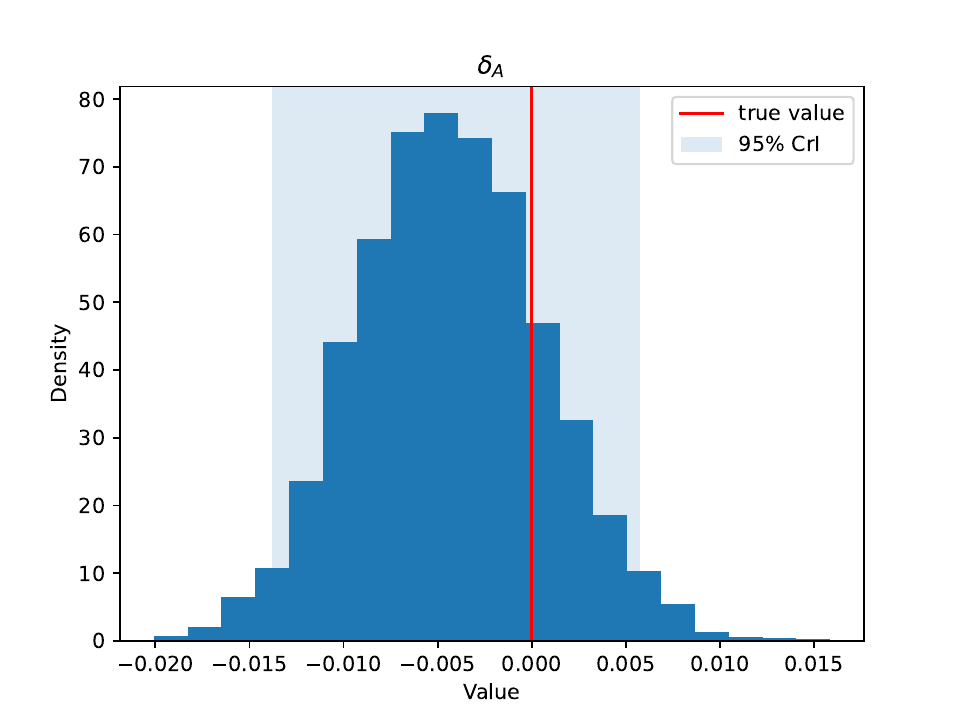}
     \end{subfigure}
     \hfill
     \begin{subfigure}[h]{0.45\textwidth}
         \centering
         \includegraphics[width=\textwidth]{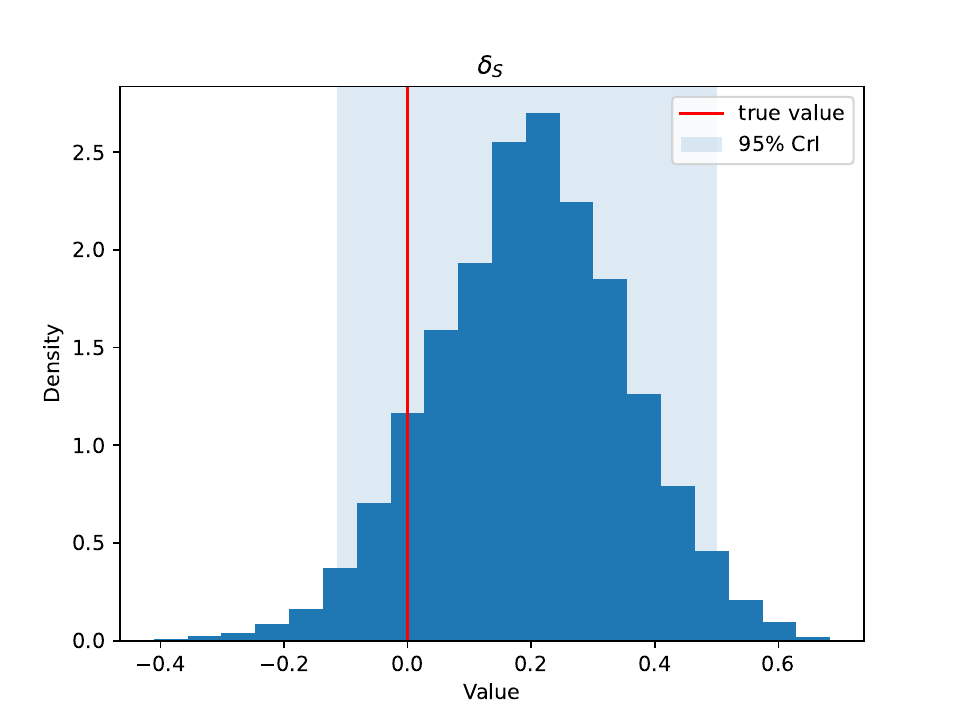}
     \end{subfigure}
     \hfill
     \caption{Histograms of the posterior samples generated for the parameters $\bm{\theta}=(\beta_G,\beta_H,\delta_A,\delta_S)$ for simulated 3GC-R \textit{E. coli} test data (using the Rippler algorithm for inference). The true value is shown by the red line and the 95\% credible interval is shown in light blue.}
     \label{sim_full_parameters}
\end{figure}

We see in Figure \ref{sim_full_parameters} and Table \ref{sim_full_table} that the true value of each parameter lies within the 95\% credible interval of the sample. However, while there is clear convergence for $\beta_G$ and $\beta_H$, with the median of the sample very close to the true value, convergence is less clear for $\delta_A$ and $\delta_S$, with the median significantly further away from the true value.

\begin{table}[h!]
    \centering
    \caption{Parameter estimates for simulated 3GC-R \textit{E. coli} test data using the Rippler algorithm for inference.}
    \begin{tabular}{ c  c  c  c  c}
        \hline Parameter &  Definition & True & Median & 95\% CrI 
        \\ \hline $\beta_G$ & global transmission rate (week$^{-1}$) & 0.1 & 0.0958 & (0.0673, 0.128) 
        \\ $\beta_H$ & relative within-household transmission & 1.5 & 1.71 & (1.15, 2.71) 
        \\ $\delta_A$ & age covariate effect (year$^{-1}$) & 0 & -0.00442 & (-0.0138, 0.00574) 
        \\ $\delta_S$ & sex covariate effect & 0 & 0.202 & (-0.114, 0.500) 
        \\ \hline
    \end{tabular}
    \label{sim_full_table}
\end{table}

\begin{figure}[h]
    \centering
    \includegraphics[width=0.85\textwidth]{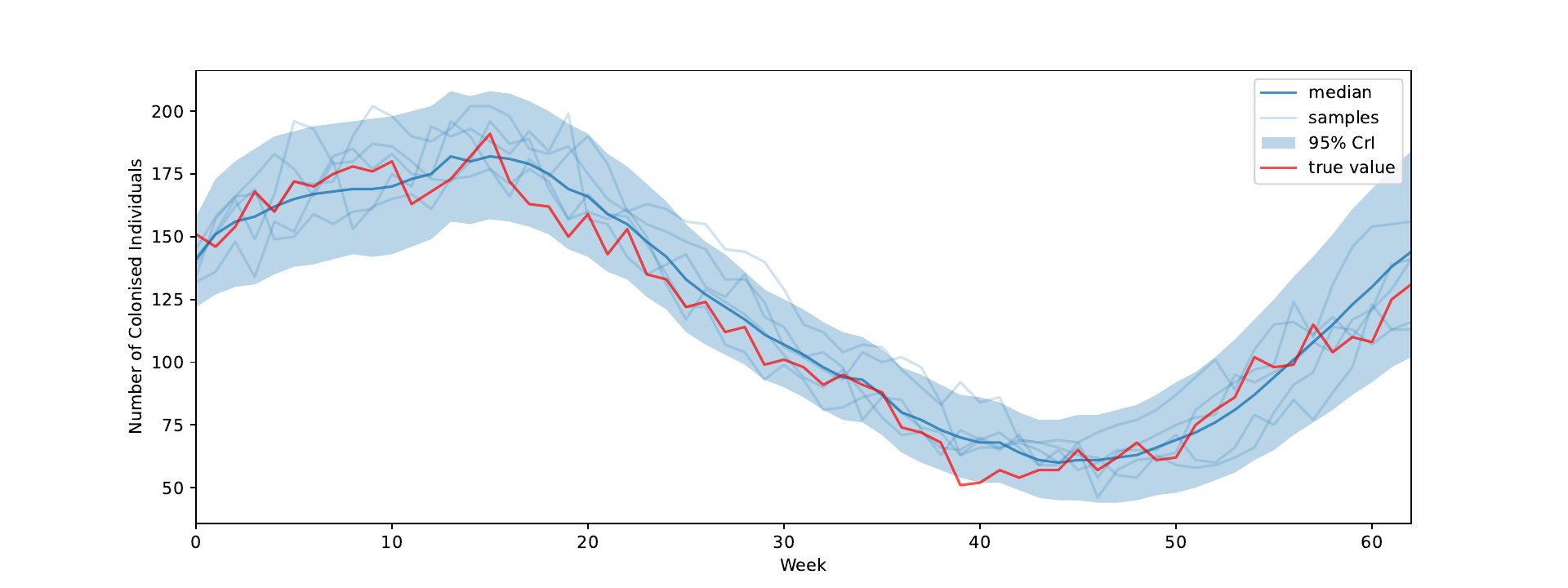}
    \caption{The number of colonised individuals over time for simulated 3GC-R \textit{E. coli} test data (using the Rippler algorithm for inference). The median value is shown by the dark blue line and 5 randomly chosen samples are shown by the faint blue lines. The true value is shown by the red line and the 95\% credible interval is shown in light blue.}
    \label{sim_full_latent}
\end{figure}

We see in Figure \ref{sim_full_latent} that the latent variable part of the inference performs well, with the true number of individuals colonised at each point in time lying mostly within the 95\% credible interval. Note that the credible interval is wider at the beginning and end of the 62 weeks than in the middle, which loosely matches with the number of test results available at each point in time.

\subsection{AntiDOTE data analysis} \label{subsection_real_data}

We now analyse the data collected from the AntiDOTE study in Blantyre, Malawi (for which $N=478$ and $T=62$). The study aimed to sample each individual four times: an initial test, and then follow-up tests one, three, and six months later. However, not all individuals could be sampled during each follow-up visit, resulting in an average of 3.47 tests per individual. In total 1659 tests were conducted (on average, 5.60\% of individuals were tested each week). A visual representation of this test data can be found in Figure S1.

The 478 individuals were split between 110 houses. All but three houses had populations between 2 and 8, with a median house size of 4. 59.8\% of the tested population was female. The ages of individuals were right-skewed, with a median age of 15.3 years and a mean age of 20.2 years. Further details can be found in Section S1 of the supplementary materials.

We run the MCMC algorithm for $K=100{,}000$ iterations, repeating the latent variable step 400 times within each iteration. The first $20{,}000$ iterations are discarded as burn-in. We let $\bm{\theta}^{(0)}=(0.5,0.5,0,0)$, and then generate $\bm{X}^{(0)}$ from our transmission model using parameters $\bm{\theta}^{(0)}$. The rate of the prior distribution for each parameter is set to $0.001$.

The de-colonisation rate, $\gamma$, is set to 0.5 per week, since the average length of colonisation is 2 weeks \citep{lester2021third}. We assume a test sensitivity of 0.8 and a test specificity of 0.95, as in \cite{rimella2023inference}. The initial probability of colonisation, $p_0$, is set to 0.35 (running the MCMC with different values of $p_0$ showed that this choice gave better agreement between the observed and predicted prevalence than higher or lower values).

\begin{figure}[h]
     \centering
     \hfill
     \begin{subfigure}[h]{0.45\textwidth}
         \centering
         \includegraphics[width=\textwidth]{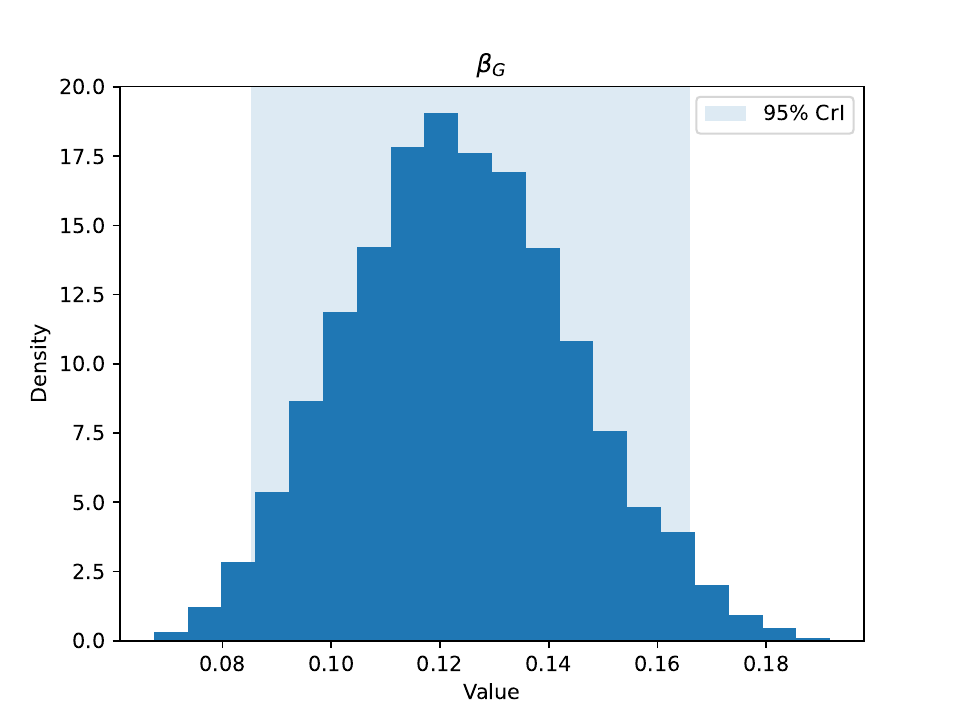}
     \end{subfigure}
     \hfill
     \begin{subfigure}[h]{0.45\textwidth}
         \centering
         \includegraphics[width=\textwidth]{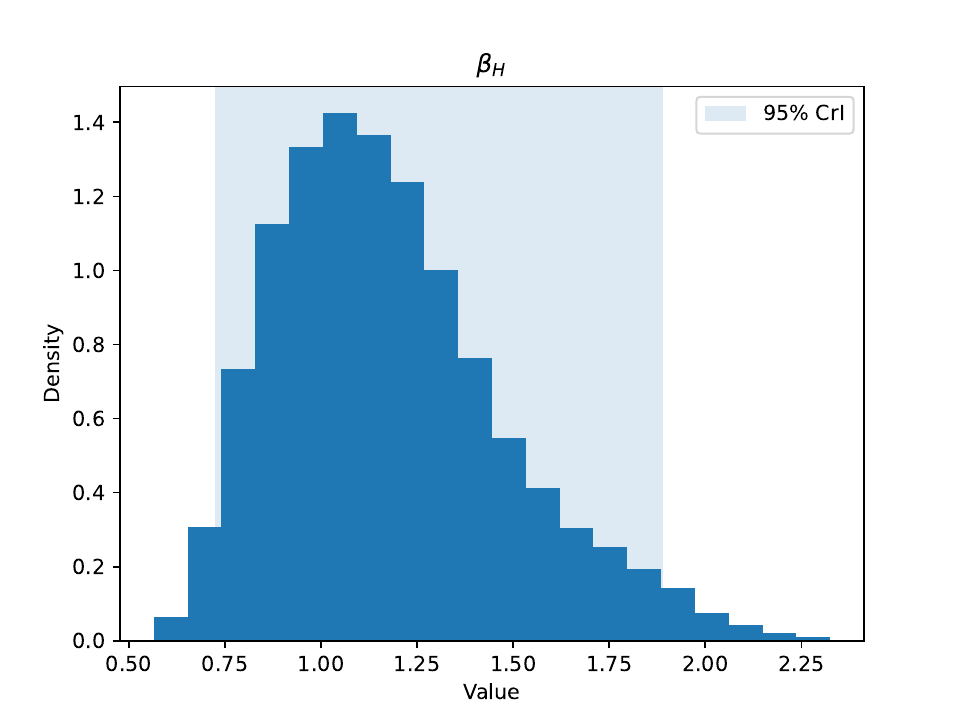}
     \end{subfigure}
     \hfill
     \\
     \hfill
     \begin{subfigure}[h]{0.45\textwidth}
         \centering
         \includegraphics[width=\textwidth]{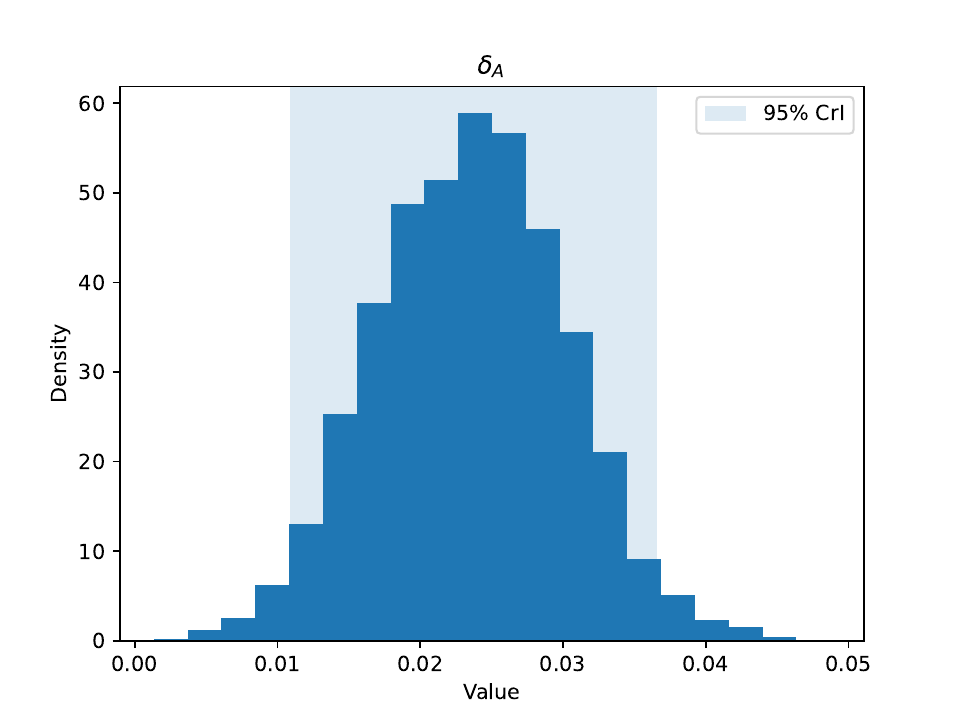}
     \end{subfigure}
     \hfill
     \begin{subfigure}[h]{0.45\textwidth}
         \centering
         \includegraphics[width=\textwidth]{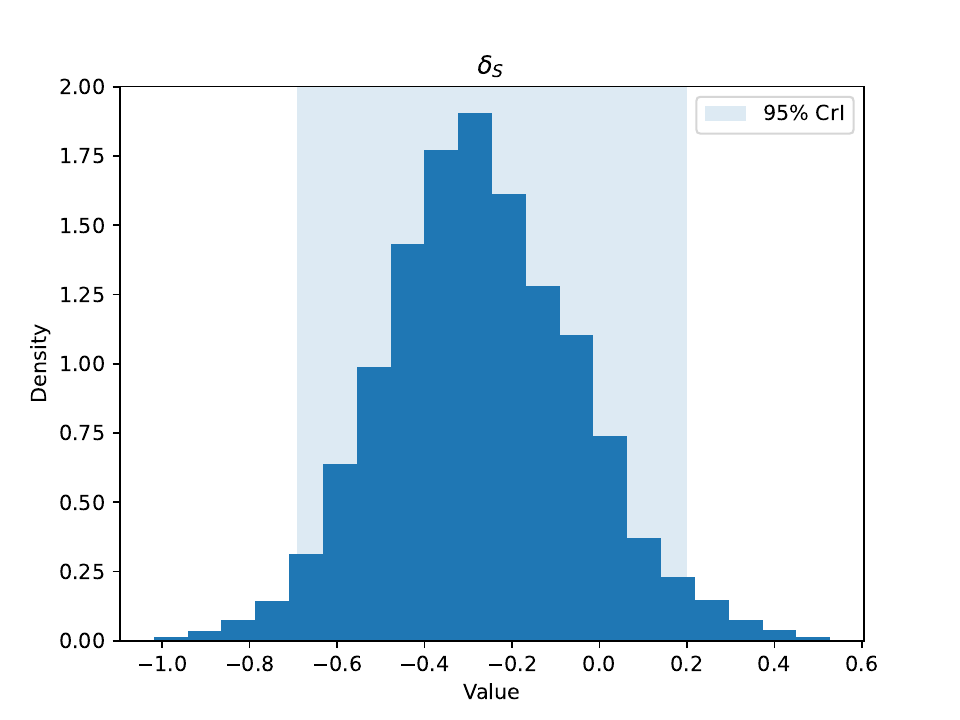}
     \end{subfigure}
     \hfill
     \caption{Histograms of the posterior samples generated for the parameters $\bm{\theta}=(\beta_G,\beta_H,\delta_A,\delta_S)$ for 3GC-R \textit{E. coli} transmission using test data from the AntiDOTE study (using the Rippler algorithm for inference). The 95\% credible interval is shown in light blue.}
     \label{real_parameters}
\end{figure}
\begin{figure}[h]
    \centering
    \includegraphics[width=0.85\textwidth]{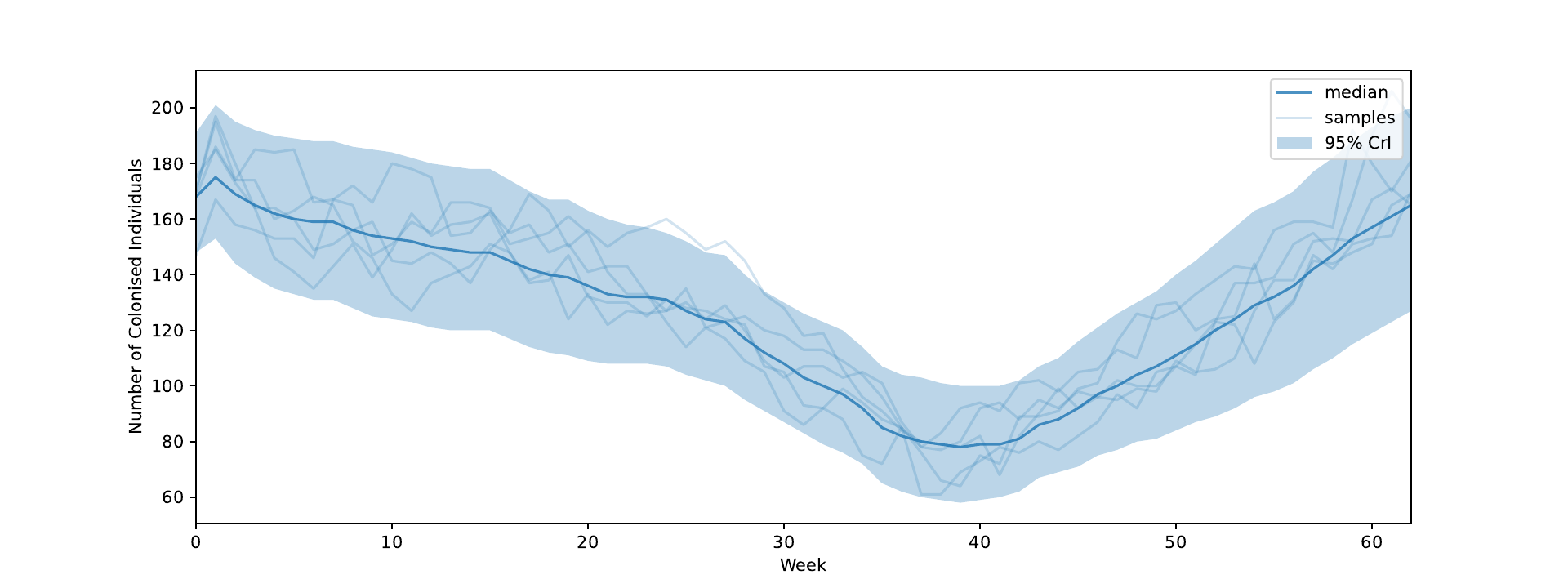}
    \caption{The number of colonised individuals over time for 3GC-R \textit{E. coli} transmission using test data from the AntiDOTE study (using the Rippler algorithm for inference). The median value is shown by the dark blue line and 5 randomly chosen samples are shown by the faint blue lines. The 95\% credible interval is shown in light blue.}
    \label{real_latent}
\end{figure}

The histogram for the posterior samples of $\beta_G$, $\beta_H$, $\delta_A$, and $\delta_S$ are shown in Figure \ref{real_parameters}, with the median and 95\% credible interval of each sample given in Table \ref{real_table}. The trace plots for the posterior samples of the parameters are shown in Figure S3. Figure \ref{real_latent} shows the 95\% credible interval for the number of colonised individuals over time.


\begin{table}[h!]
    \centering
    \caption{Parameter estimates for 3GC-R \textit{E. coli} transmission using test data from the AntiDOTE study in Malawi \citep{lester2021third} using the Rippler algorithm for inference.}
    \begin{tabular}{ c  c  c  c}
        \hline Parameter & Definition & Median & 95\% CrI
        \\ \hline $\beta_G$ & global transmission rate (week$^{-1}$) & 0.123 & (0.0853, 0.166)
        \\ $\beta_H$ & relative within-household transmission & 1.14 & (0.723, 1.89)
        \\ $\delta_A$ & age covariate effect (year$^{-1}$) & 0.0237 & (0.0109, 0.0366)
        \\ $\delta_S$ & sex covariate effect & -0.282 & (-0.691, 0.120)
        \\ \hline
    \end{tabular}
    \label{real_table}
\end{table}


We can use these posterior samples to determine the relative increase in colonisation pressure when comparing individuals inside and outside a common household. For example, we consider the time $t=20$, when the seasonality term has a value of 1.24. The colonisation pressure on some individual $j$, with no colonised individuals in their household, has a 95\% credible interval of $(0.0271,0.0634)$. When one colonised individual is moved into their household this rate increases to $(0.160,0.209)$; the relative increase in colonisation rate \[\frac{\lambda^G_j(21, \bm{x}_{20},\bm{\theta})+\lambda^H_j(21, \bm{x}_{20},\bm{\theta})}{\lambda^G_j(21, \bm{x}_{20},\bm{\theta})}\] has a 95\% credible interval of $(2.95, 6.70)$. This increases the 95\% credible interval for the probability that individual $j$ is colonised at time $t=21$ from $(0.0268,0.0614)$ to $(0.148,0.189)$. Note that the values of the probabilities are close to the values of the rates -- this is because $1-e^{-x} \approx x$ for small $x$, and so doubling the colonisation rate also (approximately) doubles the probability of colonisation. 

We can make similar calculations for any other point in time; we find that the lower bound of the 95\% credible interval of the relative increase in colonisation rate is always greater than 2. This means that, at any point in time and in at least 97.5\% of posterior samples, adding one colonised individual to an uncolonised household will more than double the colonisation pressure on individuals in that household. Clearly there is significantly greater risk of colonisation from colonised individuals in a common household than from the general population.

\section{Comparison with Existing Inference Methods} \label{method_comparison}


We now use the simulated data to compare the Rippler method to two other methods for the data-augmentation step of our MCMC: a reversible-jump MCMC method, and a FFBS method. Details of these two methods (including the inference algorithm and a simulation study) are given in sections S2 and S3 of the supplementary materials.

In the reversible-jump method, the current state of the latent variables is expressed as a sequence of events: for this model we have colonisation events (moving from state U to state C) and decolonisation events (moving from state C to state U). The proposal algorithm chooses an individual uniformly at random, and then randomly chooses to either move one of their events, add a new event to their colonisation history, or remove an event from their colonisation history. We then accept or reject this change using a Metropolis-Hastings step, taking into account the probability of the reverse move. The original reversible-jump algorithm was proposed in \cite{gibson1998estimating}, and expanded for transmission models in \cite{o1999bayesian}. In our model, individuals can become colonised multiple times, so adding or removing events requires a second event to be added or removed respectively -- therefore we modify the method using the algorithm presented in \cite{spencer2015super}.

The other method we consider is the individual forward filtering backwards sampling (iFFBS) algorithm from \cite{touloupou2020scalable}. This is an extension of the general FFBS method from \cite{carter1994gibbs} and \cite{chib1996calculating}. The FFBS algorithm samples from the conditional distribution $\pi(\bm{X}|\bm{\theta},\bm{Y})$ in two stages. First, we iteratively calculate the forward probabilities $\pi(\bm{x}_{t}|\bm{\theta},\bm{Y}_{0:i})$ from $t=0$ to $t=T$. The final probability calculated in this step is $\pi(\bm{x}_{T}|\bm{\theta},\bm{Y}_{0:T})$ -- this is the full conditional distribution for $\bm{x}_T$, and so we can use it to sample for $\bm{x}_T$. We then sample iteratively backwards, sampling $\bm{x}_t$ from $\pi(\bm{x}_t|\bm{\theta},\bm{x}_{i+1},\bm{Y}_{0:i})$ from $t=T-1$ to $t=0$. However, since $\bm{x}_t$ has $2^N$ possible states for any $t$, this method will be computationally slow, and the iFFBS method will be used instead. 

The iFFBS algorithm chooses an individual uniformly at random and performs a FFBS update on that individual only -- the latent variables for all other individuals are held constant. The forward filtering probabilities are modified from the general FFBS method to account for all other individuals being fixed. This sample is from the full conditional distribution of that individual's latent variables (given all the other latent variables), so no Metropolis-Hastings accept-reject step is required. Since this is effectively a Gibbs step and within-individual correlation in colonisation status over time is stronger than between-individual correlation, we expect iFFBS to be the best of the three algorithms for efficient exploration of the latent variable space.

We first compare the computational complexity of the three algorithms. All three methods involve calculating $\lambda_j(t, \bm{x}_{t-1},\bm{\theta})$ for, at most, each individual $j$ and time-step $t$. For the Rippler algorithm, this occurs twice: when we sample a random number matrix $\bm{U}$ that could have produced $\bm{X}$, and when we perform the `ripple step', converting the new random numbers matrix $\bm{U}^{\ast}$ into the proposed latent variables $\bm{X}^{\ast}$. For the reversible-jump method, this is when we calculate the conditional distribution of the proposed latent variables $\bm{X}^{\ast}$ (given $\bm{\theta}$ and $\bm{Y}$). For iFFBS, this is when we calculate the forward probabilities and must take into account the probability of the latent variables that are being held fixed. To calculate $\lambda_j(t, \bm{x}_{t-1},\bm{\theta})$ for all individuals at a given time-step, we must calculate $\bm{H}\bm{x}^\top_{t-1,\bm{\cdot}}$. Both $\bm{H}$ and $\bm{x}^\top_{t-1,\bm{\cdot}}$ have length $N$, so the computational complexity of all three methods scales linearly with timespan $T$ and quadratically with population size $N$.

We can also consider the computational complexity with respect to the number of states in the model $S$. In this case we have $S=2$. The computational complexity of the reversible-jump method is clearly independent of $S$; when calculating the likelihood of the proposed latent variables we calculate only the probability of the events that occur in the proposed latent variables (one per individual-time-step). For the Rippler algorithm, the computational complexity scales linearly with $S$: when both sampling the random number matrix and performing the `ripple step', we must calculate the probability of moving to each of the $S$ states (for each individual and time-step). However, in the iFFBS method one step of calculating the forward probabilities involves a sum of $S$ terms calculated $S$ times for each individual and time-step \citep{touloupou2020scalable} -- since this is a Gibbs sampler, we must calculate the probability of moving from any one of $S$ states to each of the $S$ states. Therefore the computational complexity of the iFFBS method scales quadratically with $S$. This means the overall computational complexity of the reversible-jump method is $\mathcal{O}(N^2T)$, the computational complexity of the Rippler method is $\mathcal{O}(2SN^2T)$, and the computational complexity of the iFFBS method is $\mathcal{O}(S^2N^2T)$.

We also compare the mean squared jump distance (MSJD) of the latent variables for the three methods. MSJD is defined as the average squared difference in the elements of the colonisation matrix between consecutive iterations: 
\begin{equation*}
    \text{MSJD} = \frac{\displaystyle\sum_{k=1}^K\sum_{t=0}^T\sum_{j=1}^N \left(x_{t,j}^{(k)}-x_{t,j}^{(k-1)} \right)^2}{K}.
\end{equation*} 

We found that the MSJD for the Rippler algorithm performs better than the reversible-jump method, but significantly worse than the iFFBS method. When 400 latent updates are performed within each MCMC iteration (so that there is sufficient mixing of the latent variables between each parameter update), the MSJD is 247 for the reversible-jump algorithm, 1710 for the Rippler algorithm, and 4340 for the iFFBS algorithm. This is roughly what we expect -- the iFFBS algorithm (which changes the whole colonisation history of each individual updated) is proposing bigger changes than the Rippler algorithm (which changes part of the colonisation history of at least one individual but potentially more), and the Rippler algorithm is proposing bigger changes than the reversible-jump algorithm (which changes just one colonisation/de-colonisation period for one individual). The Rippler and iFFBS algorithms are more successful because they exploit the transmission model structure in some way, whereas the reversible-jump algorithm only takes into account the transmission model structure during the Metropolis-Hastings accept-reject step. However, only the iFFBS algorithm takes into account the observation model during the proposal step (with the reversible-jump and Rippler algorithms using the observation model likelihood during the accept-reject step). This places the Rippler algorithm halfway between the other two algorithms, using the transmission model but not the observation model during the proposal step.


Consider a proposed update to the latent variables using the Rippler method. Let $t^{\ast}$ be the time at which we alter one value in $\bm{U}$. This change to $\bm{U}$ will `ripple through' the epidemic, potentially changing the latent variables for any time $t>t^{\ast}$. Since $t^{\ast}$ is chosen uniformly from $\{0,\dots,T\}$, changes will occur more often to latent variables later in the epidemic history.

\begin{figure}[h]
     \centering
     \begin{subfigure}[h]{0.32\textwidth}
         \centering
         \includegraphics[width=\textwidth]{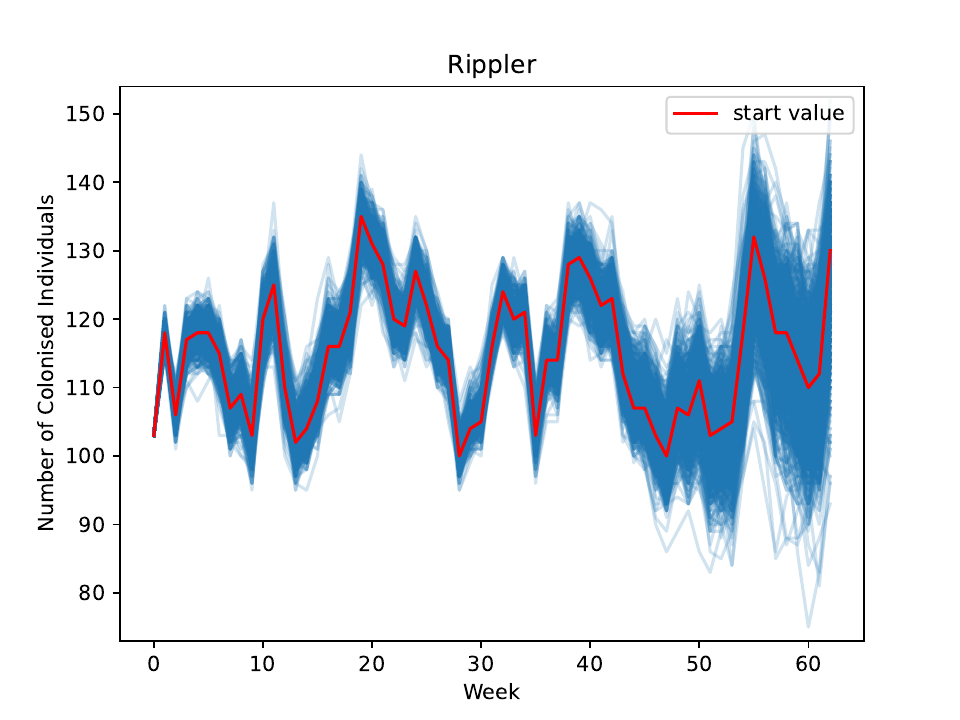}
     \end{subfigure}
     \begin{subfigure}[h]{0.32\textwidth}
         \centering
         \includegraphics[width=\textwidth]{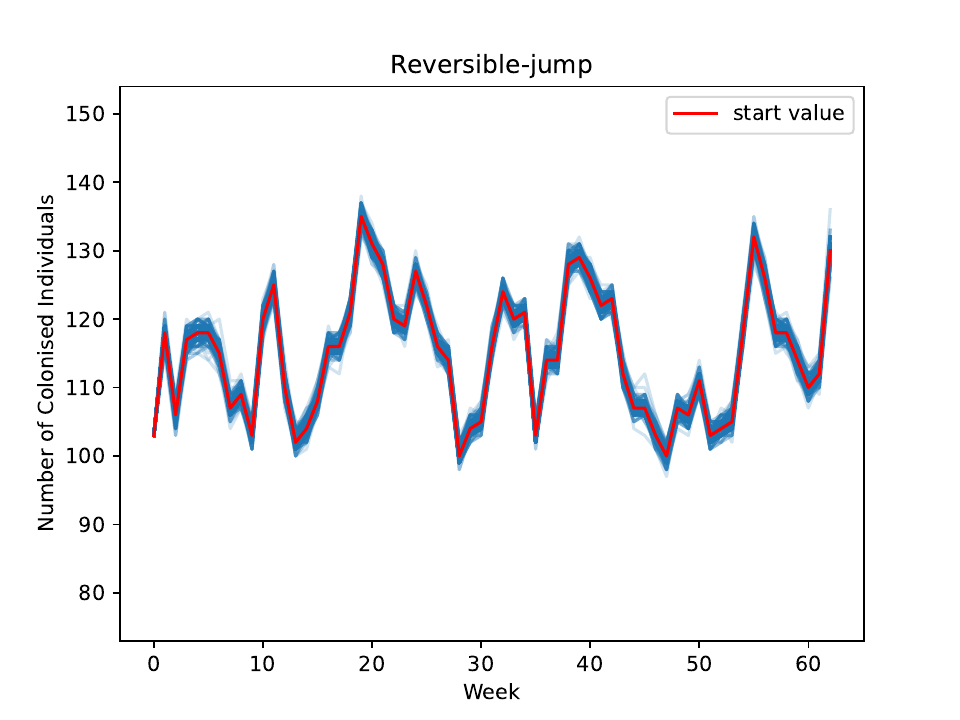}
     \end{subfigure}
     \begin{subfigure}[h]{0.32\textwidth}
         \centering
         \includegraphics[width=\textwidth]{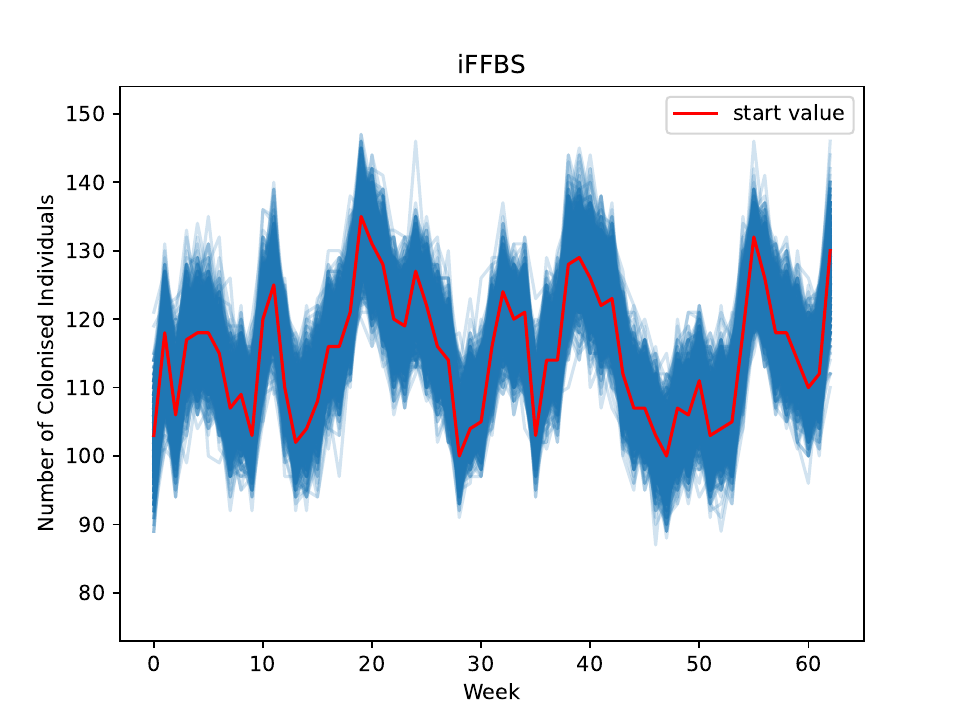}
     \end{subfigure}
     \caption{Moves of 1000 proposed updates to the latent variables, each using the Rippler (left), reversible-jump (middle), or iFFBS (right) algorithms 100 times. The starting value of each move is shown by the red line.}
     \label{updates_accepted_all}
\end{figure}

Figure \ref{updates_accepted_all} shows the moves of 1000 proposed updates to the latent variables (starting from the same latent variables every time, with each using the Rippler algorithm 100 times). The seasonality term has been removed to avoid any possible interference with the acceptance rate of the moves. The results when using the other two methods are also shown in Figure \ref{updates_accepted_all} for comparison. We can see that Rippler updates generally move more than reversible-jump updates, but less than iFFBS updates. In addition, we see that as the epidemic progresses there are both more and larger Rippler moves away from the current latent variable value -- we believe this is because ripples that start early have had time to move significantly far away by the end of the timeframe.

\begin{figure}[h]
     \centering
     \hfill
     \begin{subfigure}[h]{0.45\textwidth}
         \centering
         \includegraphics[width=\textwidth]{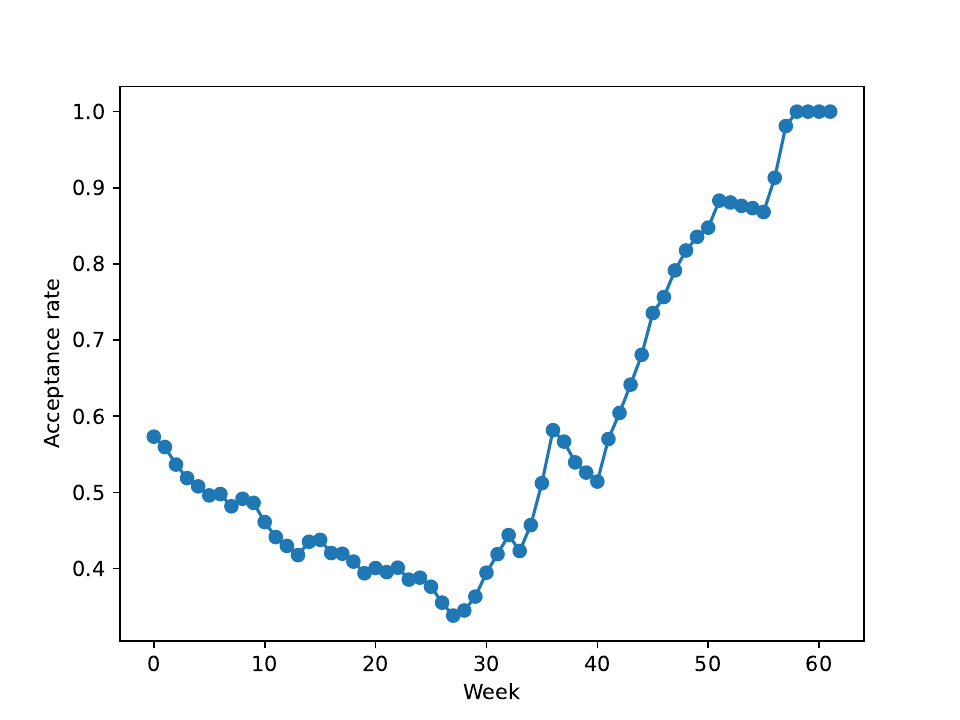}
     \end{subfigure}
     \hfill
     \begin{subfigure}[h]{0.45\textwidth}
         \centering
         \includegraphics[width=\textwidth]{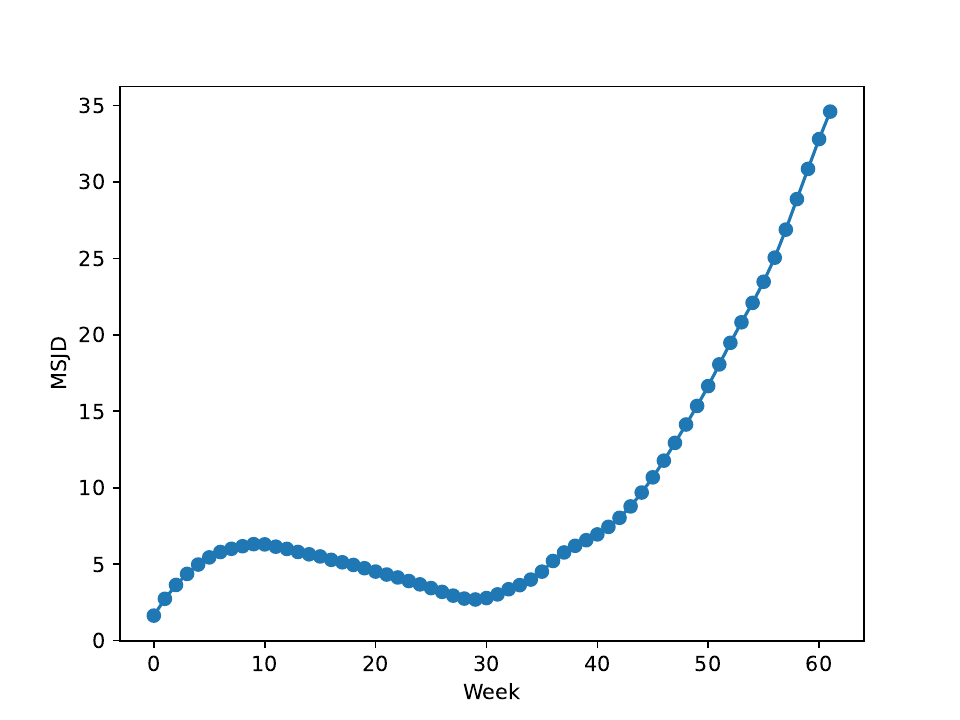}
     \end{subfigure}
     \hfill
     \caption{Acceptance rate by time of ripple start (left) and MSJD for each point in time across all individuals (right).}
     \label{acceptance_MSJD_time}
\end{figure}

We see in Figure \ref{acceptance_MSJD_time} (left) that the acceptance rate for the latent variable update changes based on what time-step the ripple starts from (we have also removed the seasonality term here). We see that the acceptance rate significantly increasing from $t=28$ onwards. We believe this is because starting the ripple from a later time-step results in a smaller change to the latent variables, and so the change is more likely to be accepted.

This means that the time-step of a latent variable change being proposed has two effects: ripples that are started early move further away from the current trajectory as time increases, and ripples that are started late get accepted more often. Combining these two effects, we see in Figure \ref{acceptance_MSJD_time} (right) that the MSJD significantly increases for the latent variables in the second half of the epidemic. In the future we would like to improve the Rippler algorithm such that the mixing of the latent variables is not dependent on time -- one alternative would be to dampen the size of any ripple as it progressed (so that the size of all ripples would be similar).

\begin{figure}[h]
     \centering
     \hfill
     \begin{subfigure}[h]{0.45\textwidth}
         \centering
         \includegraphics[width=\textwidth]{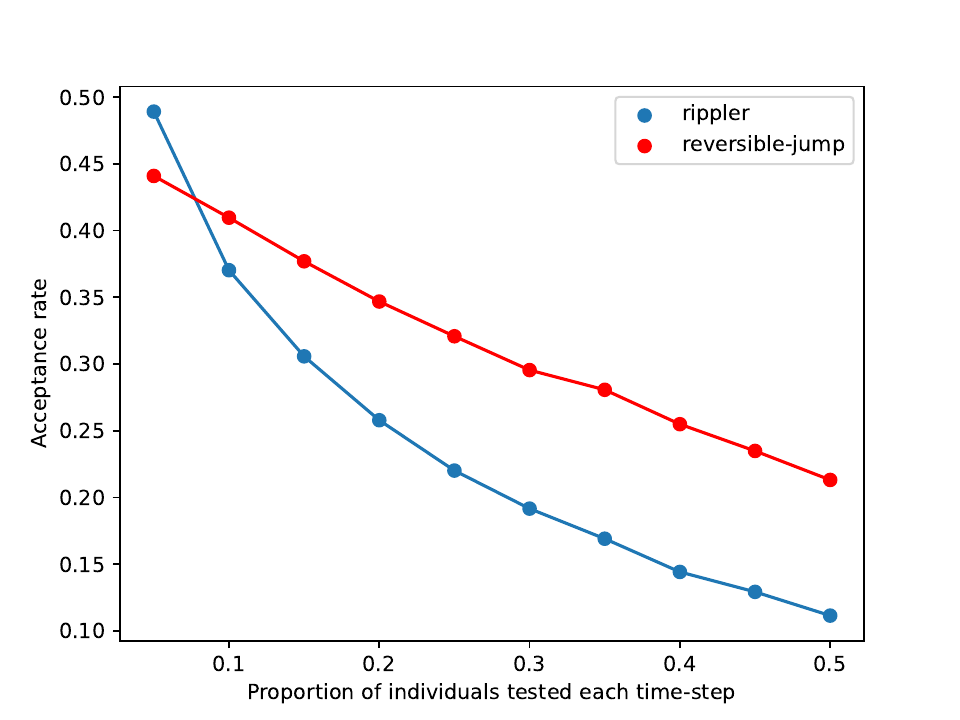}
     \end{subfigure}
     \hfill
     \begin{subfigure}[h]{0.45\textwidth}
         \centering
         \includegraphics[width=\textwidth]{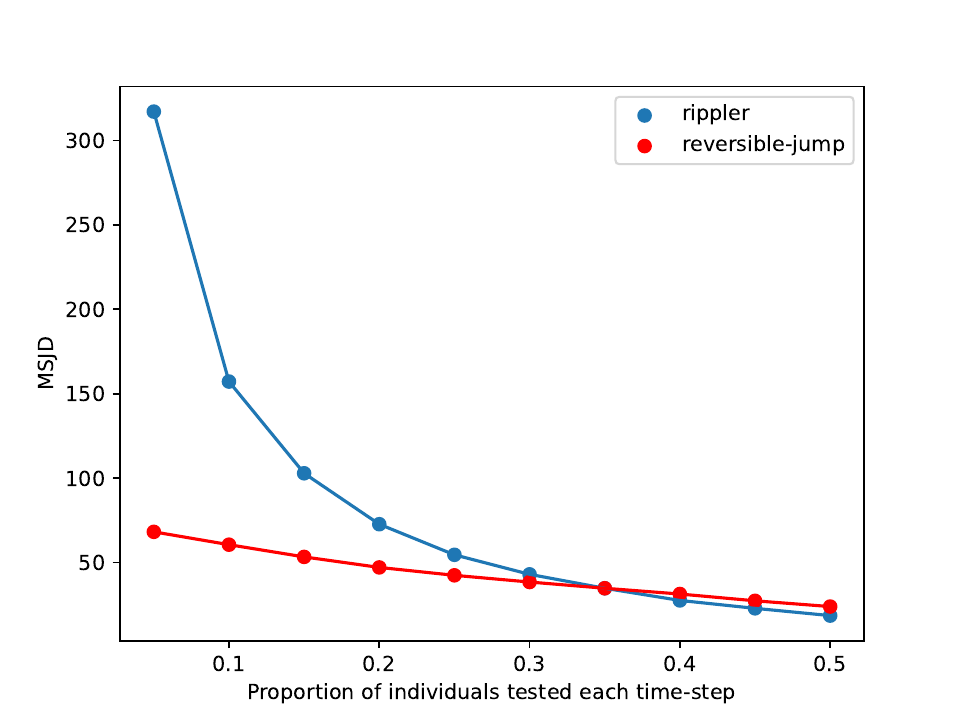}
     \end{subfigure}
     \hfill
     \caption{Changes to the acceptance rate (left) and MSJD (right) as the number of tests increases, for both the Rippler and reversible-jump algorithms. The Rippler is shown by the blue line and the reversible-jump is shown by the red line. The iFFBS algorithm is not included as there is no Metropolis-Hastings accept-reject step (it has MSJD significantly greater than either other method for all numbers of tests).}
     \label{test_prop}
\end{figure}

We also see in Figure \ref{acceptance_MSJD_time} that the acceptance rate, before increasing dramatically after $t=28$, is decreasing over time. We believe this is because the number of tests conducted per week is increasing over this time period; initially we only have the tests conducted on the first visit to households, then both first and second visits, and then the first, second and third visits. Later the number of tests conducted per week decreases, but at this point we believe the effect of the smaller ripples in increasing the acceptance rate is greater than the effect of the number of tests conducted. 

We have tested how the overall acceptance rate and mixing of the Rippler algorithm changes as the number of tests changes. Figure \ref{test_prop} (left) shows that for both the Rippler and reversible-jump algorithms the acceptance rate decreases as the number of tests increases. We also see in Figure \ref{test_prop} (right) that the MSJD increases for both methods as the number of tests decreases (this makes sense, since having fewer tests means the latent variable changes are less constrained). The MSJD for the Rippler algorithm increases faster than for the reversible-jump algorithm, due to the Rippler algorithm generally proposing bigger moves. Note that for real studies we typically only have an average of 5-10\% of individuals tested each time-step, where we see that the Rippler algorithm mixes significantly better than the reversible-jump algorithm.

Our model for 3GC-R \textit{E. coli} has $S=2$, and so the computational complexity for the Rippler and iFFBS methods are equal. Therefore the greater MSJD of the iFFBS method also means the iFFBS method has a greater MSJD per unit computational cost. However, as the number of states in a model increases, the computational complexity of the iFFBS method grows quadratically, while the computational complexity of the Rippler algorithm grows linearly (the ratio of the two computational costs is $S/2$). This means that the iFFBS method is unfeasible for models with a large number of states, and that for these models the MSJD per unit computational cost is greater for the Rippler method.

\section{Discussion}
\label{discussion}

In this article, we have developed a new Bayesian inference method for discrete-time individual-level stochastic epidemic models that uses the transmission model structure in the proposal of new latent states: the Rippler algorithm. Note that this method can also be extended to population-level models -- simulating from the transmission model using binomial draws, rather than Bernoulli. In a model with no branches (where there is at most one possible compartment an individual can transition to at any given point), we only require one random number per time-step. Any model with branches \citep{overton2022epibeds} requires further generalisation of the Rippler algorithm.

We have tested the Rippler algorithm on simulated data, finding that it can recover the model parameters and latent variables, and then applied it to AMR data from Malawi to estimate the community and household transmission rates for 3GC-R \textit{E. coli} bacteria. We found that within-household colonisation status is more important in determining community risk than overall community prevalence. We note, however, that the sampling of households was sparse, so we could not implement a full spatial model, which would have given us a more detailed picture of the role of proximity of infectious individuals, such as in \cite{rimella2023inference}.

We have compared the Rippler algorithm to two other existing methods for inference on partially observed stochastic epidemics: reversible-jump MCMC and iFFBS, the latter being the state-of-the-art for inference for individual-based stochastic epidemic models. The Rippler algorithm represents an improvement on the standard reversible-jump MCMC approach \citep{o1999bayesian, spencer2015super} to data augmentation for stochastic epidemic models in terms of the mixing of the MCMC; MSJD per iteration is greater for the Rippler algorithm than reversible-jump MCMC. Although the iFFBS approach \citep{touloupou2020scalable} outperforms the Rippler algorithm on MSJD for this model for similar computation time, we conjecture the Rippler algorithm will perform comparatively better for models with greater numbers of compartments (due to the computational cost of the Rippler algorithm scaling linearly with the number of compartments, while that of iFFBS scales with the number of compartments squared).




\section*{Author contributions statement}


J.N., L.C., C.J.~wrote and reviewed the manuscript. R.L., G.R.~reviewed the manuscript. R.L., J.N., L.C., C.J.~designed the AntiDOTE model. J.N.~wrote the code and analysed the results. R.L., W.B., N.F.~carried out the AntiDOTE study and collected the data. L.C., C.J.~supervised the research presented in this paper. R.L., N.F., C.J.~obtained the funding for the research.

\section*{Acknowledgments}

This paper is based on work completed while James Neill was part of the EPSRC funded STOR-i centre for doctoral training EP/S022252/1. This work was supported by the Malawi-Liverpool Wellcome Trust Major Overseas Programme Core Grant (grant number 206545/Z/17/Z). Rebecca Lester was supported by a Wellcome Trust Clinical PhD Fellowship (University of Liverpool block award grant number 203919/Z/16/Z). The authors thank Chris Sherlock for his valuable suggestions. The authors thank the Data Management department and the Laboratory staff at the Malawi-Liverpool Wellcome Programme.

\section*{Supplementary Material}

The Supplementary Material includes the inference algorithms and simulation studies for the reversible-jump MCMC and iFFBS methods, as well as further details on the Rippler algorithm. The Python code used to run the simulation and inference, as well as the AntiDOTE dataset, can be found at \url{https://github.com/neilljn/AntiDOTE_Methods}.

\bibliography{reference.bib}

\end{document}


\maketitle

\renewcommand{\thefigure}{S\arabic{figure}}
\renewcommand{\theequation}{S\arabic{equation}}
\renewcommand{\thetable}{S\arabic{table}}
\renewcommand{\thealgorithm}{S\arabic{algorithm}}
\renewcommand{\thesection}{S\arabic{section}}
\renewcommand{\theproposition}{S\arabic{proposition}}

\section{Rippler Algorithm}

\subsection{AntiDOTE Data Analysis}

\begin{figure}[H]
     \centering
     \hfill
     \begin{subfigure}[h]{0.45\textwidth}
         \centering
         \includegraphics[width=\textwidth]{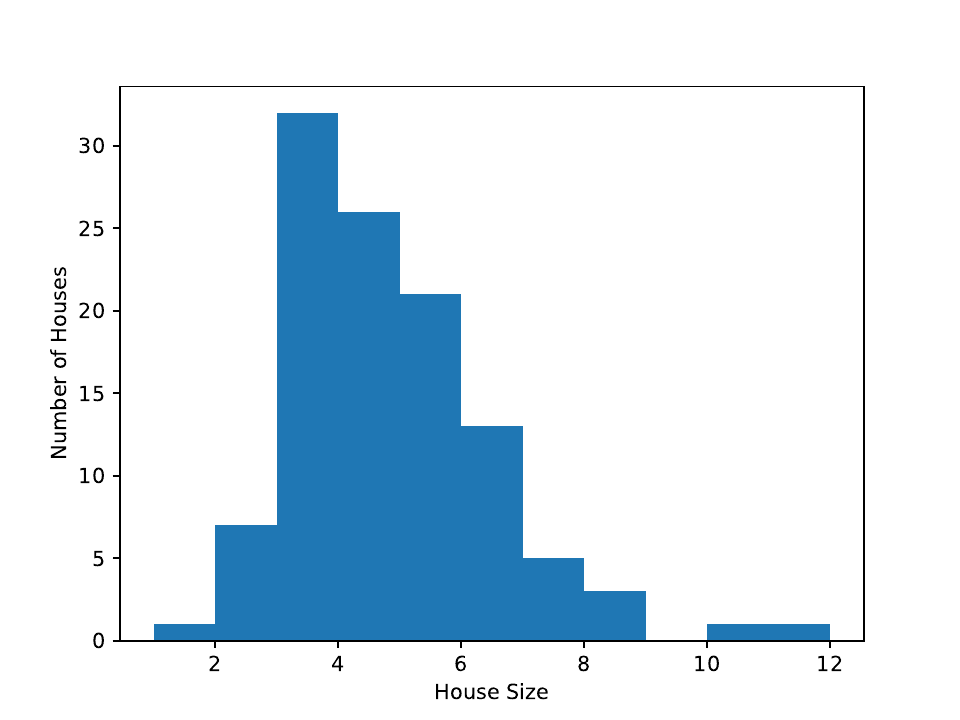}
     \end{subfigure}
     \hfill
     \begin{subfigure}[h]{0.45\textwidth}
         \centering
         \includegraphics[width=\textwidth]{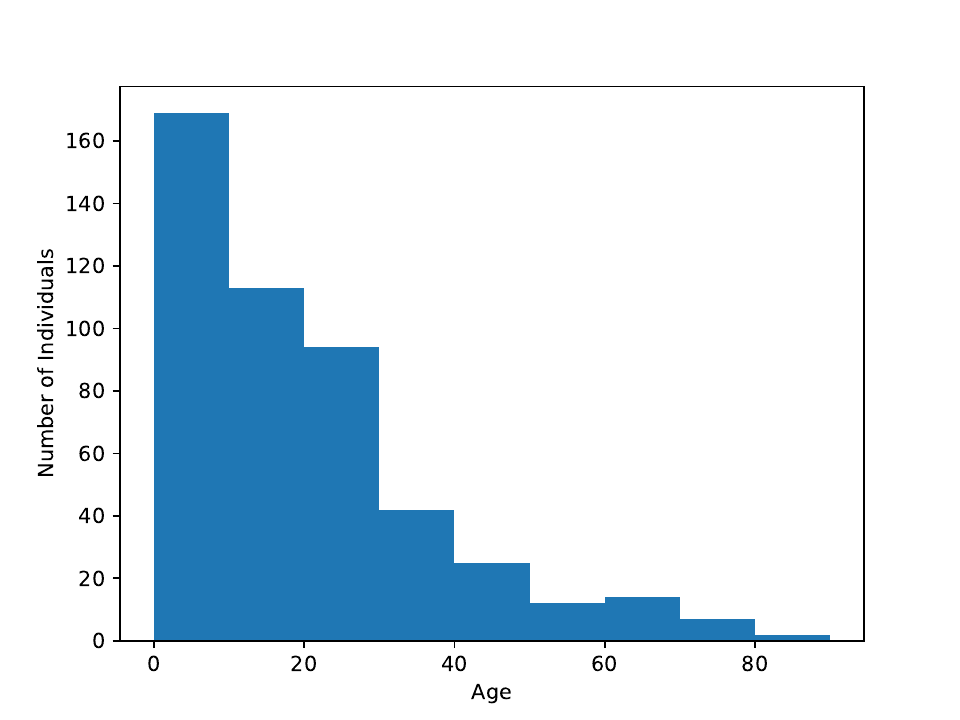}
     \end{subfigure}
     \hfill
     \caption{Household size (left) and age data (right) from the AntiDOTE study.}
     \label{h_and_age}
\end{figure}

\begin{figure}[H]
    \centering
    \hfill
    \begin{subfigure}[h]{0.45\textwidth}
        \centering
        \includegraphics[width=\textwidth]{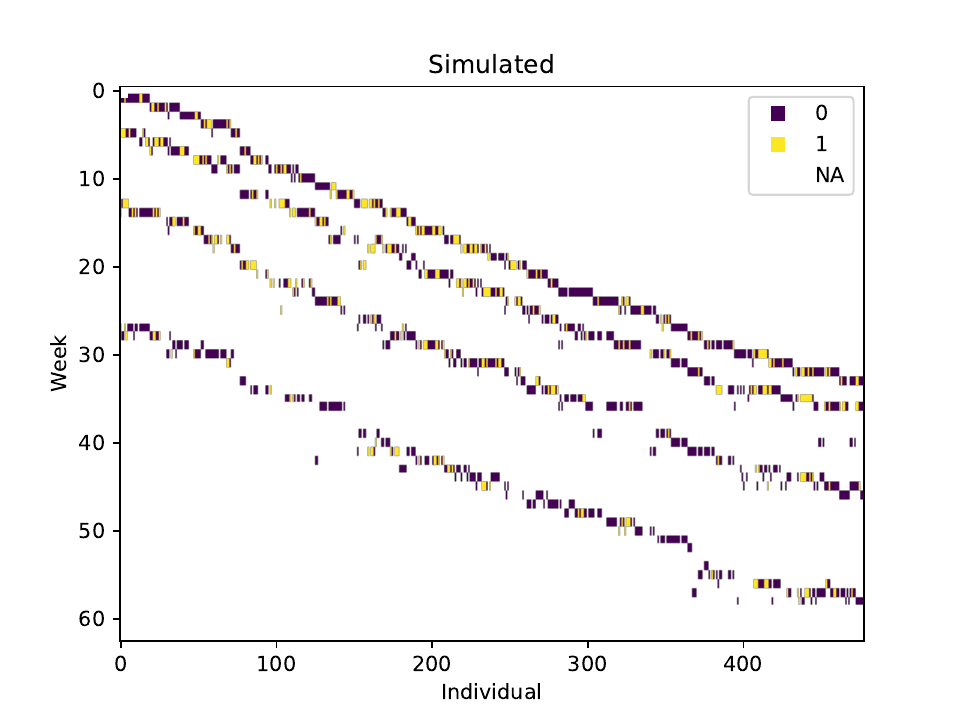}
    \end{subfigure}
    \hfill
    \begin{subfigure}[h]{0.45\textwidth}
        \centering
        \includegraphics[width=\textwidth]{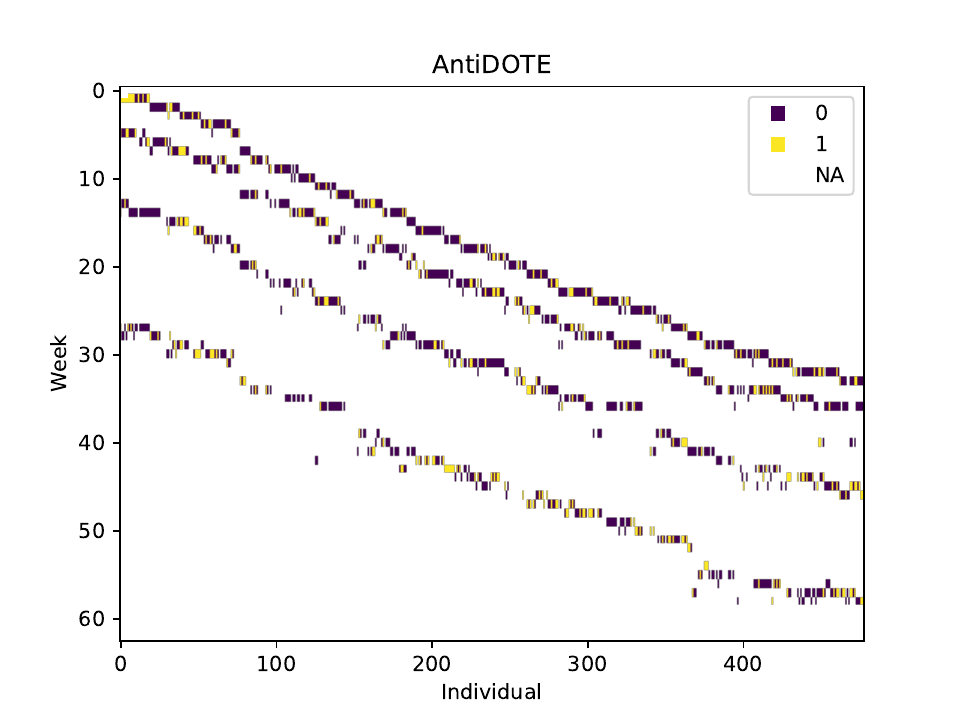}
    \end{subfigure}
    \hfill
    \caption{Test results for 3GC-R \textit{E. coli} using simulated test data (left) and test data from the AntiDOTE study (right). Yellow and purple squares represent positive and negative test results respectively.}
    \label{test_results}
\end{figure}

In the AntiDOTE study, the 478 individuals who were sampled are split between 110 houses, with the distribution of house sizes shown in Figure \ref{h_and_age} (left). We see from the histogram in Figure \ref{h_and_age} (right) that the ages of individuals (at the time they were first tested) were right-skewed, with a a majority of individuals under the age of 18. The test results for each individual over the 62 weeks are shown in Figure \ref{test_results}, with 5.60\% of entries containing a sample. We can clearly see when each individual was initially sampled, and then follow-up tests one, three, and six months later.

\begin{figure}[H]
     \centering
     \hfill
     \begin{subfigure}[h]{0.45\textwidth}
         \centering
         \includegraphics[width=\textwidth]{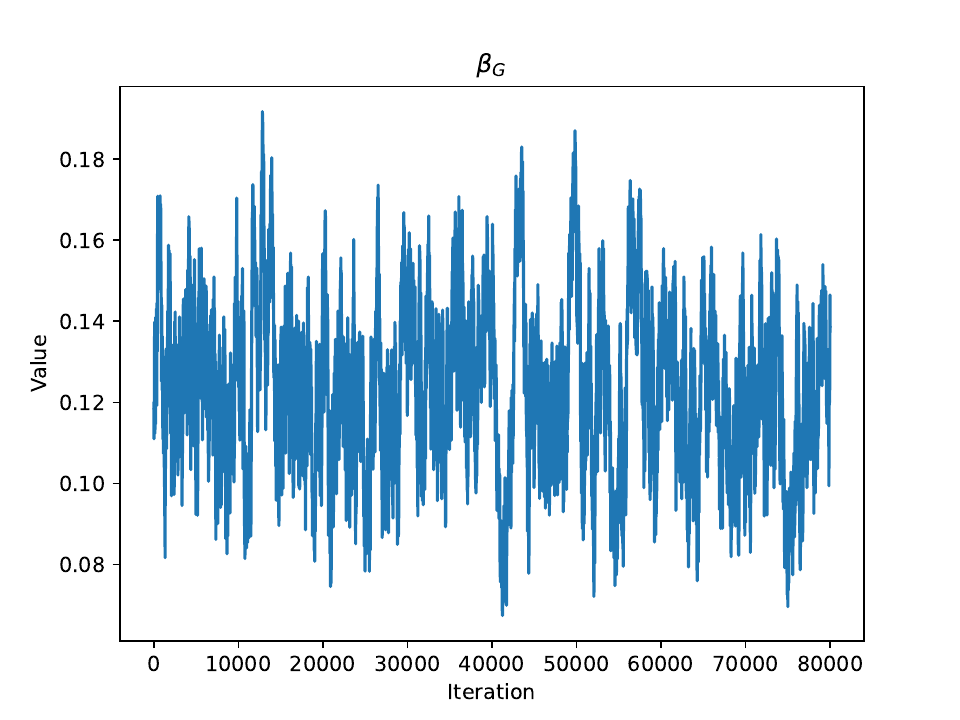}
     \end{subfigure}
     \hfill
     \begin{subfigure}[h]{0.45\textwidth}
         \centering
         \includegraphics[width=\textwidth]{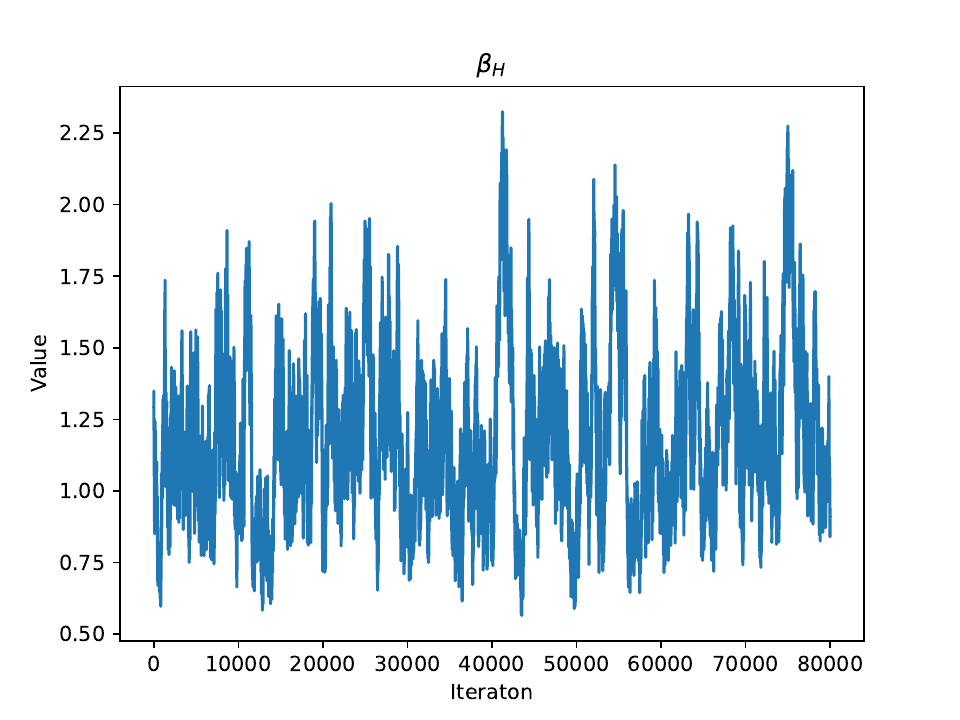}
     \end{subfigure}
     \hfill
     \\
     \hfill
     \begin{subfigure}[h]{0.45\textwidth}
         \centering
         \includegraphics[width=\textwidth]{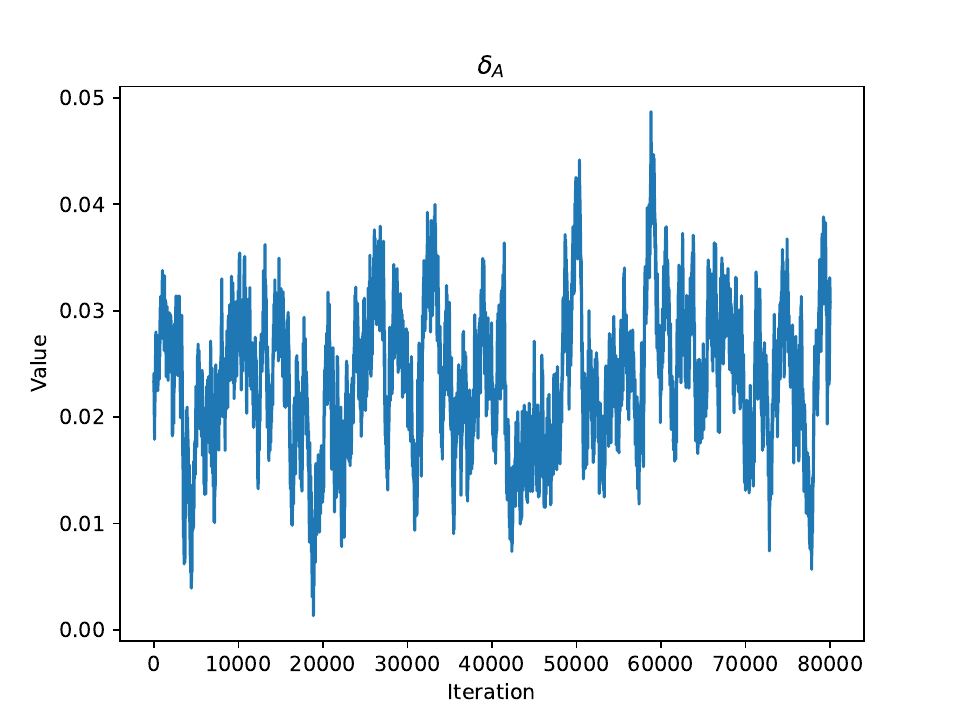}
     \end{subfigure}
     \hfill
     \begin{subfigure}[h]{0.45\textwidth}
         \centering
         \includegraphics[width=\textwidth]{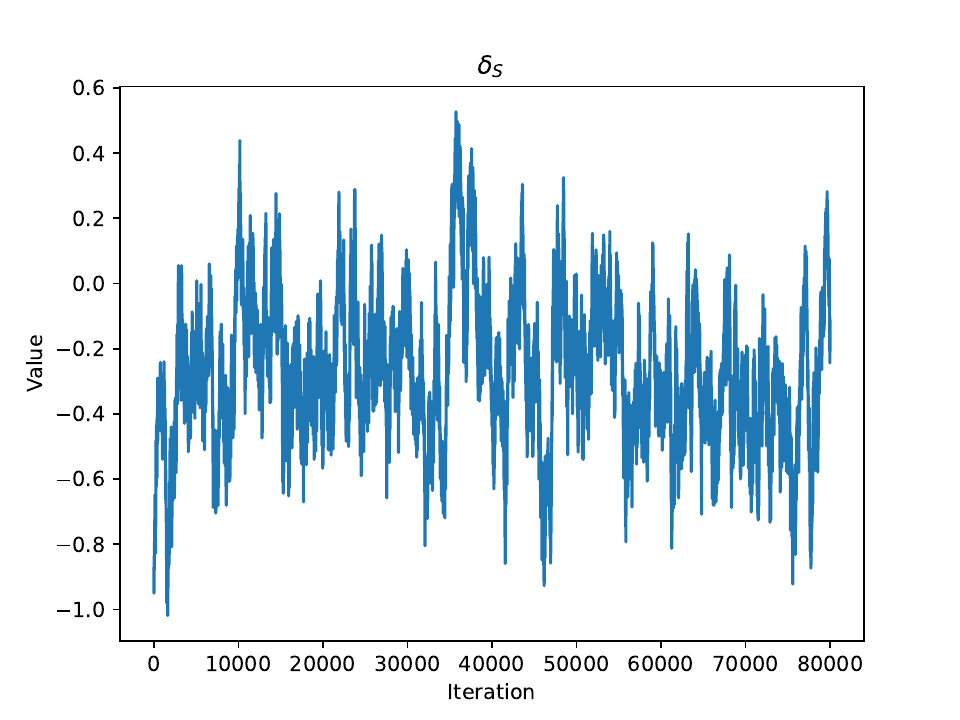}
     \end{subfigure}
     \hfill
     \caption{Trace plots of the posterior samples generated for the parameters $\bm{\theta}=(\beta_G,\beta_H,\delta_A,\delta_S)$ for 3GC-R \textit{E. coli} transmission using test data from the AntiDOTE study using the Rippler algorithm.}
     \label{real_parameters_trace}
\end{figure}

The parameter trace plots when using the Rippler algorithm for inference on the AntiDOTE test data are shown in Figure \ref{real_parameters_trace}. For full details of the AntiDOTE data analysis see the main text.

\subsection{Simulation Study}

\begin{figure}[H]
    \centering
    \hfill
    \begin{subfigure}[h]{0.45\textwidth}
        \centering
        \includegraphics[width=\textwidth]{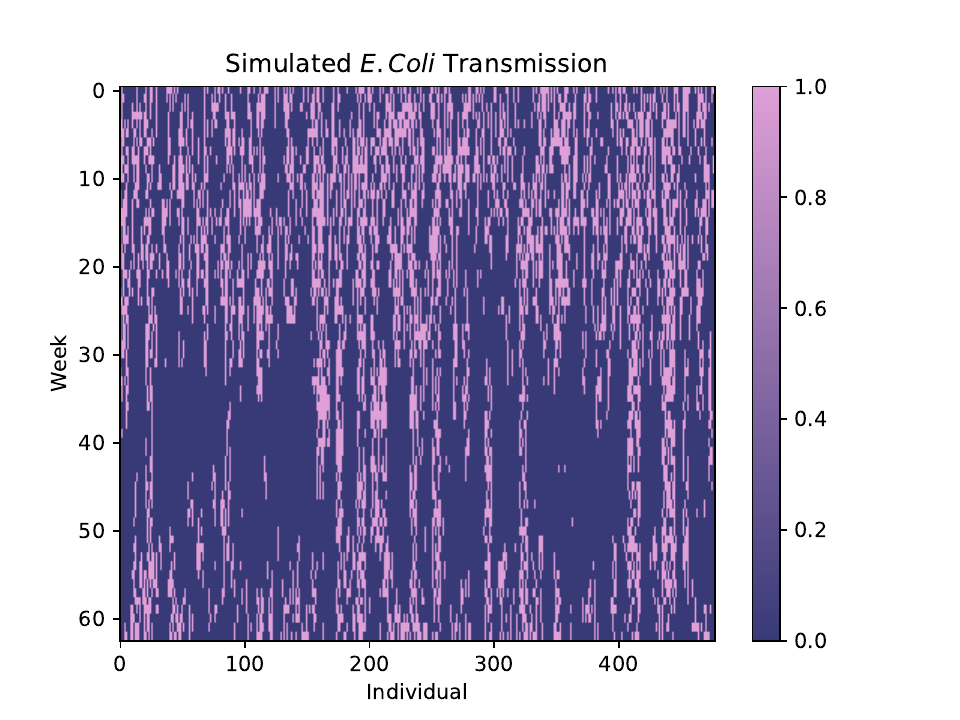}
    \end{subfigure}
    \hfill
    \begin{subfigure}[h]{0.45\textwidth}
        \centering
        \includegraphics[width=\textwidth]{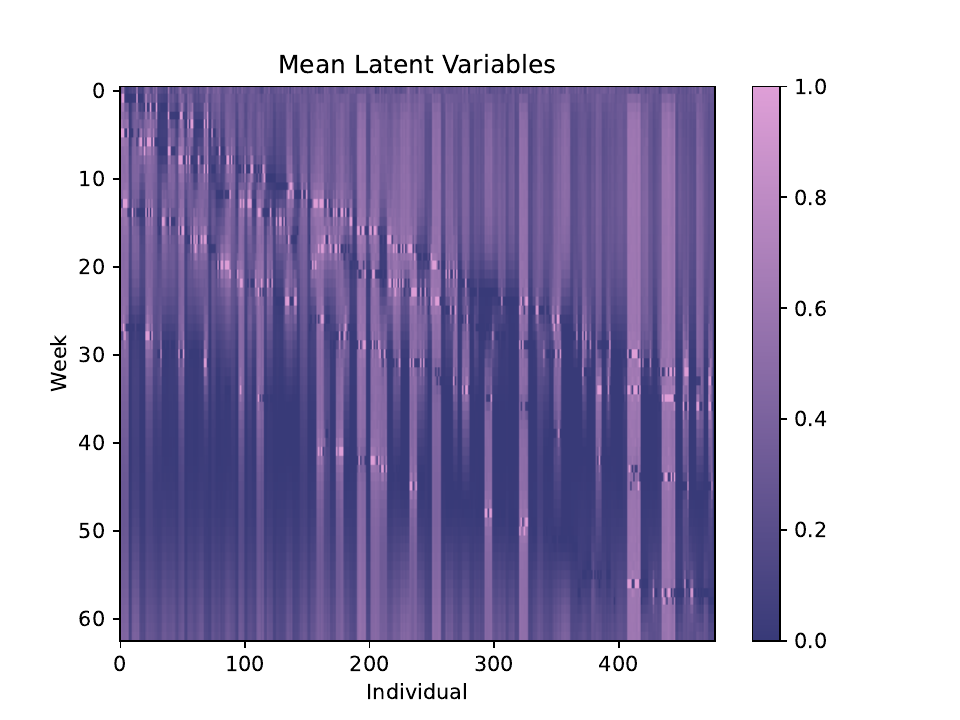}
    \end{subfigure}
    \hfill
    \caption{The underlying ``true'' simulation from the transmission model (left) and the mean value of each latent variable (right) for simulated 3GC-R \textit{E. coli} test data using the Rippler algorithm.}
    \label{latent_variables}
\end{figure}

The mean value of each latent variable when using the Rippler algorithm for inference on the simulated test data is shown in Figure \ref{latent_variables}, along with the underlying ``true'' simulation from the transmission model. We see that the Rippler algorithm is broadly able to identify the latent variables, with greater mean values in regions with more colonised individuals/time-points. In particular, the few households that are often colonised during the dry season are identified. We also see the four times each individual is tested as spots where the mean latent variables are close to 0 or 1; the shape of the bright spots closely matches Figure \ref{test_results}. This pattern does not appear on the underlying colonisation data because only the transmission model was used to simulate this data, not the observation model.

\begin{figure}[H]
     \centering
     \hfill
     \begin{subfigure}[h]{0.45\textwidth}
         \centering
         \includegraphics[width=\textwidth]{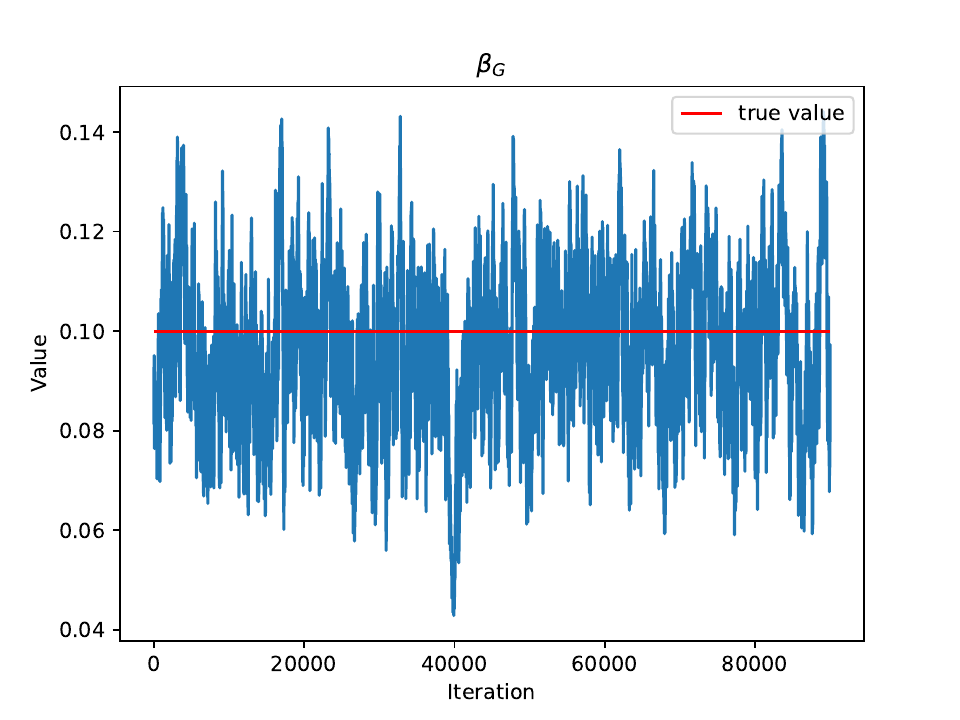}
     \end{subfigure}
     \hfill
     \begin{subfigure}[h]{0.45\textwidth}
         \centering
         \includegraphics[width=\textwidth]{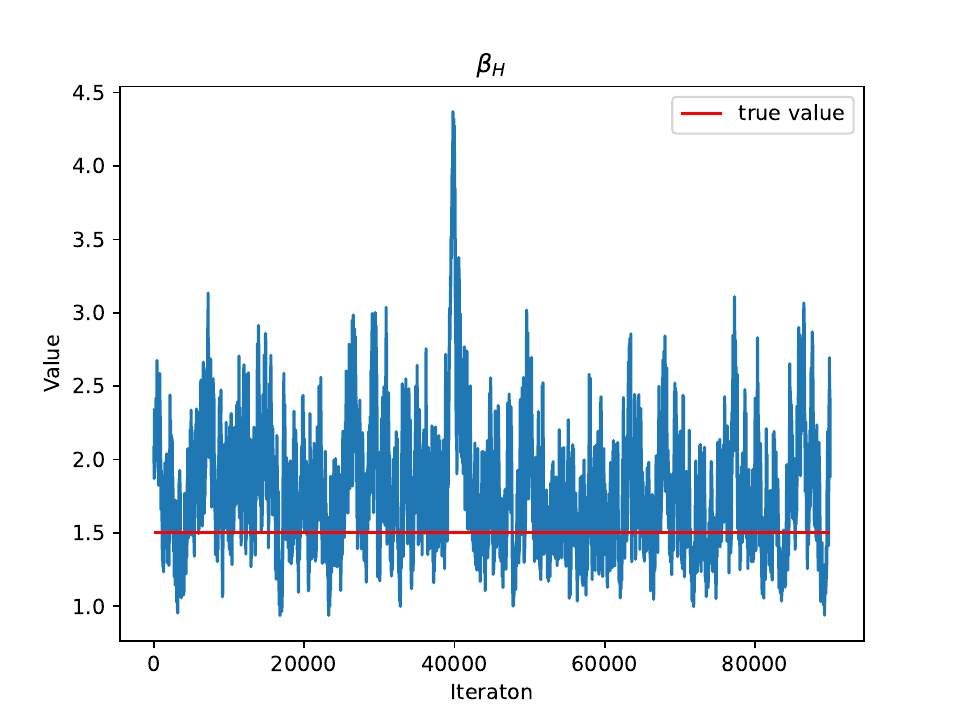}
     \end{subfigure}
     \hfill
     \\
     \hfill
     \begin{subfigure}[h]{0.45\textwidth}
         \centering
         \includegraphics[width=\textwidth]{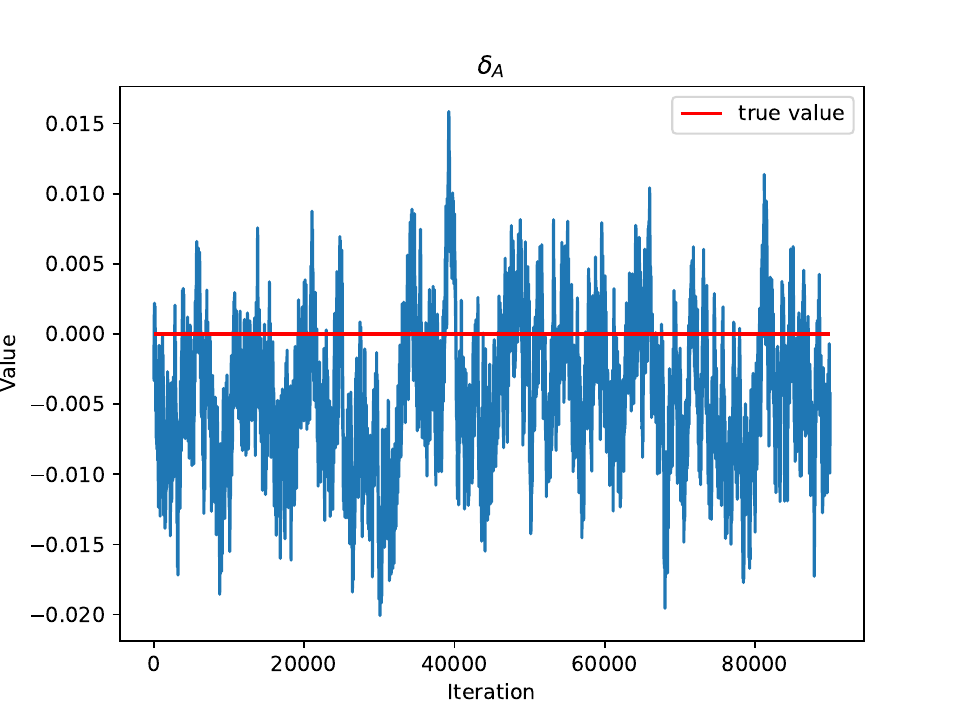}
     \end{subfigure}
     \hfill
     \begin{subfigure}[h]{0.45\textwidth}
         \centering
         \includegraphics[width=\textwidth]{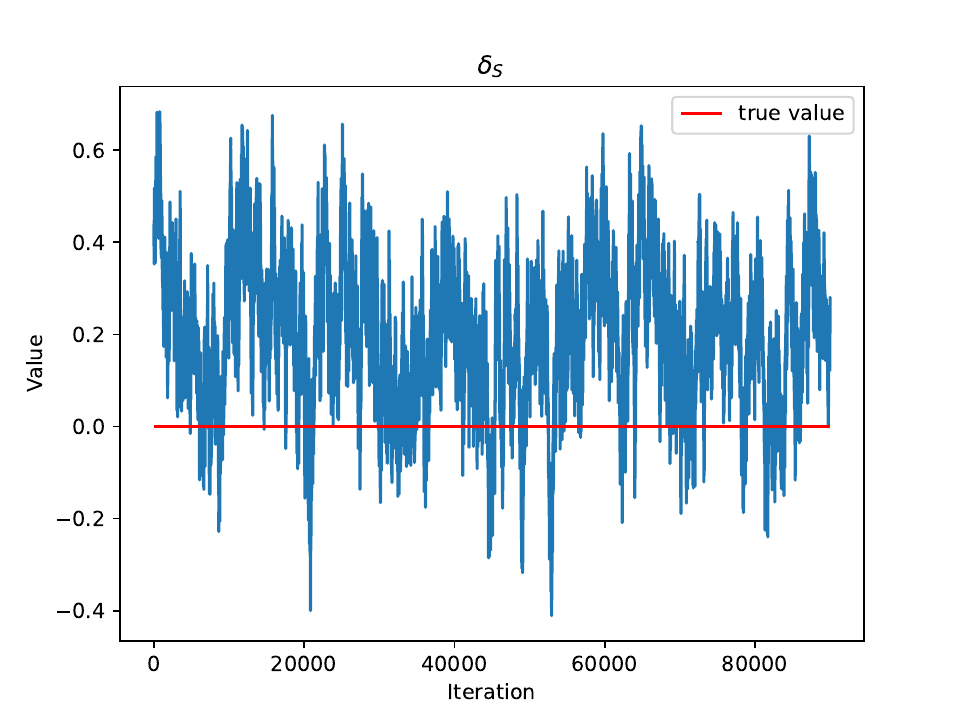}
     \end{subfigure}
     \hfill
     \caption{Trace plots of the posterior samples generated for the parameters $\bm{\theta}=(\beta_G,\beta_H,\delta_A,\delta_S)$ for simulated 3GC-R \textit{E. coli} test data using the Rippler algorithm. The true value is shown by the red line.}
     \label{sim_full_parameters_trace}
\end{figure}

Figure \ref{sim_full_parameters_trace} shows the parameter trace plots when using the Rippler algorithm for inference on the simulated test data. For full details of the simulation study using the Rippler algorithm see the main text.

\subsection{Rippler Tuning}

We also tested how the acceptance rate and mixing of the Rippler algorithm changes as we vary the number of elements of the random number matrix that are altered in the proposal step. Let the number of elements in the random number matrix to alter be $K''$. Then we follow Algorithm 4 as normal, but repeat the process in steps 7-9 $K''$ times. This means we must also change the proposal part of the $\alpha(\bm{X},\bm{X}^\ast)$ calculation in step 11 to \[q(\bm{U}^{\ast}|\bm{\theta},\bm{U}) = \frac{1}{\left(\sum_{t=0}^T\sum_{j=1}^N (1+a_{t,j}-b_{t,j})\right)^{K''}}.\]

\begin{figure}[H]
     \centering
     \hfill
     \begin{subfigure}[h]{0.45\textwidth}
         \centering
         \includegraphics[width=\textwidth]{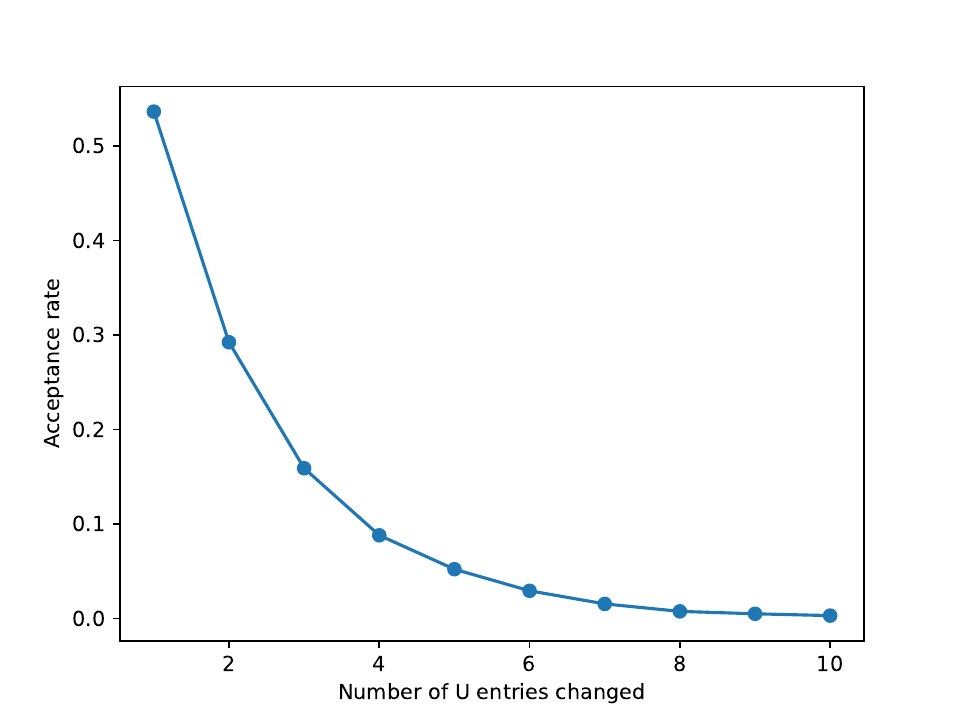}
     \end{subfigure}
     \hfill
     \begin{subfigure}[h]{0.45\textwidth}
         \centering
         \includegraphics[width=\textwidth]{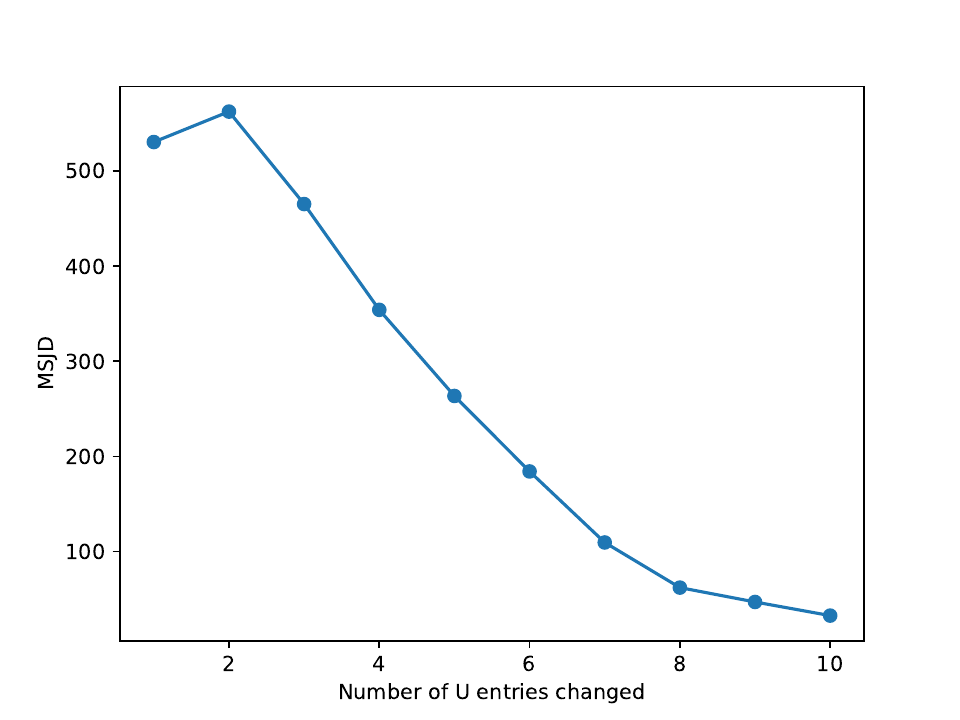}
     \end{subfigure}
     \hfill
     \caption{Changes to the acceptance rate (left) and MSJD (right) as the number of altered elements of the random number matrix in the Rippler proposal increases.}
     \label{U_changes}
\end{figure}

We see how this affects the MSJD in Figure \ref{U_changes} (left), with the MSJD maximised when 2 elements of $\bm{U}$ are changed in each step. In Figure \ref{U_changes} (right), we see that the acceptance rate decreases as the the number of altered elements of $\bm{U}$ increases; the greater difference between $\bm{U}$ and $\bm{U}^{\ast}$ leads to a greater difference between $\bm{X}$ and $\bm{X}^{\ast}$, and so it makes sense that this causes the acceptance rate to decrease. The size of changes to $\bm{X}$ that are likely to be accepted depends on the density of the observed data; we expect that with sparser observation data bigger moves to $\bm{X}$ would be more likely to be accepted, and so a greater number of altered elements of $\bm{U}$ would be optimal to maximise the MSJD.

\section{Reversible-jump MCMC} \label{appendix_rjmcmc}

\subsection{Inference Algorithm}

We use the method from Spencer et al.~(2015) to perform reversible-jump MCMC. As with the Rippler algorithm, we use a Metropolis-within-Gibbs approach to update our parameters and latent variables separately. We use the same method of updating the parameters $\bm{\theta}$ as in the Rippler algorithm (an adaptive random walk Metropolis algorithm). The initial conditions are updated separately from the rest of the latent variables, so we now derive both $\pi(\bm{x}_0|\bm{\theta},\bm{X}_{1:T},\bm{Y})$ and $\pi(\bm{X}_{1:T}|\bm{\theta},\bm{x}_0,\bm{Y})$. 

By Bayes' theorem we have \begin{equation*}
    \pi(\bm{\theta},\bm{X}|\bm{Y}) \propto \pi(\bm{Y}|\bm{X},\bm{\theta})\pi(\bm{X}|\bm{\theta})\pi(\bm{\theta}). \label{theta,X|Y_2}
\end{equation*} We know that $\pi(\bm{X}|\bm{\theta},\bm{Y}) \propto \pi(\bm{\theta},\bm{X}|\bm{Y})$. Let $N_0(\bm{X})=\sum_{j=1}^Nx_{0,j}$. Then from the main text we have \begin{align*} \begin{split}
    \pi(\bm{X}|\bm{\theta},\bm{Y}) &\propto s_e^{TP(\bm{X},\bm{Y})} (1-s_e)^{FN(\bm{X},\bm{Y})} s_p^{FP(\bm{X},\bm{Y})} (1-s_p)^{TN(\bm{X},\bm{Y})}
    \\ &~ \hspace{0.5cm} \prod_{(t,j) \in UC(\bm{X})} (1-e^{-\lambda_j(t, \bm{x}_{t-1},\bm{\theta})\Delta}) \prod_{(t,j) \in UU(\bm{X})} e^{-\lambda_j(t, \bm{x}_{t-1},\bm{\theta})\Delta}
    \\ &~ \hspace{0.5cm} (1-e^{-\gamma\Delta})^{|CU(\bm{X})|}~e^{-\gamma\Delta|CC(\bm{X})|}~p_0^{N_0(\bm{X})}~(1-p_0)^{N-N_0(\bm{X})}. 
\end{split} \end{align*}

This means we have \begin{align} \begin{split}
    \pi(\bm{x}_0|\bm{\theta},\bm{X}_{1:T},\bm{Y})
    &\propto \prod_{(t,j) \in UC(\bm{X}_{0:1})} (1-e^{-\lambda_j(t, \bm{x}_{t-1},\bm{\theta})\Delta}) \prod_{(t,j) \in UU(\bm{X}_{0:1})} e^{-\lambda_j(t, \bm{x}_{t-1},\bm{\theta})\Delta}
    \\ &~ \hspace{0.5cm} (1-e^{-\gamma\Delta})^{|CU(\bm{X}_{0:1})|}~e^{-\gamma\Delta|CC(\bm{X}_{0:1})|}~p_0^{N_0(\bm{X})}~(1-p_0)^{N-N_0(\bm{X})}, \label{X_0|theta,X_1:T,Y}
\end{split} \end{align} and \begin{align} \begin{split}
    \pi(\bm{X}_{1:T}|\bm{\theta},\bm{x}_0,\bm{Y}) &\propto s_e^{TP(\bm{X},\bm{Y})} (1-s_e)^{FN(\bm{X},\bm{Y})} s_p^{FP(\bm{X},\bm{Y})} (1-s_p)^{TN(\bm{X},\bm{Y})}
    \\ &~ \hspace{0.5cm} \prod_{(t,j) \in UC(\bm{X})} (1-e^{-\lambda_j(t, \bm{x}_{t-1},\bm{\theta})\Delta}) \prod_{(t,j) \in UU(\bm{X})} e^{-\lambda_j(t, \bm{x}_{t-1},\bm{\theta})\Delta}
    \\ &~ \hspace{0.5cm} (1-e^{-\gamma\Delta})^{|CU(\bm{X})|}~e^{-\gamma\Delta|CC(\bm{X})|}. \label{X_1:T|theta,X_0,Y}
\end{split} \end{align}

We propose a new value $\bm{x}_0^{\ast}$ by randomly choosing one individual $j$ and setting $x_{0,j}^{\ast} = 1-x_{0,j}$. For all other individuals $k$, we keep $x_{0,k}^{\ast}=x_{0,k}$. The probability of any one move is $1/N$, so the proposal ratio is \begin{equation*}
    \frac{q(\bm{x}_{0}|\bm{x}_0^{\ast})}{q(\bm{x}_0^{\ast}|\bm{x}_0)} = \frac{1/N}{1/N} = 1.
\end{equation*} This means the acceptance probability is \begin{equation*}
    \alpha(\bm{x}_0,\bm{x}_0^{\ast}) = \min\left\{1,\frac{\pi(\bm{x}^{\ast}_0|\bm{\theta},\bm{X}_{1:T},\bm{Y})}{\pi(\bm{x}_0|\bm{\theta},\bm{X}_{1:T},\bm{Y})}\right\},
\end{equation*} which we calculate using Equation \ref{X_0|theta,X_1:T,Y}.

We propose a new value $\bm{X}^{\ast}_{1:T}$ by proposing alterations to the colonisation and recovery events in the epidemic. At each step, we make one of three possible changes: move a colonisation/recovery time, add a colonisation/recovery time, or remove a colonisation/recovery time. 

Which of the three possible changes to propose is chosen randomly (probability $1/3$), and is proposed for a random individual $j$ (probability $1/N$). For the chosen individual $j$, let $(t_1,\dots,t_E)$ be the times when colonisation or recovery occur (where $E$ is the total number of colonisation and recovery events). We limit the sizes of the changes to the duration of colonisation/recovery periods to some value $m$, chosen to maximise the efficiency of the algorithm.

When moving an event, we take the following steps: \begin{enumerate}
    \item Randomly choose $e$ from $\{1,\dots,E\}$. We will be proposing a new time for event $e$.
    \item If $e=1$, then let $t_0=0$. If $e=E$, then let $t_{E+1}=T+1$.
    \item Let $F=\{t_{e-1}+1,\dots,t_{e+1}-1\}$, all possible times between the previous and next events.
    \item Let $G=\{t_e-m,\dots,t_e+m\}$, all possible times within the update limit.
    \item Then let $H = (F \cap G) \backslash \{t_e\}$, the set of possible times to move event $e$ to.
    \item Randomly choose $t_e^{\ast}$ from $H$ (this is the new time for event $e$).
    \item If $t_e^{\ast}>t_e$, then let $x^{\ast}_{t,j}=1-x_{t,j}$ for $t \in \{t_e+1,\dots,t_e^{\ast}\}$. If $t_e^{\ast}<t_e$, then let $x^{\ast}_{t,j}=1-x_{t,j}$ for $t \in \{t_e^{\ast},\dots,t_e-1\}$. All other entries in $\bm{X}_{1:T}^{\ast}$ are equal to $\bm{X}_{1:T}$.
\end{enumerate} This means our proposed $\bm{X}_{1:T}^{\ast}$ occurs with probability $1/(3NE|H|)$.

When adding an event, we take the following steps: \begin{enumerate}
    \item Let $A=\{t: x_{t-1,j}=x_{t,j}=x_{t+1,j}\}$, the set of acceptable new event times.
    \item Randomly choose $t_a$ from $A$. This is the time where a new event will be added.
    \item Let $e'=\min\{e : t_e>t_a\}$, the next event after time $t_a$.
    \item Let $B=\{t_a+1,\dots,\min\{t_{e'}-1,t_a+m\}\}$, the set of acceptable event times to reverse the newly added event.
    \item Randomly choose $t_b$ from $B$. This is the time when another new event will be added.
    \item Let $x^{\ast}_{t,j}=1-x_{t,j}$ for $t \in \{t_a,\dots,t_b-1\}$. All other entries in $\bm{X}_{1:T}^{\ast}$ are equal to $\bm{X}_{1:T}$.
\end{enumerate} This means our proposed $\bm{X}_{1:T}^{\ast}$ occurs with probability $1/(3N|A||B|)$.

When removing an event, we take the following steps: \begin{enumerate}
    \item Let $R=\{e : t_e + m \geq t_{e+1}\}$, the set of events we can remove.
    \item Randomly choose $r$ from $R$. This is the event we will remove, along with event $r+1$.
    \item Let $x^{\ast}_{t,j}=1-x_{t,j}$ for $t \in \{t_r,\dots,t_{r+1}-1\}$. All other entries in $\bm{X}_{1:T}^{\ast}$ are equal to $\bm{X}_{1:T}$.
\end{enumerate} This means our proposed $\bm{X}_{1:T}^{\ast}$ occurs with probability $1/(3N|R|)$.

The reverse step to a move event is another move event, so the proposal ratio is \begin{equation*}
    Q = \frac{q(\bm{X}_{1:T}|\bm{X}_{1:T}^{\ast})}{q(\bm{X}_{1:T}^{\ast}|\bm{X}_{1:T})} = \frac{1/(3NE|H^{\ast}|)}{1/(3NE|H|)} = \frac{|H|}{|H^{\ast}|}.
\end{equation*}

The reverse step to an add event is a remove event, so the proposal ratio is \begin{equation*}
    Q = \frac{q(\bm{X}_{1:T}|\bm{X}_{1:T}^{\ast})}{q(\bm{X}_{1:T}^{\ast}|\bm{X}_{1:T})} = \frac{1/(3N|R^{\ast}|)}{1/(3N|A||B|)} = \frac{|A||B|}{|R^{\ast}|}.
\end{equation*}

The reverse step to a remove event is another add event, so the proposal ratio is \begin{equation*}
    Q = \frac{q(\bm{X}_{1:T}|\bm{X}_{1:T}^{\ast})}{q(\bm{X}_{1:T}^{\ast}|\bm{X}_{1:T})} = \frac{1/(3N|A^{\ast}||B^{\ast}|)}{1/(3N|R|)} = \frac{|R|}{|A^{\ast}||B^{\ast}|}.
\end{equation*}

Then the acceptance rate is \begin{equation*}
    \alpha(\bm{X}_{1:T}, \bm{X}_{1:T}^{\ast}) = \min\left\{1,\frac{\pi(\bm{X}_{1:T}^{\ast}|\bm{\theta},\bm{x}_0,\bm{Y})}{\pi(\bm{X}_{1:T}|\bm{\theta},\bm{x}_0,\bm{Y})}Q\right\},
\end{equation*} which we calculate using Equation \ref{X_1:T|theta,X_0,Y}.

Each move in $\bm{X}_{1:T}$ will only change the latent variables of one individual (for no more than $m$ time-steps). Hence we run the latent variable update step many times within each overall MCMC iteration (as in the Rippler algorithm). The full method is shown in Algorithm \ref{alg_block}.

\begin{algorithm}[h!]
\caption{Reversible-jump MCMC}
\label{alg_block}
\begin{algorithmic}[1]
\Require Number of iterations $K$, number of iterations for latent updates $K'$, starting values $\bm{\theta}^{(0)},\bm{X}^{(0)},\bm{\Sigma}^{(0)},\kappa^{(0)}$, escape probability $\varepsilon$, prior distribution hyperparameters $\bm{\mu}$, recovery rate $\gamma$, initial probability of colonisation $p_0$, test data $\bm{Y}$, maximum block update size $m$.
\State Let $\bm{\theta}=\bm{\theta}^{(0)},\bm{X}=\bm{X}^{(0)},\bm{\Sigma}=\bm{\Sigma}^{(0)},\kappa=\kappa^{(0)}$.
\For{$k \in \{1,\dots,K\}$}
    \State With probability $1-\varepsilon$, sample $\bm{\theta}^{\ast} \sim \text{Normal}(\bm{\theta},\kappa^2\bm{\Sigma})$. Otherwise, sample ${\bm{\theta}^{\ast} \sim \text{Normal}(\bm{\theta},\bm{\Sigma}^{(0)})}$.
    \State Calculate \[\alpha(\bm{\theta},\bm{\theta}^{\ast}) = \min\left\{1,\frac{\pi(\bm{\theta}^{\ast}|\bm{X},\bm{Y})}{\pi(\bm{\theta}|\bm{X},\bm{Y})}\right\}.\]
    \State Let $\bm{\theta}=\bm{\theta}^{\ast}$ with probability $\alpha(\bm{\theta},\bm{\theta}^{\ast})$. Otherwise keep $\bm{\theta}$ constant.
    \State Update $\kappa$ and $\bm{\Sigma}$.
    \State Let $\bm{x}_{0}^{\ast}=\bm{x}_0$. Randomly choose $j \in \{1,\dots,N\}$ and let $x_{0,j}^{\ast}=1-x_{0,j}$.
    \State Calculate \[\alpha(\bm{x}_0,\bm{x}_0^{\ast}) = \min\left\{1,\frac{\pi(\bm{x}^{\ast}_0|\bm{\theta},\bm{X}_{1:T},\bm{Y})}{\pi(\bm{x}_0|\bm{\theta},\bm{X}_{1:T},\bm{Y})}\right\}.\]
    \State Let $\bm{x}_0=\bm{x}_0^{\ast}$ with probability $\alpha(\bm{x}_0,\bm{x}_0^{\ast})$. Otherwise keep $\bm{x}_0$ constant.
    \For{$k' \in \{1,\dots,K'\}$}
        \State Randomly choose an individual $j \in \{1,\dots,N\}$.
        \State Randomly choose between the following options: move, add, remove.
        \State Follow the respective method on individual $j$ to generate $\bm{X}_{1:T}^{\ast}$ and $Q$.
        \State Calculate \[\alpha(\bm{X}_{1:T}, \bm{X}_{1:T}^{\ast}) = \min\left\{1,\frac{\pi(\bm{X}_{1:T}^{\ast}|\bm{\theta},\bm{x}_0,\bm{Y})}{\pi(\bm{X}_{1:T}|\bm{\theta},\bm{x}_0,\bm{Y})}Q\right\}.\]
        \State Let $\bm{X}_{1:T}=\bm{X}_{1:T}^{\ast}$ with probability $\alpha(\bm{X}_{1:T}, \bm{X}_{1:T}^{\ast})$. Otherwise keep $\bm{X}$ constant.
    \EndFor
\EndFor
\end{algorithmic}
\end{algorithm}

\subsection{Simulation Study}

We now assess the performance of the reversible-jump algorithm by attempting to recover the parameters for the simulated data. We use the same data as when we performed parameter recovery using the Rippler algorithm in the main text.

We run the MCMC algorithm for $K=100{,}000$ iterations, repeating the latent variable step 400 times within each iteration. The first $10{,}000$ iterations are discarded as burn-in. We let $\bm{\theta}^{(0)}=(0.5,0.5,0,0)$, and then generate $\bm{X}^{(0)}$ from our transmission model using parameters $\bm{\theta}^{(0)}$. The rate of the prior distribution for each parameter is set to $0.001$, such that the prior is uninformative. These are all the same values as when we ran the Rippler algorithm in the main text (to make the comparison as fair as possible). The maximum block update size is set to $m=4$; we are assuming a recovery rate of $\gamma=0.5$, so more than 85\% of colonisation periods will be less than or equal to 4 weeks.

\begin{figure}[H]
     \centering
     \hfill
     \begin{subfigure}[h]{0.45\textwidth}
         \centering
         \includegraphics[width=\textwidth]{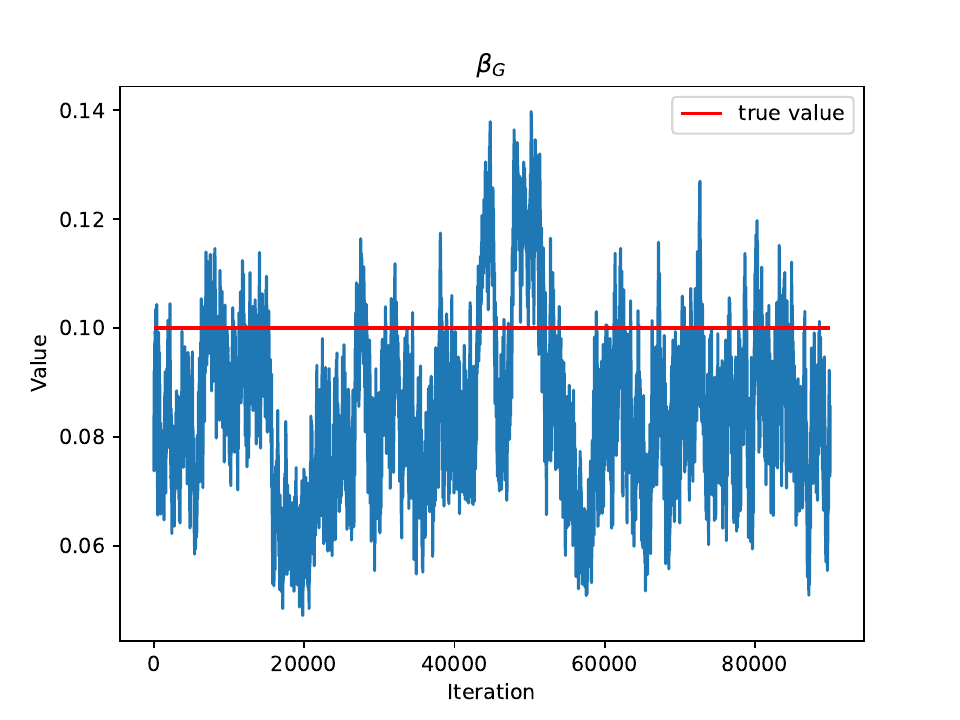}
     \end{subfigure}
     \hfill
     \begin{subfigure}[h]{0.45\textwidth}
         \centering
         \includegraphics[width=\textwidth]{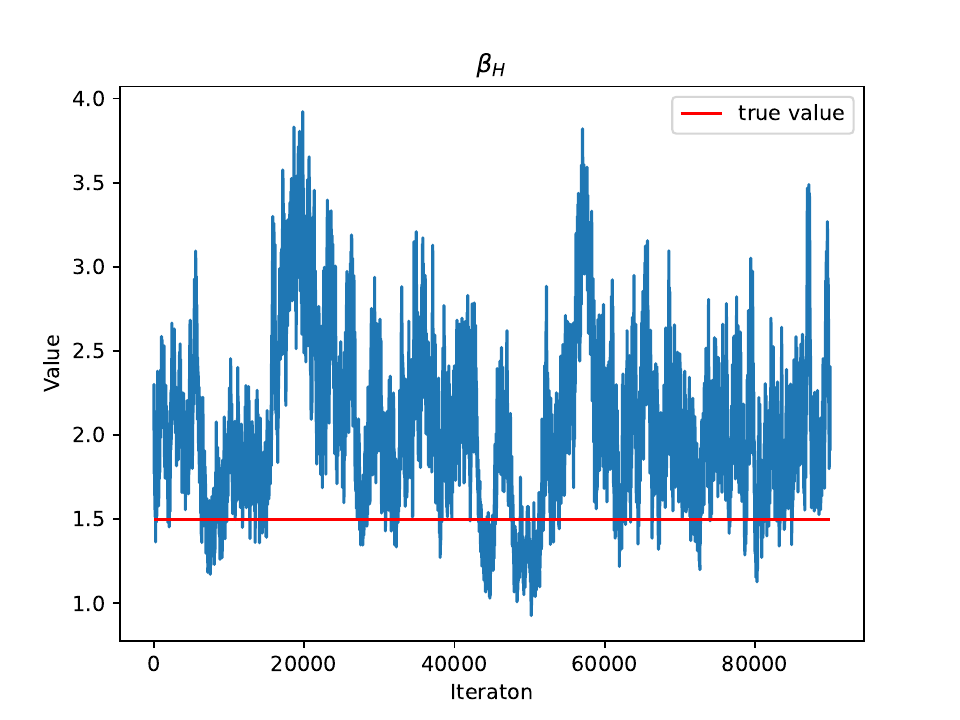}
     \end{subfigure}
     \hfill
     \\
     \hfill
     \begin{subfigure}[h]{0.45\textwidth}
         \centering
         \includegraphics[width=\textwidth]{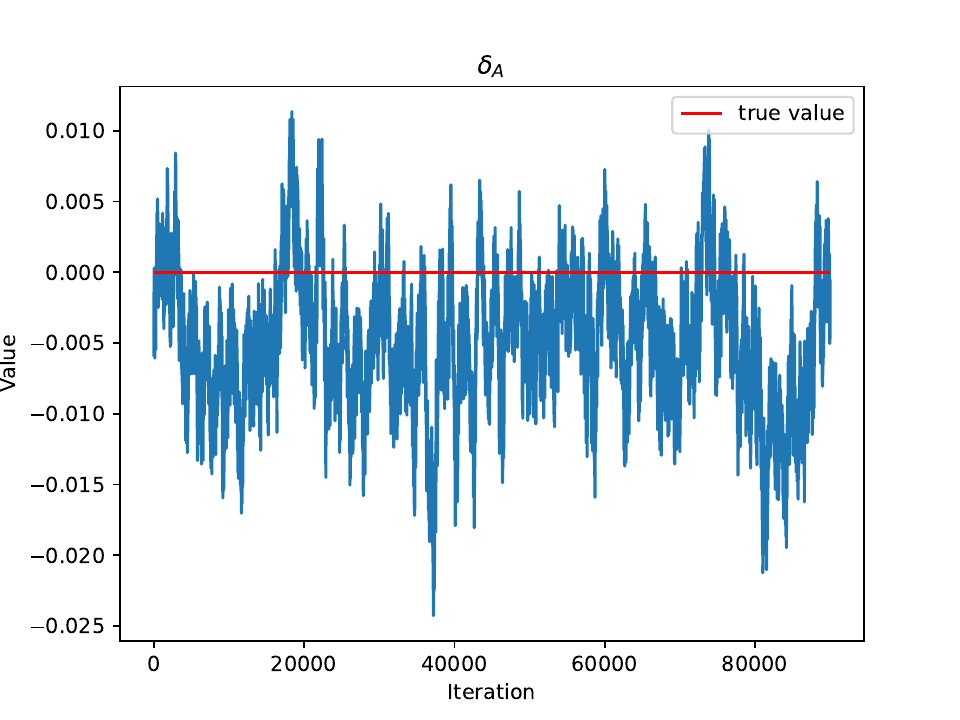}
     \end{subfigure}
     \hfill
     \begin{subfigure}[h]{0.45\textwidth}
         \centering
         \includegraphics[width=\textwidth]{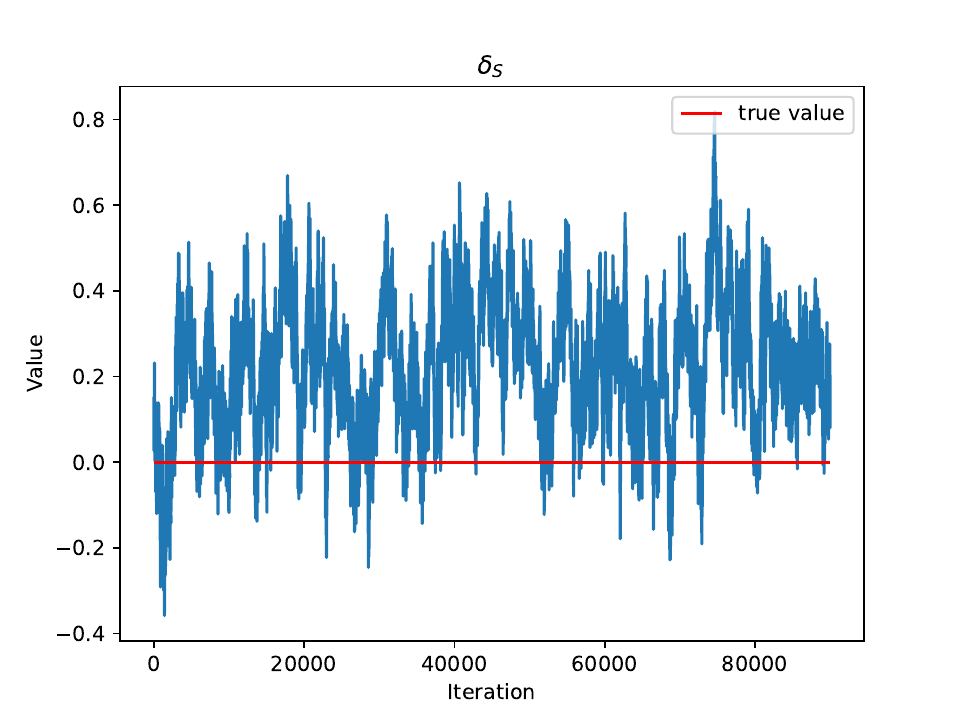}
     \end{subfigure}
     \hfill
     \caption{Trace plots of the posterior samples generated for the parameters $\bm{\theta}=(\beta_G,\beta_H,\delta_A,\delta_S)$ for simulated 3GC-R \textit{E. coli} test data (using the reversible-jump algorithm for inference). The true value is shown by the red line.}
     \label{sim_full_parameters_block_trace}
\end{figure}

\begin{figure}[H]
     \centering
     \hfill
     \begin{subfigure}[h]{0.45\textwidth}
         \centering
         \includegraphics[width=\textwidth]{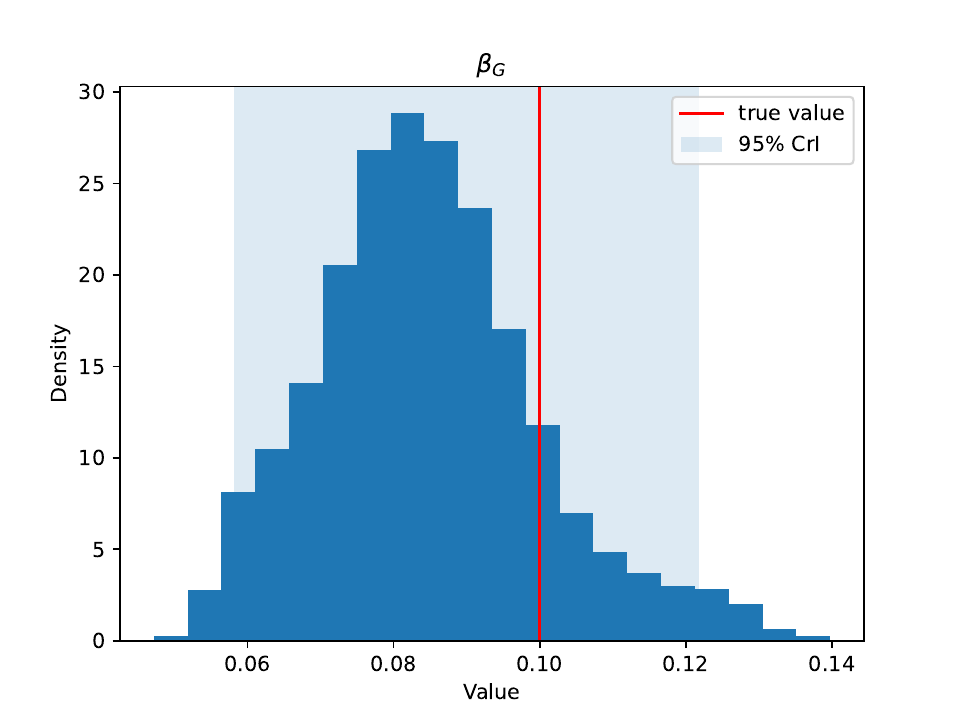}
     \end{subfigure}
     \hfill
     \begin{subfigure}[h]{0.45\textwidth}
         \centering
         \includegraphics[width=\textwidth]{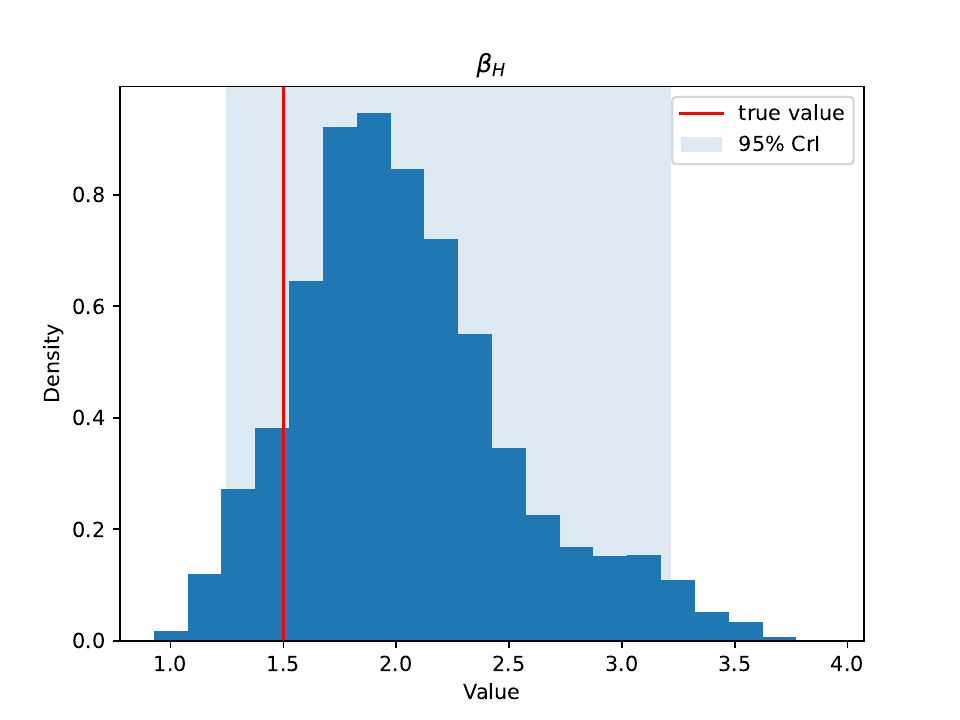}
     \end{subfigure}
     \hfill
     \\
     \hfill
     \begin{subfigure}[h]{0.45\textwidth}
         \centering
         \includegraphics[width=\textwidth]{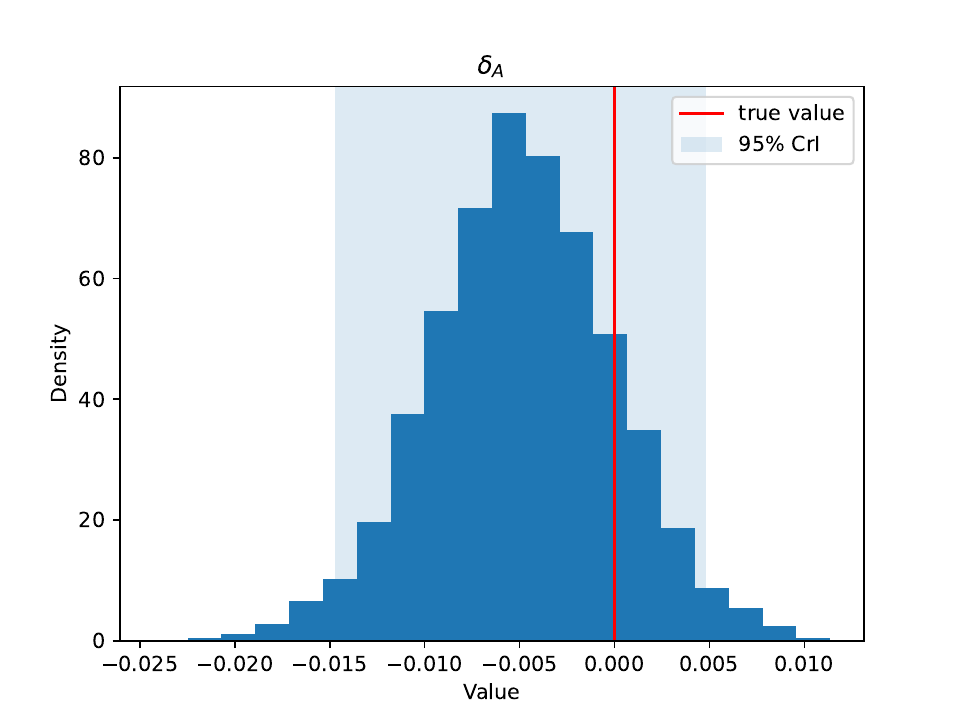}
     \end{subfigure}
     \hfill
     \begin{subfigure}[h]{0.45\textwidth}
         \centering
         \includegraphics[width=\textwidth]{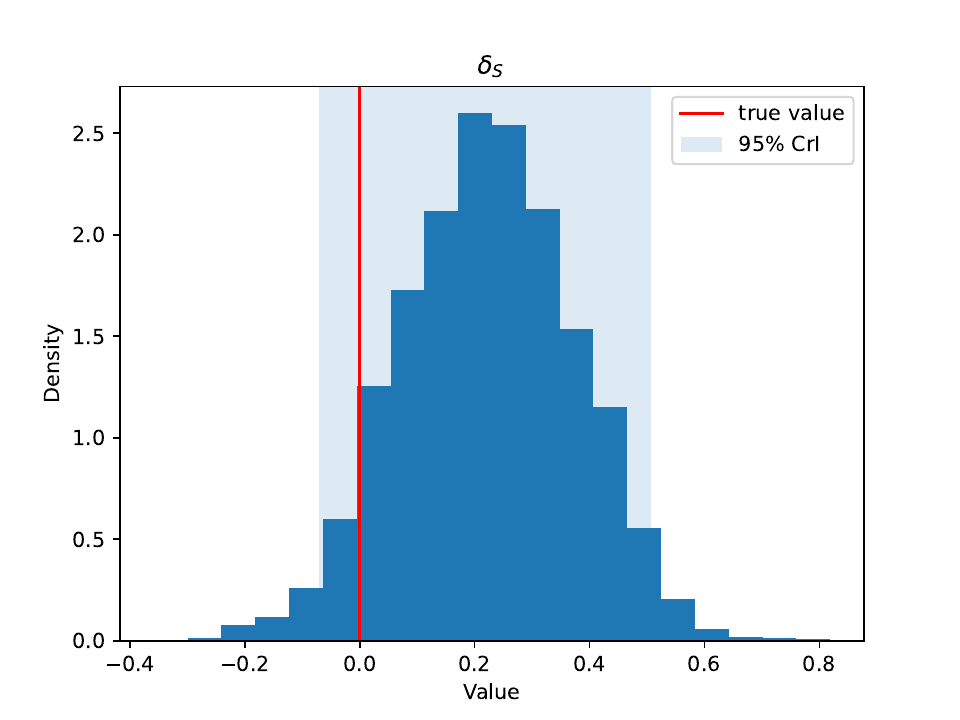}
     \end{subfigure}
     \hfill
     \caption{Histograms of the posterior samples generated for the parameters $\bm{\theta}=(\beta_G,\beta_H,\delta_A,\delta_S)$ for simulated 3GC-R \textit{E. coli} test data using the reversible-jump algorithm (Algorithm \ref{alg_block}). The true value is shown by the red line and the 95\% credible interval is shown in light blue.}
     \label{sim_full_parameters_block}
\end{figure}

\begin{figure}[H]
    \centering
    \includegraphics[width=0.85\textwidth]{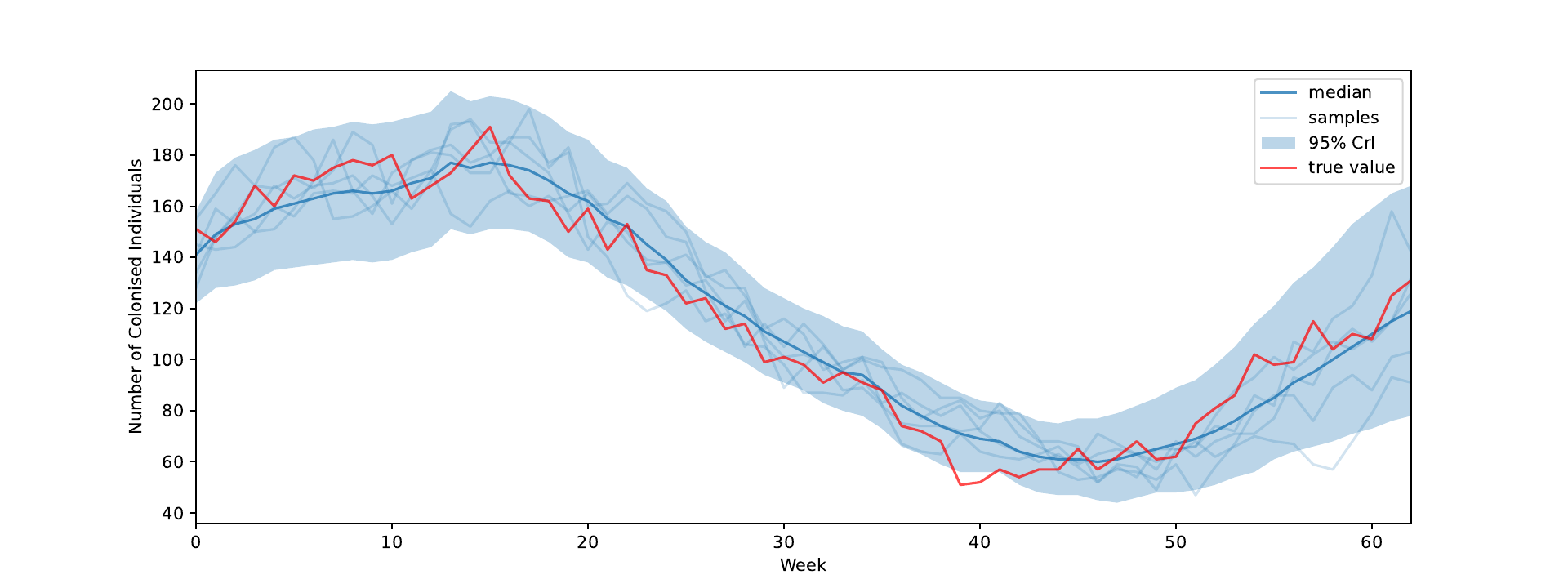}
    \caption{The number of colonised individuals over time for simulated 3GC-R \textit{E. coli} test data inferred using the reversible-jump algorithm. The median value is shown by the dark blue line and 5 randomly chosen samples are shown by the faint blue lines. The true value is shown by the red line and the 95\% credible interval is shown in light blue.}
    \label{sim_full_latent_block}
\end{figure}

The parameter trace plots and histograms are shown in Figures \ref{sim_full_parameters_block_trace} and \ref{sim_full_parameters_block} respectively. We see that the true value of each parameter lies within the 95\% credible interval of the sample. Details are given in Table \ref{sim_full_table_block}. Similarly to the results with the Rippler algorithm, the median of the sample is very close to the true value for $\beta_G$ and $\beta_H$, but significantly further away for $\delta_A$ and $\delta_S$. The central 95\% credible interval for the number of individuals colonised at each point in time is shown in Figure \ref{sim_full_latent_block}, with the true value mostly lying within the interval. We can see from the trace plots of the parameters that the mixing is worse than when using the Rippler algorithm. The higher autocorrelation is because of the slow movement in the latent variable space when using the reversible-jump algorithm. A potential fix for this would be increasing the number of reversible-jump updates per MCMC iteration, although this has the cost of increasing the runtime.

\begin{table}[H]
    \centering
    \small
    \caption{Parameter estimates for simulated 3GC-R \textit{E. coli} test data using the reversible-jump algorithm for inference.}
    \begin{tabular}{ c  c  c  c  c}
        \hline Parameter &  Definition & True & Median & 95\% CrI 
        \\ \hline $\beta_G$ & global transmission rate (week$^{-1}$) & 0.1 & 0.0836 & (0.0581, 0.122) 
        \\ $\beta_H$ & relative within-household transmission & 1.5 & 1.98 & (1.25, 3.22) 
        \\ $\delta_A$ & age covariate effect (year$^{-1}$) & 0 & -0.00490 & (-0.0147, 0.00480) 
        \\ $\delta_S$ & sex covariate effect & 0 & 0.224 & (-0.0717, 0.507) 
        \\ \hline
    \end{tabular}
    \label{sim_full_table_block}
\end{table}

\section{iFFBS} \label{appendix_iFFBS}

\subsection{Inference Algorithm}

We use the method from Touloupou et al.~(2020) to perform individual forward filtering backwards sampling (iFFBS). As with the Rippler algorithm, we update our parameters and latent variables separately: a Metropolis-within-Gibbs step for the parameters, and a Gibbs step for the latent variables. We use the same method of updating the parameters $\bm{\theta}$ as in the Rippler algorithm (adaptive random walk Metropolis).

iFFBS is an extension of the standard FFBS algorithm: instead of sampling all of the latent variables, the latent variables for one individual are sampled conditionally on the latent variables for all other individuals (which are held constant). We randomly choose an individual $j$ to have their latent variables updated, and let $\bm{X}_{0:T,-j}$ consist of the latent variables for all individuals other than $j$. We sample directly from the conditional distribution $\pi(\bm{X}_{0:T,j}|\bm{\theta},\bm{X}_{0:T,-j},\bm{Y})$, so we do not need to perform a Metropolis-Hastings accept-reject step. 

From Bayes' theorem we have \begin{align*} \begin{split}
    \pi(\bm{X}_{0:T,j}|\bm{\theta},\bm{X}_{0:T,-j},\bm{Y}) &= \pi(X_{T,j}|\bm{\theta},\bm{X}_{0:T,-j},\bm{Y}_{0:T,j}) 
    \\ &~ \hspace{0.7cm} \prod_{t=0}^{T-1} \pi(X_{t,j}|\bm{\theta},X_{t+1,j},\bm{X}_{0:t+1,-j},\bm{Y}_{0:t,j})
    \\ &\propto \pi(X_{T,j}|\bm{\theta},\bm{X}_{0:T,-j},\bm{Y}_{0:T,j}) 
    \\ &~ \hspace{0.7cm} \prod_{t=0}^{T-1} \pi(X_{t+1,j}|\bm{\theta},\bm{X}_t) \pi(X_{t,j}|\bm{\theta},\bm{X}_{0:t+1,-j},\bm{Y}_{0:t,j}),
\end{split} \end{align*} using the fact that the state of any individual at time $t+1$ is determined based on the state of all individuals at time $t$. We will iteratively calculate the probabilities $\pi(X_{t,j}|\bm{\theta},\bm{X}_{0:t+1,-j},\bm{Y}_{0:t,j})$ forwards through time, and then combine this term with $\pi(X_{t+1,j}|\bm{\theta},\bm{X}_t)$ to sample backwards through time. Note that from our transmission model, equation (2) in the main text, we have \begin{equation}
    \mathbb{P}(X_{t+1,j}=x_{t+1,j}|\bm{\theta},X_{t,j}=x_{t,j},\bm{X}_{t,-j}) = \begin{cases}
        e^{-\lambda_j(t+1,\bm{x}_t,\bm{\theta})\Delta} &~~~~~~\text{ if } x_{t,j}=0,x_{t+1,j}=0,
        \\ 1-e^{-\lambda_j(t+1,\bm{x}_t,\bm{\theta})\Delta} &~~~~~~\text{ if } x_{t,j}=0,x_{t+1,j}=1,
        \\ 1-e^{-\gamma\Delta} &~~~~~~\text{ if } x_{t,j}=1,x_{t+1,j}=0,
        \\ e^{-\gamma\Delta} &~~~~~~\text{ if } x_{t,j}=1,x_{t+1,j}=1,
    \end{cases} \label{X_t+1|X_t}
\end{equation} for any $t \in \{0,\dots,T-1\}$ and $j \in \{1,\dots,N\}$. 

First, we calculate the forward probabilities for $t=0$. We have \begin{equation}
    \mathbb{P}(X_{0,j}=x_{0,j}|\bm{\theta},\bm{X}_{0:1,-j}) \propto \mathbb{P}(X_{0,j}=x_{0,j})\prod_{i=1,i \neq j}^N \pi(X_{1,i}=x_{1,i}|\bm{\theta},X_{0,i}=x_{0,i},\bm{X}_{0,-i}). \label{forward_0}
\end{equation} The first term is the probability of the initial conditions, which is \begin{equation*}
    \mathbb{P}(X_{0,j}=x_{0,j}) = \begin{cases}
        p_0 &~~~~~~\text{ if } x_{0,j}=1,
        \\ 1-p_0 &~~~~~~\text{ if } x_{0,j}=0,
    \end{cases}
\end{equation*} and the second term is the transition probabilities for all individuals other than $j$ at time $t=1$, which we can calculate using Equation \ref{X_t+1|X_t}. We calculate these probabilities up to some constant of proportionality using these equations for $x_{0,j} \in \{0,1\}$, and then normalise such that the probabilities add to 1.

Next, we calculate the forward probabilities iteratively for $t \in \{1,\dots,T-1\}$. We have \begin{align} \begin{split}
    \mathbb{P}(X_{t,j}=x_{t,j}|\bm{\theta},\bm{X}_{0:t+1,-j},\bm{Y}_{0:t,j}) &\propto \mathbb{P}(Y_{t,j}=y_{t,j}|X_{t,j}=x_{t,j})
    \\ &~ \hspace{0.5cm} \prod_{i=1,i \neq j}^N \pi(X_{t+1,i}=x_{t+1,i}|\bm{\theta},X_{t,i}=x_{t,i},\bm{X}_{t,-i})
    \\ &~ \hspace{0.5cm} \mathbb{P}(X_{t,j}=x_{t,j}|\bm{\theta},\bm{X}_{1:t,-j},\bm{Y}_{1:t-1,j}).
\end{split} \label{forward_t} \end{align} The first term is the probability of the test result given the true state of the individual, which is \begin{equation*}
    \mathbb{P}(Y_{t,j}=y_{t,j}|X_{t,j}=x_{t,j}) = \begin{cases}
        s_e &~~~~~~\text{ if } y_{t,j}=1, x_{t,j}=1,
        \\ 1-s_p &~~~~~~\text{ if } y_{t,j}=1, x_{t,j}=0,
        \\ 1-s_e &~~~~~~\text{ if } y_{t,j}=0, x_{t,j}=1,
        \\ s_p &~~~~~~\text{ if } y_{t,j}=0, x_{t,j}=0,
        \\ 1 &~~~~~~\text{ if } y_{t,j} \text{ is undefined}.
    \end{cases}
\end{equation*} Note that we let this term equal 1 if there is no test result at this time, so the term has no effect on the calculation of the probabilities. The second term is the transition probabilities for all individuals other than $j$ at time $t+1$, which we can calculate using Equation \ref{X_t+1|X_t}. By the law of total probability, the third term can be expanded into \begin{align*} \begin{split}
    &\mathbb{P}(X_{t,j}=x_{t,j}|\bm{\theta},\bm{X}_{1:t,-j},\bm{Y}_{1:t-1,j}) 
    \\ &= \mathbb{P}(X_{t_j}=x_{t,j}|\bm{\theta},X_{t-1,j}=0,\mathbf{X}_{t-1,-j})\mathbb{P}(X_{t-1,j}=0|\bm{\theta},\bm{X}_{0:t,-j},\bm{Y}_{0:t-1,j})
    \\ &~ \hspace{0.5cm} + \mathbb{P}(X_{t_j}=x_{t,j}|\bm{\theta},X_{t-1,j}=1,\mathbf{X}_{t-1,-j})\mathbb{P}(X_{t-1,j}=1|\bm{\theta},\bm{X}_{0:t,-j},\bm{Y}_{0:t-1,j}),
\end{split} \end{align*} which we can calculate using Equation \ref{X_t+1|X_t} and using the forward probabilities from the previous time-step. Again, we calculate these probabilities up to some constant of proportionality using these equations for $x_{t,j} \in \{0,1\}$, and then normalise such that the probabilities add to 1.

The final forward probability calculation we need to determine is for $t=T$. We have \begin{align} \begin{split}
    \mathbb{P}(X_{T,j}=x_{T,j}|\bm{\theta},\bm{X}_{0:T,-j},\bm{Y}_{0:T,j}) &\propto \mathbb{P}(Y_{T,j}=y_{T,j}|X_{T,j}=x_{T,j})
    \\ &~ \hspace{0.5cm} \mathbb{P}(X_{T,j}=x_{T,j}|\bm{\theta},\bm{X}_{1:T,-j},\bm{Y}_{1:T-1,j}).
\end{split} \label{forward_T} \end{align} Calculating the forward probabilities is the same as for $t \in \{1,\dots,T-1\}$, but with one term removed.

Now we can simulate new values for $\bm{X}_{0:T,j}$ backwards through time. First, we let \[p=\mathbb{P}(X_{T,j}=1|\bm{\theta},\bm{X}_{0:T,-j},\bm{Y}_{0:T,j}),\] and then sample $x_{T,j}^{\ast} \sim \text{Bernoulli}(p)$.

Then for $t \in \{T-1,\dots,0\}$, we calculate \begin{align} \begin{split}
    \mathbb{P}(X_{t,j}=x_{t,j}|\bm{\theta},X_{t+1,j}=x_{t+1,j}^{\ast},\bm{X}_{0:t+1,-j},\bm{Y}_{0:t,j}) &\propto  \mathbb{P}(X_{t+1,j}=x_{t+1,j}^{\ast}|\bm{\theta},X_{t,j}=x_{t,j},\bm{X}_{t,-j})
    \\ &~ \hspace{0.5cm} \mathbb{P}(X_{t,j}=x_{t,j}|\bm{\theta},\bm{X}_{0:t+1,-j},\bm{Y}_{0:t,j}).
\end{split} \label{backward_t} \end{align} The first term is calculated using Equation \ref{X_t+1|X_t}, the second term is the forward probability for time $t$ -- we calculate for $x_{t,j} \in \{0,1\}$ and then normalise such that the probabilities add to 1. Next, we let \[p=\mathbb{P}(X_{t,j}=1|\bm{\theta},X_{t+1,j}=x_{t+1,j}^{\ast},\bm{X}_{0:t+1,-j},\bm{Y}_{0:t,j}),\] and then sample $x_{t,j}^{\ast} \sim \text{Bernoulli}(p)$.

We now have new latent variables $\bm{X}_{0:T,j}=\bm{x}_{0:T,j}^{\ast}$. Each iFFBS update will only change the latent variables of one individual, so we will run the latent variable update step many times within each overall MCMC iteration (as in the Rippler algorithm). The full method is shown in Algorithm \ref{alg_iFFBS}.

\begin{algorithm}[h!]
\caption{iFFBS}
\label{alg_iFFBS}
\begin{algorithmic}[1]
\Require Number of iterations $K$, number of iterations for latent updates $K'$,starting values $\bm{\theta}^{(0)},\bm{X}^{(0)},\bm{\Sigma}^{(0)},\kappa^{(0)}$, escape probability $\varepsilon$, prior distribution hyperparameters $\bm{\mu}$, tuning parameter $\lambda$, recovery rate $\gamma$, initial probability of colonisation $p_0$, covariate data $\bm{c}$, test data $\bm{Y}$.
\State Let $\bm{\theta}=\bm{\theta}^{(0)},\bm{X}=\bm{X}^{(0)},\bm{\Sigma}=\bm{\Sigma}^{(0)},\kappa=\kappa^{(0)}$.
\For{$k \in \{1,\dots,K\}$}
    \State With probability $1-\varepsilon$, sample $\bm{\theta}^{\ast} \sim \text{Normal}(\bm{\theta},\kappa^2\bm{\Sigma})$. Otherwise, sample ${\bm{\theta}^{\ast} \sim \text{Normal}(\bm{\theta},\bm{\Sigma}^{(0)})}$.
    \State Calculate \[\alpha(\bm{\theta},\bm{\theta}^{\ast}) = \min\left\{1,\frac{\pi(\bm{\theta}^{\ast}|\bm{X},\bm{Y})}{\pi(\bm{\theta}|\bm{X},\bm{Y})}\right\}.\]
    \State Let $\bm{\theta}=\bm{\theta}^{\ast}$ with probability $\alpha(\bm{\theta},\bm{\theta}^{\ast})$. Otherwise keep $\bm{\theta}$ constant.
    \State Update $\kappa$ and $\bm{\Sigma}$.
    \For{$k' \in \{1,\dots,K'\}$}
        \State Randomly choose an individual $j \in \{1,\dots,N\}$.
        \State Calculate forward probabilities for $x_{0,j} \in \{0,1\}$ using Equation \ref{forward_0}.
        \For{$t \in \{1,\dots,T-1\}$}
            \State Calculate forward probabilities for $x_{t,j} \in \{0,1\}$ using Equation \ref{forward_t}.
        \EndFor
        \State Calculate forward probabilities for $x_{T,j} \in \{0,1\}$ using Equation \ref{forward_T}.
        \State Let $p=\mathbb{P}(X_{T,j}=1|\bm{\theta},\bm{X}_{0:T,-j},\bm{Y}_{0:T,j})$.
        \State Sample $x_{T,j}^{\ast} \sim \text{Bernoulli}(p)$.
        \For{$t \in \{T-1,\dots,0\}$}
            \State Calculate backwards probabilities for $x_{t,j} \in \{0,1\}$ using Equation \ref{backward_t}.
            \State Let $p=\mathbb{P}(X_{t,j}=1|\bm{\theta},X_{t+1,j}=x_{t+1,j}^{\ast},\bm{X}_{0:t+1,-j},\bm{Y}_{0:t,j})$.
            \State Sample $x_{t,j}^{\ast} \sim \text{Bernoulli}(p)$.
        \EndFor
        \State Let $\bm{X}_{0:T,j}=\bm{x}_{0:T,j}^{\ast}$.
    \EndFor
\EndFor
\end{algorithmic}
\end{algorithm}

\subsection{Simulation Study}

We now assess the performance of the iFFBS algorithm by attempting to recover the parameters for the simulated data. We use the same data as when we performed parameter recovery using the Rippler algorithm in the main text.

We run the MCMC algorithm for $K=100{,}000$ iterations, repeating the latent variable (iFFBS) step 400 times within each iteration. The first $10{,}000$ iterations are discarded as burn-in. We let $\bm{\theta}^{(0)}=(0.5,0.5,0,0)$, and then generate $\bm{X}^{(0)}$ from our transmission model using parameters $\bm{\theta}^{(0)}$. The rate of the prior distribution for each parameter is set to $0.001$. For consistency we use all the same values as when we ran the Rippler algorithm in the main text.

\begin{figure}[H]
     \centering
     \hfill
     \begin{subfigure}[h]{0.45\textwidth}
         \centering
         \includegraphics[width=\textwidth]{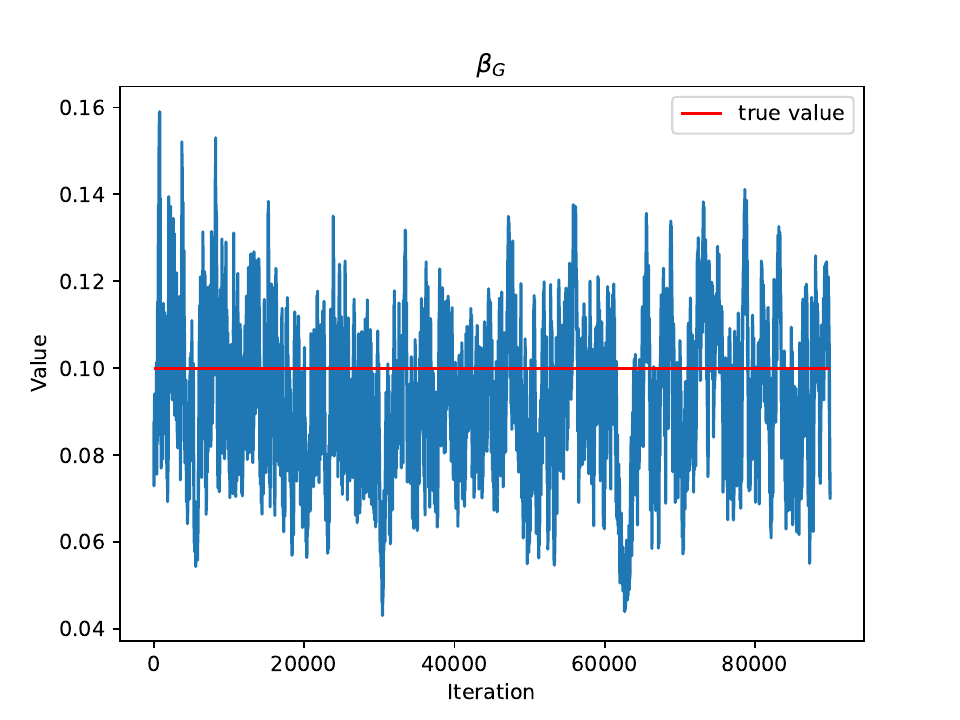}
     \end{subfigure}
     \hfill
     \begin{subfigure}[h]{0.45\textwidth}
         \centering
         \includegraphics[width=\textwidth]{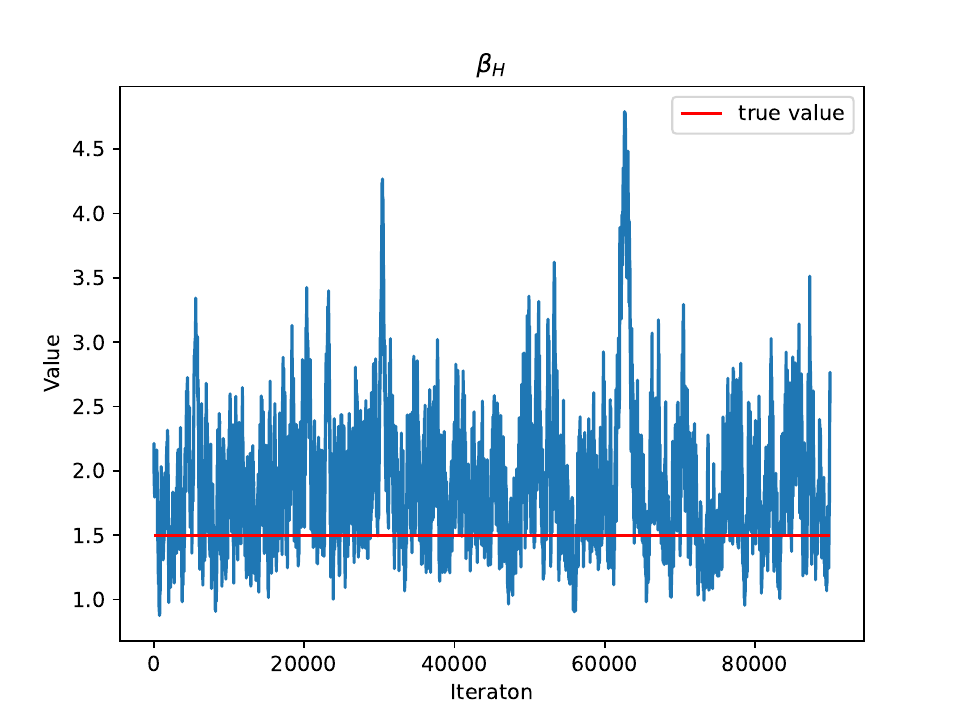}
     \end{subfigure}
     \hfill
     \\
     \hfill
     \begin{subfigure}[h]{0.45\textwidth}
         \centering
         \includegraphics[width=\textwidth]{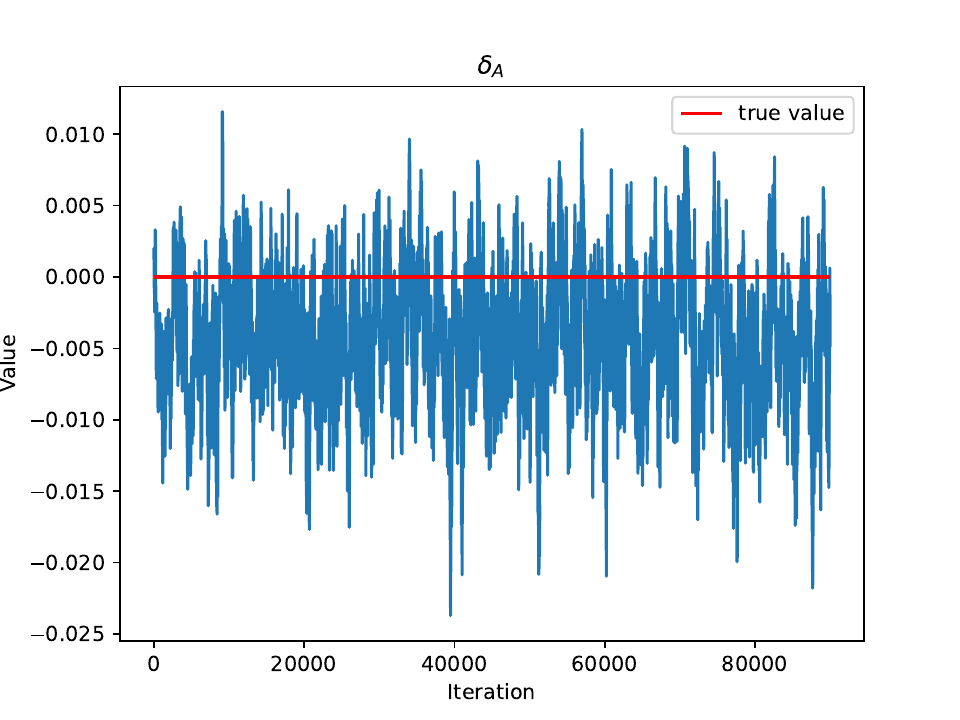}
     \end{subfigure}
     \hfill
     \begin{subfigure}[h]{0.45\textwidth}
         \centering
         \includegraphics[width=\textwidth]{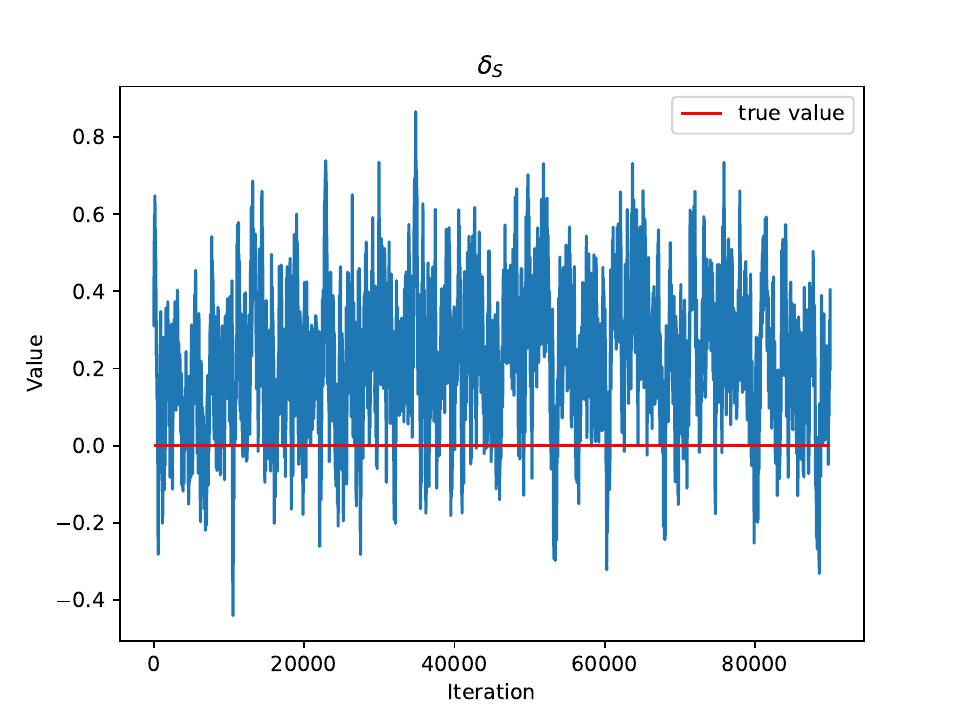}
     \end{subfigure}
     \hfill
     \caption{Trace plots of the posterior samples generated for the parameters $\bm{\theta}=(\beta_G,\beta_H,\delta_A,\delta_S)$ for simulated 3GC-R \textit{E. coli} test data using the iFFBS algorithm (Algorithm \ref{alg_iFFBS}). The true value is shown by the red line.}
     \label{sim_full_parameters_iFFBS_trace}
\end{figure}

\begin{figure}[H]
     \centering
     \hfill
     \begin{subfigure}[h]{0.45\textwidth}
         \centering
         \includegraphics[width=\textwidth]{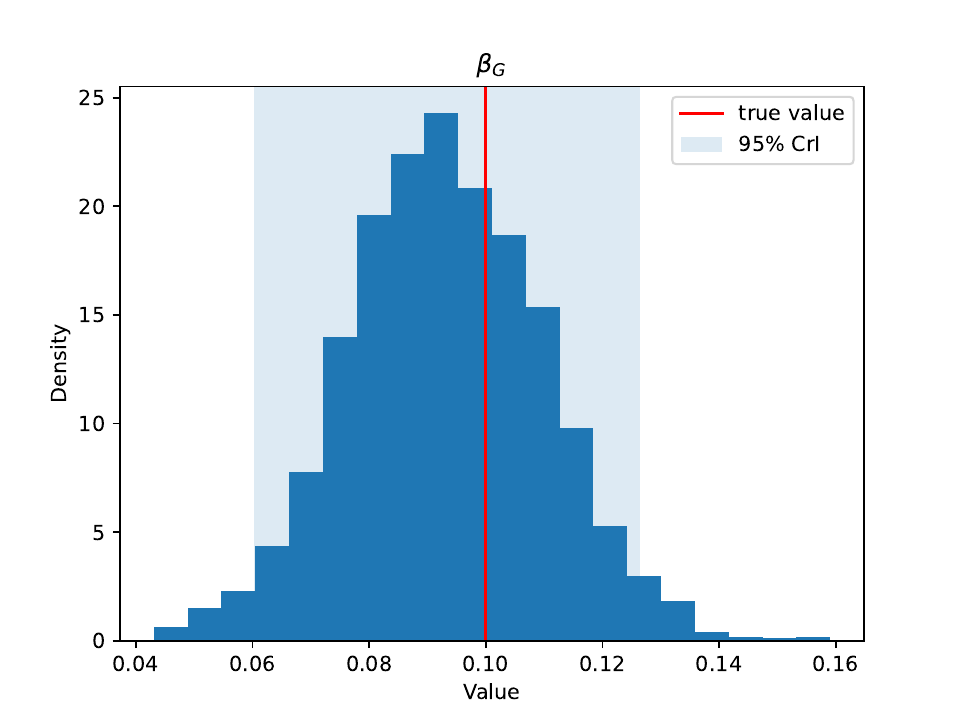}
     \end{subfigure}
     \hfill
     \begin{subfigure}[h]{0.45\textwidth}
         \centering
         \includegraphics[width=\textwidth]{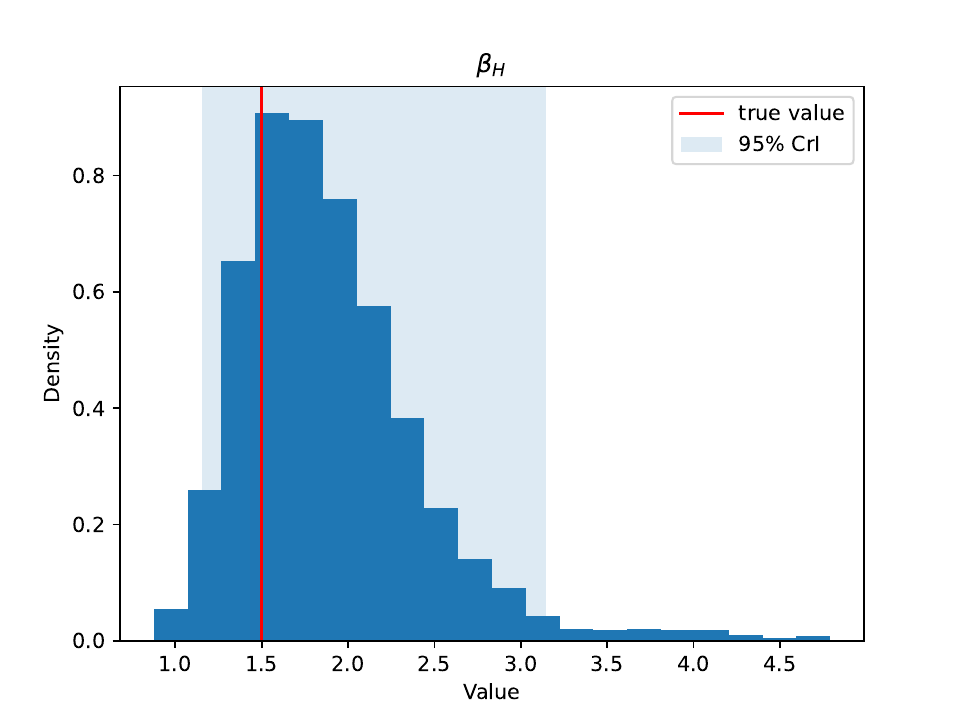}
     \end{subfigure}
     \hfill
     \\
     \hfill
     \begin{subfigure}[h]{0.45\textwidth}
         \centering
         \includegraphics[width=\textwidth]{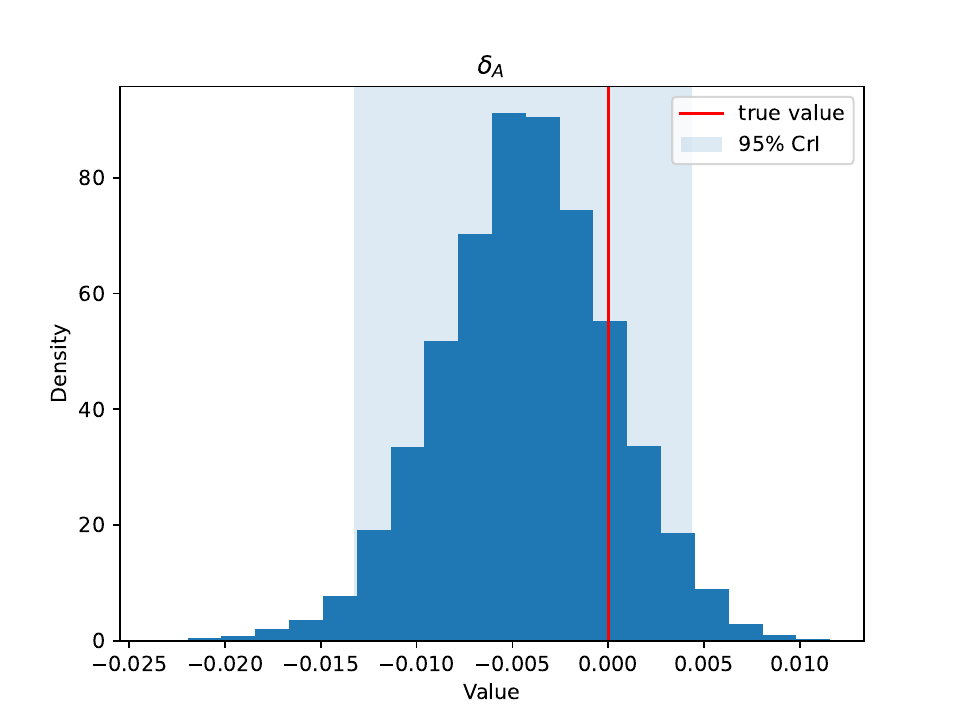}
     \end{subfigure}
     \hfill
     \begin{subfigure}[h]{0.45\textwidth}
         \centering
         \includegraphics[width=\textwidth]{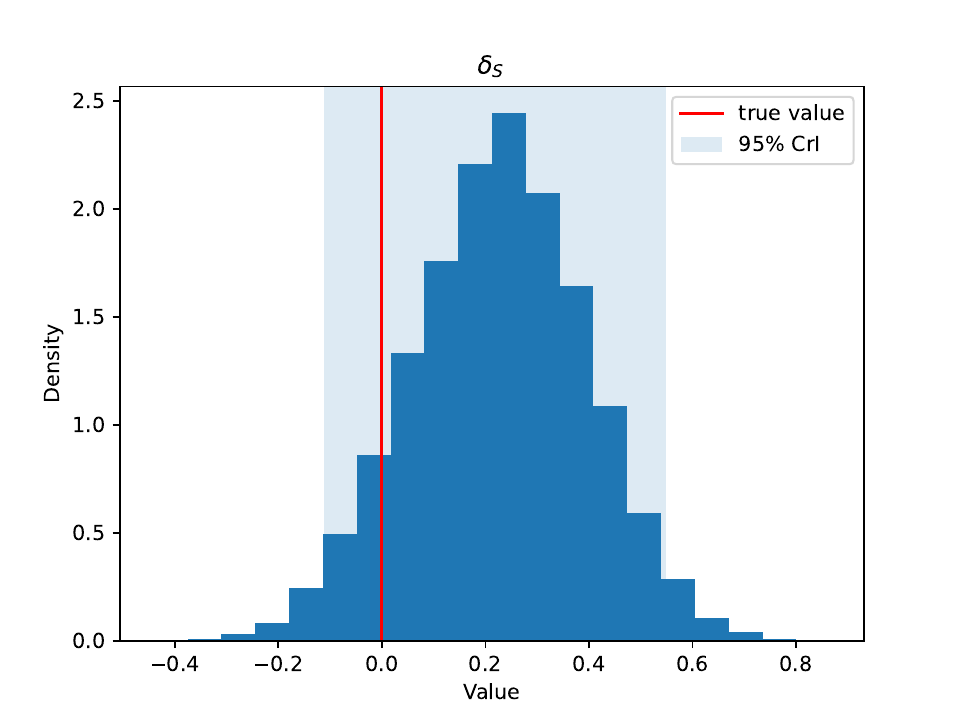}
     \end{subfigure}
     \hfill
     \caption{Histograms of the posterior samples generated for the parameters $\bm{\theta}=(\beta_G,\beta_H,\delta_A,\delta_S)$ for simulated 3GC-R \textit{E. coli} test data using the iFFBS algorithm. The true value is shown by the red line and the 95\% credible interval is shown in light blue.}
     \label{sim_full_parameters_iFFBS}
\end{figure}

\begin{figure}[H]
    \centering
    \includegraphics[width=0.85\textwidth]{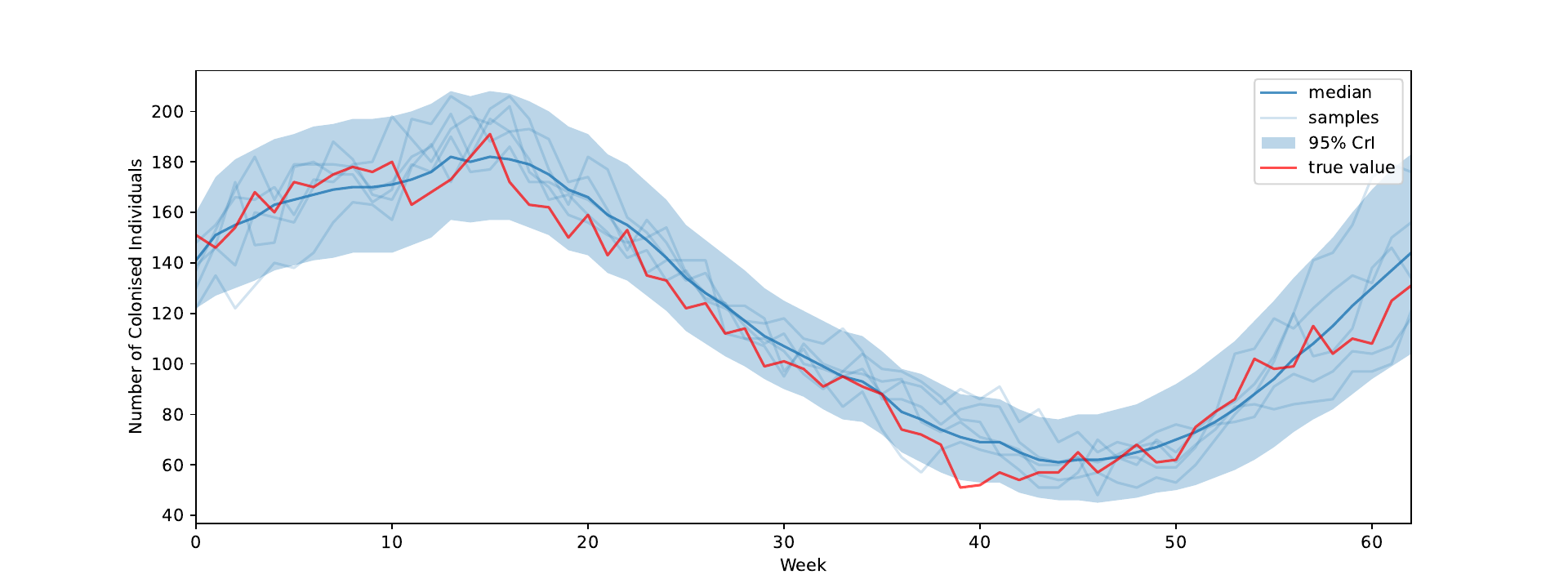}
    \caption{The number of colonised individuals over time for simulated 3GC-R \textit{E. coli} test data inferred using the iFFBS algorithm. The median value is shown by the dark blue line and 5 randomly chosen samples are shown by the faint blue lines. The true value is shown by the red line and the 95\% credible interval is shown in light blue.}
    \label{sim_full_latent_iFFBS}
\end{figure}

The parameter trace plots and histograms are shown in Figures \ref{sim_full_parameters_iFFBS_trace} and \ref{sim_full_parameters_iFFBS} respectively. Details are given in Table \ref{sim_full_table_iFFBS}. Clearly from the trace plots we see that the mixing is better than when using the reversible-jump algorithm. We see that the true value of each parameter lies within the 95\% credible interval of the sample. Similarly to the results with the Rippler algorithm, the median of the sample is very close to the true value for $\beta_G$ and $\beta_H$, but significantly further away for $\delta_A$ and $\delta_S$. The central 95\% credible interval for the number of individuals colonised at each point in time is shown in Figure \ref{sim_full_latent_iFFBS}, with the true value mostly lying within the interval. We can see from the trace plots of the parameters that the mixing is at least as good as when using the Rippler algorithm.

\begin{table}[H]
    \centering
    \small
    \caption{Parameter estimates for simulated 3GC-R \textit{E. coli} test data using the iFFBS algorithm for inference.}
    \begin{tabular}{ c  c  c  c  c}
        \hline Parameter &  Definition & True & Median & 95\% CrI 
        \\ \hline $\beta_G$ & global transmission rate (week$^{-1}$) & 0.1 & 0.0927 & (0.0602, 0.126) 
        \\ $\beta_H$ & relative within-household transmission & 1.5 & 1.81 & (1.15, 3.14) 
        \\ $\delta_A$ & age covariate effect (year$^{-1}$) & 0 & -0.00425 & (-0.0132, 0.00438) 
        \\ $\delta_S$ & sex covariate effect & 0 & 0.230 & (-0.111, 0.549) 
        \\ \hline
    \end{tabular}
    \label{sim_full_table_iFFBS}
\end{table}

